\documentclass{article}

\usepackage{graphicx}
\usepackage{amsmath}
\usepackage{amsthm}
\usepackage{amssymb}
\usepackage[margin=1in]{geometry}
\usepackage{xcolor}
\usepackage{bm}
\usepackage{hyperref}
\usepackage{subcaption}
\usepackage{algorithm}
\usepackage{algpseudocode}
\usepackage{booktabs}
\usepackage{siunitx}
\usepackage{adjustbox}
\usepackage[numbers,sort&compress]{natbib}
\usepackage{enumitem}
\usepackage{float}

\newtheorem{remark}{Remark}

\renewcommand{\textcolor}[2]{#2}

\title{A comparison of Markov Chain Monte Carlo algorithms for Bayesian inference of constitutive models}
\author{Aricia Rinkens$^{1}$, Rodrigo L. S. Silva$^{1,2}$, Erik Quaeghebeur$^{2}$, Nick Jaensson$^{1}$,
\\
Clemens Verhoosel$^{1}$
\\
\\
$^{1}$Department of Mechanical Engineering
\\
$^{2}$Department of Mathematics and Computer Science
\\
\\
Eindhoven University of Technology, The Netherlands}
\date{}

\begin{document}

\maketitle

\begin{abstract}
    Employing Bayesian inference to calibrate constitutive model parameters has grown substantially in recent years. Among the available techniques, Markov Chain Monte Carlo (MCMC) sampling remains one of the most widely used approaches for estimating the posterior distribution. Nevertheless, the selection of a specific MCMC algorithm is often driven by practical considerations, such as software availability or prior user experience. To support sampler selection, we present a comparison of three prominent samplers in the context of two distinct physical systems: a thermal conduction system and a viscous flow system. Calibration data are obtained through tailor-made experimental setups. \textcolor{black}{We use the Kullback-Leibler (KL) divergence, which quantifies the statistical distance between the sampled posterior and the reference ('true') posterior, as a measure of convergence to compare the performance of the following MCMC sampling methods: the Metropolis-Hastings (MH) sampler, the Affine Invariant Stretch Move (AISM) sampler, and the No-U-Turn Sampler (NUTS).} We study how this metric correlates to heuristic indicators such as the Gelman-Rubin diagnostic and the effective sample size. In addition, we assess the samplers’ computational effort in terms of required number of model evaluations. \textcolor{black}{Based on the results, we find that the heuristic convergence and performance indicators provide a good qualitative measure for KL-divergence for both systems.} Regarding computational effort, the NUTS is net beneficial for the viscous flow system, as the high effective sample size outweighs the additional effort required for gradient-based proposal generation. For the thermal conduction system, which involves more expensive model evaluations, \textcolor{black}{the NUTS is not advantageous}. Thus, the computational efficiency of gradient evaluations is an important argument in sampler selection.
\end{abstract}

\section{Introduction}\label{sec:1}
\textcolor{black}{
In science and engineering, mathematical and physical models are used as simplified representations of real‑world systems. These models originate from fundamental conservation laws such as the conservation of mass and momentum. To make such models complete, these governing laws must be complemented with parametrized relations that describe material behavior, known as constitutive relations. In fluid mechanics, they link stress to strain rate; in solid mechanics, they relate stress to strain; and in thermal analysis, they connect heat flux to temperature gradients. Unlike the conservation laws, which hold universally, constitutive relations introduce uncertainty into the modeling process.}

The uncertainty surrounding constitutive modeling is rooted in the absence of one universally applicable constitutive relation. The applicability of a constitutive relation is highly context-dependent, both through the material under consideration and through the imposed modeling requirements (accuracy, solution method, etc.). As a result, a plethora of constitutive relations is available, spanning a wide range of complexities in terms of the number of model parameters.

The selection and parameter determination of constitutive relations is typically based on well-defined laboratory experiments. For example, shear-flow based rheological measurements are used to determine the constitutive relations for viscous flows. Due to the ever-increasing complexity of engineering systems, the representativeness of the well-defined conditions under which the constitutive relations have been determined in the laboratory is subject to uncertainty. This can compromise the objectivity of the modeling process.

\textcolor{black}{
To address these challenges, the Bayesian inference framework offers an alternative approach that captures the inherent uncertainty in engineering systems in a comprehensive manner. It does so by treating model parameters as probabilistic quantities and updating prior beliefs about these parameters as new data become available} \cite{Kaipio2006, Ozisik2018}. This provides a systematic approach to combine expert opinions with observational evidence. Bayesian inference also quantifies the probability that a model can explain an experimental observation, thereby objectifying constitutive model selection \cite{mackay2003information}.

Although Bayesian inference is well-established in probabilistic modeling \cite{Kaipio2006}, it is still in its infancy in the context of constitutive modeling. In recent years, it has received an increase in interest, for example, for the determination of the conductivity in thermal problems \cite{Silva-unpublish, ramos2022simultaneous}, for the determination of rheological parameters in non-Newtonian fluid mechanics \cite{Rinkens2023, Kim2019}, and for constitutive modeling in elastic materials \cite{Rappel2020}, biomaterials \cite{Akintunde2019} and elasto-viscoplastic materials \cite{Chakraborty2021}.

Bayes' rule gives a probability distribution for the parameters conditional on experimental data -- referred to as the \emph{posterior} -- as the normalized product of the \emph{prior} probability distribution for the parameters and the \emph{likelihood} that a particular parameter set can explain the data. Although evaluating the posterior is conceptually straightforward, in practice it is often problematic. Namely, analytical solutions can only be obtained for specific choices of the prior and posterior \cite{Kaipio2006}. Also, direct evaluation of the posterior on structured grids is subject to the curse of dimensionality and becomes computationally impractical when more than a few parameters are to be inferred \cite{Kaipio2006}. Therefore, both analytical and direct evaluation methods are very limiting in the context of constitutive modeling.

Markov Chain Monte Carlo (MCMC) methods, which evaluate Bayes' rule through probabilistic sampling of the (high-dimensional) parameter domain, have been proven to be viable in the context of constitutive modeling \cite{honarmandi2019uncertainty, adeli2019parameter}. On account of their success, different variants of MCMC methods have been developed over the past decades \cite{Metropolis1953, Hastings1970, Goodman2010, Hoffman2014}, each with their own characteristics. Often, researchers choose an MCMC sampler based on pragmatic arguments, such as software availability and experience. This pragmatism may render their choice sub-optimal in terms of sampler robustness, accuracy and efficiency.

The most prominent MCMC samplers have been documented well in the literature, including feature comparisons and benchmark studies \cite{Cowles1996, ballnus2017comprehensive, valderrama2019mcmc, nishio2019performance, Roy2020}. In terms of comprehensive comparisons, in particular the recent work of \citet{Allison2014} is noteworthy. By considering the number of likelihood evaluations as the primary performance indicator, a range of prominent MCMC samplers is assessed based on a synthetically manufactured problem.

While the insights from such comparative studies are a valuable source of information for selecting an MCMC sampler, the settings in which comparisons have been performed differ from the application scenarios considered in constitutive modeling in two ways. First, the considered synthetically generated data sets are not fully representative of the experimental data acquired for constitutive model determination, especially in terms of noisiness and lack of structure in the data sets. Second, the models considered in constitutive modeling are in general significantly more complex than those considered in the current comparison studies, both in terms of computational effort and in terms of model structure (which may, e.g., impact the possibility to evaluate parametric gradients).

Our contribution aims to further support the MCMC solver selection process in the context of engineering problems, especially for constitutive modeling. We systematically compare a selection of prominent MCMC solvers in the context of two real-world physical systems. Each system pertains to a different physical domain and has its own distinct characteristics in terms of experimental data and mathematical-physical model. The first experiment considers the heating of a column of paraffin wax, for which the accompanying model involves a computationally (relatively) demanding finite element solution procedure. The inference problem pertains to both the thermal conductivity property of the paraffin wax and to the heat transfer coefficients associated with the surfaces of the column. The second experiment considers a Newtonian fluid being compressed between two parallel plates, for which, under non-restrictive assumptions, a semi-analytical solution can be obtained \cite{Engmann2005, Rinkens2023}. The inference problem here pertains to the fluid viscosity and experimental uncertainties such as the loading conditions. The data gathered using the tailored setups are made available to enable reproduction \cite{rinkens_2025_15784954}.

For our comparison, we investigate three prominent MCMC samplers: the Metropolis-Hastings (MH) sampler \cite{Metropolis1953, Hastings1970}, the Affine Invariant Stretch Move (AISM) sampler \cite{Goodman2010} and the No-U-Turn Sampler (NUTS) \cite{Hoffman2014}. The selection of these samplers is based on our experience in the context of a range of problems in science and engineering. In line with standard practices, we base our comparison study on quantitative measures. Next to heuristic convergence and performance indicators, such as the Gelman-Rubin ($\hat{R}$) diagnostic, we also study the convergence of these samplers through the Kullback--Leibler (KL) divergence to quantify the statistical distance between the sampled and the `true' posterior \cite{Kullback1951}. Through our comparison methodology we optimize the objectivity of our study, although we fully acknowledge that, in complex comparisons like these, there is always an inherent degree of subjectivity. Therefore, in general, our results are not intended to be interpreted as direct advice for the selection of a particular sampler. Instead, our study aims to clarify the various considerations and their relative importance in the context of constitutive modeling.

We structure this paper as follows. In \autoref{sec:2} we discuss the modeling procedure and experiments of the thermal conduction and viscous flow systems. The input for the Bayesian inference framework for both physical systems, being the likelihood function and the prior, is considered in \autoref{sec:3}. In \autoref{sec:4} we introduce the convergence and performance indicators, followed by the MCMC sampling methods. We present the posterior results of the thermal conduction and viscous flow systems in \autoref{sec:5}, where we also compare the three MCMC sampling methods using the convergence and performance indicators. In \autoref{sec:6} we present our conclusions and recommendations.
\section{Physical systems}\label{sec:2}
To study the performance of the samplers on realistic data sets we consider two physical systems, viz., a thermal conduction system and a viscous flow system. In this section we introduce the setups and experiments, as well as the models used to mimic their behavior. In \autoref{sec:thermalsystem} we first introduce the thermal conduction system, after which the viscous flow system is discussed in \autoref{sec:viscoussystem}.

\subsection{Thermal conduction system}\label{sec:thermalsystem}
\begin{figure}
    \centering
    \includegraphics[scale=0.6]{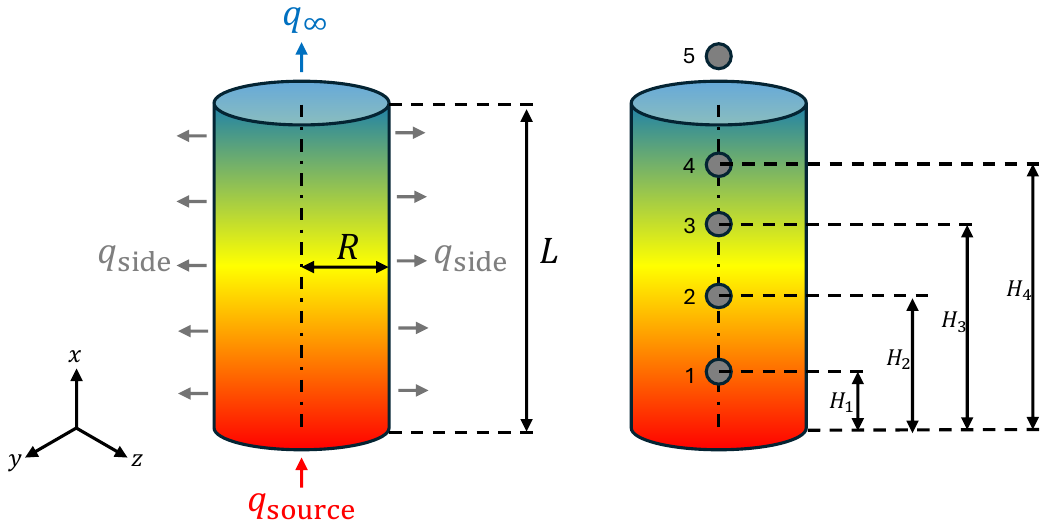}
    \caption{Schematic of the thermal conduction system.
    The temperature sensors are denoted by the numbers 1 to 5.}
    \label{fig:thermal_schem}
\end{figure}

We consider a heat conduction problem in which a column of conductive material is heated from the bottom (\autoref{fig:thermal_schem}). The radius and height of the column are denoted by $R$ and $L$, respectively. The base plate is made of a highly conductive material, of which the temperature is regulated at $T_{\rm source}$. On the sides, the column is insulated, while at the top the conductive material is in contact with the air, which has a time-dependent ambient temperature $T_{\infty}(t)$, where $t$ denotes time. The evolution of the temperate through the column is represented by $T(\bm{x},t)$, where $\bm{x}$ is the position vector.

\subsubsection{Experiment}
\begin{figure}
    \centering
    \includegraphics[width=0.35\linewidth]{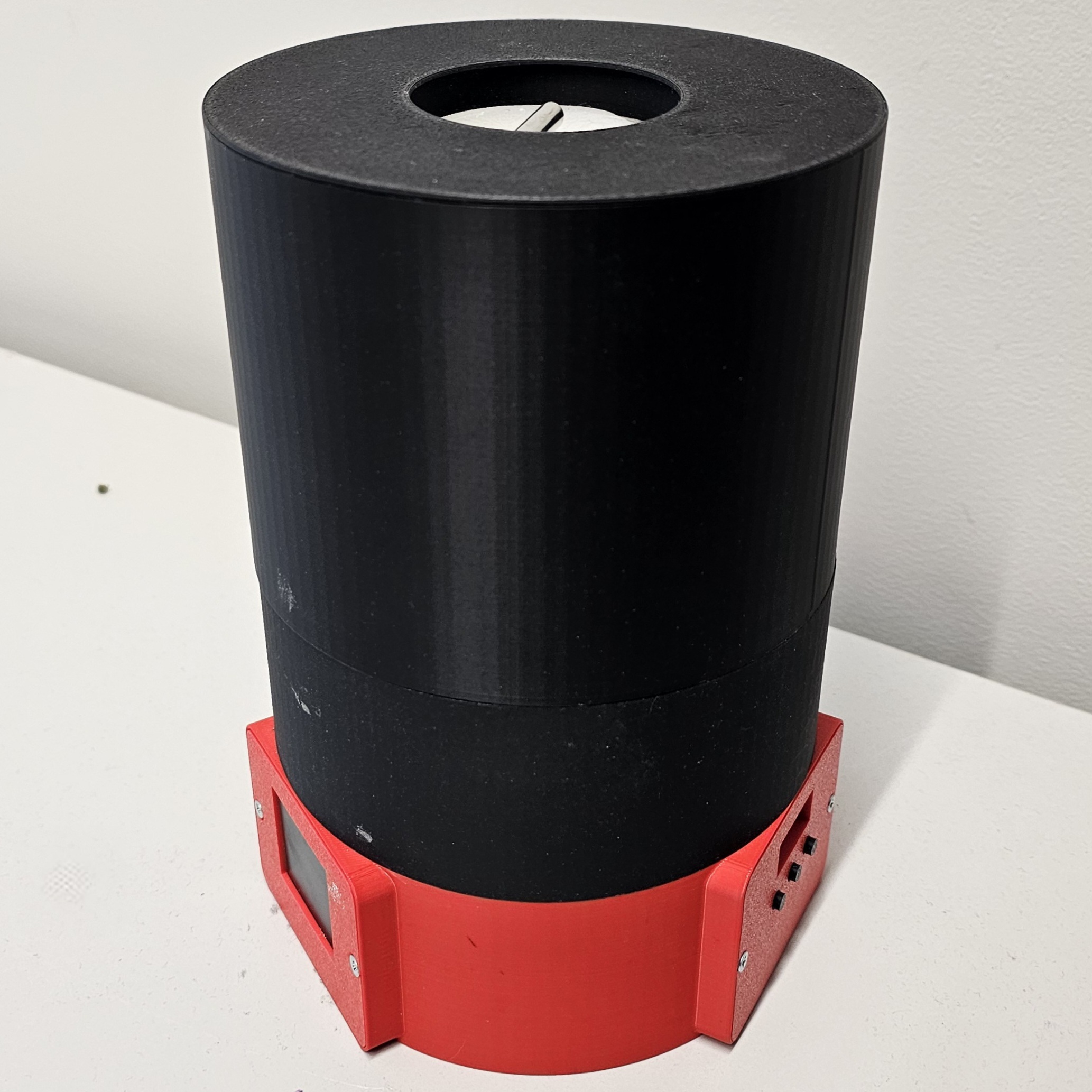}
    \hfil
    \includegraphics[width=0.55\linewidth]{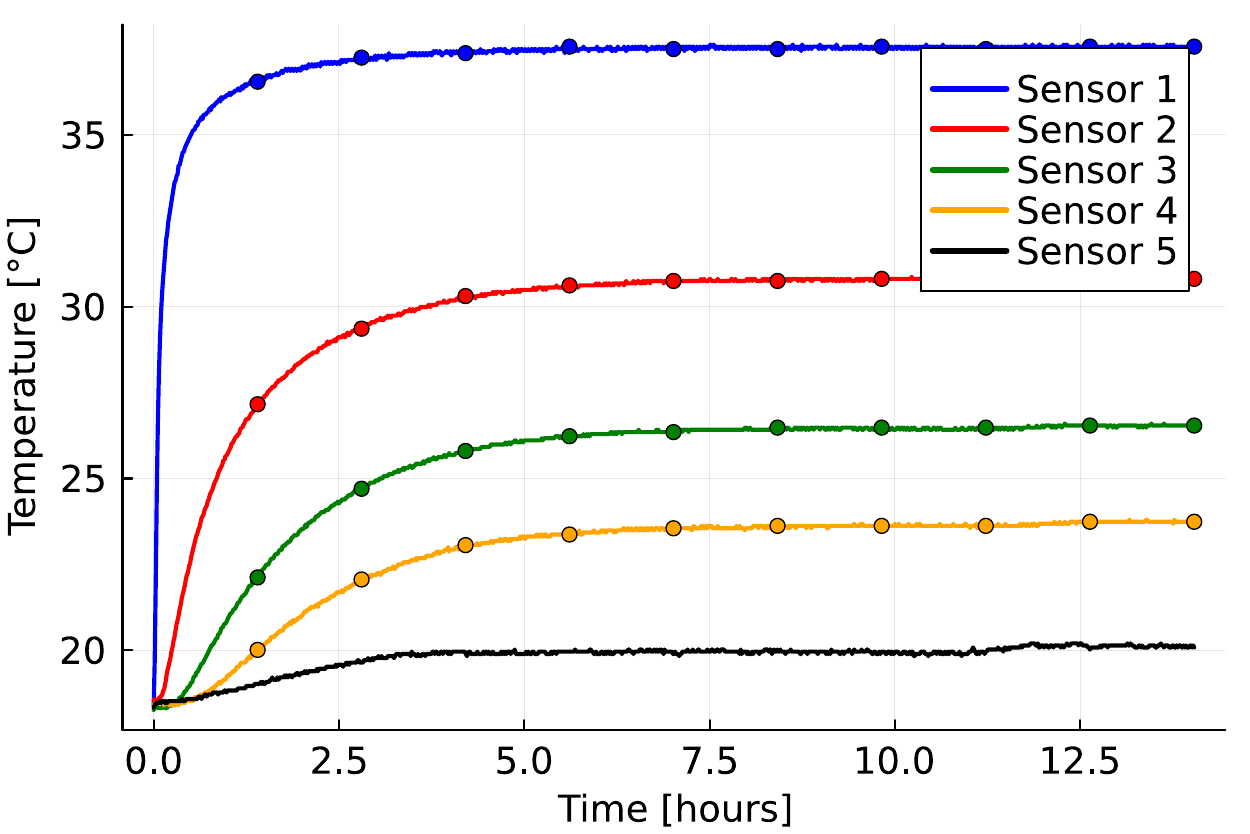}
    \caption{(\textbf{left}) Experimental thermal conduction setup.
    (\textbf{right}) Temperature measurements obtained with the five sensors.}
    \label{fig:thermal_experiment_a}
\end{figure}

In \autoref{fig:thermal_experiment_a} we show our experimental setup, which has been manufactured using fused deposition modeling (FDM) printing with polylactic acid (PLA). A column of paraffin wax with a radius of \SI{28.6}{\mm} and a height of \SI{93.0}{\mm} is contained within a thin polyvinyl chloride (PVC) tube. The brass base plate is heated by a \SI{60}{\W} Peltier element which is regulated by a W1209 control module with a temperature sensor integrated in the base plate.

The PVC tube is insulated on the outside by a thick layer of styrofoam. Throughout the experiment, the temperature of the paraffin is kept below the melting temperature of approximately $58 - \SI{60}{\degreeCelsius}$, such that \textcolor{black}{the} material does not exhibit phase changes. The temperature inside the paraffin column is registered by four DS18B20 temperature sensors, positioned along the cylinder axis. Their respective distances from the base plate are $H_1 = \SI{5.0}{\mm}$, $H_2 = \SI{25.8}{\mm}$, $H_3 = \SI{45.0}{\mm}$ and $H_4 = \SI{66.5}{\mm}$. An additional temperature sensor is positioned right above the paraffin to monitor the ambient air temperature. All sensors were calibrated prior to assembly.
        
The temperatures of all sensors are recorded approximately for 14 hours, with a total of \num{32057} measurements per sensor. We extract a minimal experimental data set from these measurements by selecting 10 (approximately) equally spaced measurements for each sensor (\autoref{fig:thermal_experiment_a}). The temperature arrays corresponding to each of the four sensors submerged in the paraffin are concatenated into a single observation array, $\bm{y}$, containing a total of 40 temperature observations. The registered temperature of the ambient is used as input for the model. The experiment was reproduced four times, but with minor variations in the ambient temperature ($\pm \SI{2}{\degreeCelsius}$). The measurement with the smallest variation in ambient temperature is used in the remainder of this work, as this is convenient for the interpretation of the inference results.

\subsubsection{Model}
We model the temperature evolution in the paraffin by means of the unsteady heat equation \cite{ozisik1993}, which in the absence of internal heat sources and sinks reads
\begin{equation}
    \rho c_p \frac{\partial T(\bm{x},t)}{\partial t}  + \nabla \cdot \boldsymbol{q}(\bm{x},t) = 0, \label{eq:thermal:balancelaw}
\end{equation}
where $\rho$ and $c_p$ are the mass density and specific heat of the paraffin, respectively, and $\bm{q}$ is the heat flux vector.
We use Fourier's law to model the constitutive behavior of the solid paraffin as 
\begin{equation}
    \bm{q}(\bm{x},t) = -k \nabla T(\bm{x},t), \label{eq:thermal:constitutive}
\end{equation}
where $k$ is the isotropic and homogeneous thermal conductivity. The physical properties $\rho$, $c_p$, and $k$, are assumed constant.

We denote the magnitude of the heat flux vectors perpendicular to the boundaries of the paraffin as $q_\text{source}$, $q_\infty$ and $q_\text{side}$ (\autoref{fig:thermal_schem}). The heat exchange at the boundaries is modeled with Newton's law of cooling as
\begin{subequations}
\begin{align}
    q_\text{source}(\bm{x},t) &= h_\text{source}[ T(\bm{x},t) - T_{\rm source} ],\\
    q_\infty(\bm{x},t) &= h_{\infty}[ T(\bm{x},t) - T_{\infty}(t) ],\\
    q_\text{side}(\bm{x},t) &= h_\text{side}[T(\bm{x},t) - T_{\infty}(t) ],
\end{align}
\end{subequations}
where $h_{\rm source}$, $h_{\infty}$ and $h_{\rm side}$ are the heat transfer coefficients corresponding to the brass--paraffin interface, the paraffin--air interface and the insulation material, respectively.
        
Since the paraffin column is well insulated and the top and bottom boundary conditions are applied uniformly, the temperature distribution in a cross-section of the column can be assumed to be homogeneous. Integration of \autoref{eq:thermal:balancelaw} over the cross-section then yields the one-dimensional heat equation
\begin{equation}
    \rho c_p \frac{\partial T(x,t)}{\partial t}  - k \frac{{\partial}^2 T(x,t)}{\partial x^2} +  \frac{2 h_{\rm side}}{R}[T(x,t) - T_{\infty}(t)] = 0,
    \label{eq:thermal:balancelawoneD}
\end{equation}
where the temperature field, $T(x,t)$, now only depends on the vertical coordinate $x$ and time $t$.
This differential equation is complemented by the boundary conditions
\begin{subequations}
    \begin{align}
        k \frac{\partial T(0,t)}{{\partial}x} & = h_{\rm source}[T(0,t) - T_{\rm source}], \label{eq:therma11DBC1}
        \\
        k \frac{\partial T(L,t)}{{\partial}x} & = h_{\infty}[T_{\infty}(t) - T(L,t)], \label{eq:therma11DBC2}
    \end{align}
\end{subequations}
and the initial condition $T(x,0) = T_{\infty}(0)$.
        
We discretize \autoref{eq:thermal:balancelawoneD} in space using uniformly-spaced linear finite elements. Based on a grid convergence analysis of the spatial domain, which is omitted here for brevity, we employ 25 finite elements. We use backward-Euler time integration with a time step of $\Delta t = \SI{20}{\s}$, which satisfies the mesh-size dependent bounds of \citet{szabo2008discretization}. Since the time instances of the solution strategy do not coincide with the sampling intervals of the experimental data, we linearly interpolate the model results at the sampling times.

\subsection{Viscous flow system}\label{sec:viscoussystem}
\begin{figure}
\centering
\includegraphics[width=0.35\textwidth]{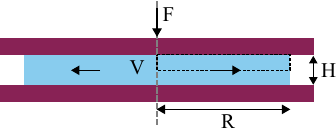}
\caption{Schematic of the squeeze flow, where two parallel plates (dark-red) compress a fluid (light blue). The model domain is denoted by the dashed black box. \textcolor{black}{Reproduced from \citet{rinkens2025comparison}.}\label{squeezeflow_schem}}
\end{figure}

\textcolor{black}{For the viscous flow system we consider the squeeze flow problem originally studied by \citet{Rinkens2023}, in which a fluid is compressed between two parallel plates(\autoref{squeezeflow_schem}).} The volume of the fluid, which is assumed to be incompressible, is denoted by $V$. The force exerted on the system, which includes the weight of the top plate, is denoted by $F$. Under the action of this force, the fluid front moves outward, i.e., the radius $R(t)$ increases, and the layer height $ H(t) = V / (\pi R(t)^2)$ decreases. The moving fluid front is in contact with the ambient air. Capillary forces at this fluid-air interface result in a pressure jump.

\subsubsection{Experiment}
\begin{figure}
\centering
\subfloat{\includegraphics[width=0.35\textwidth]{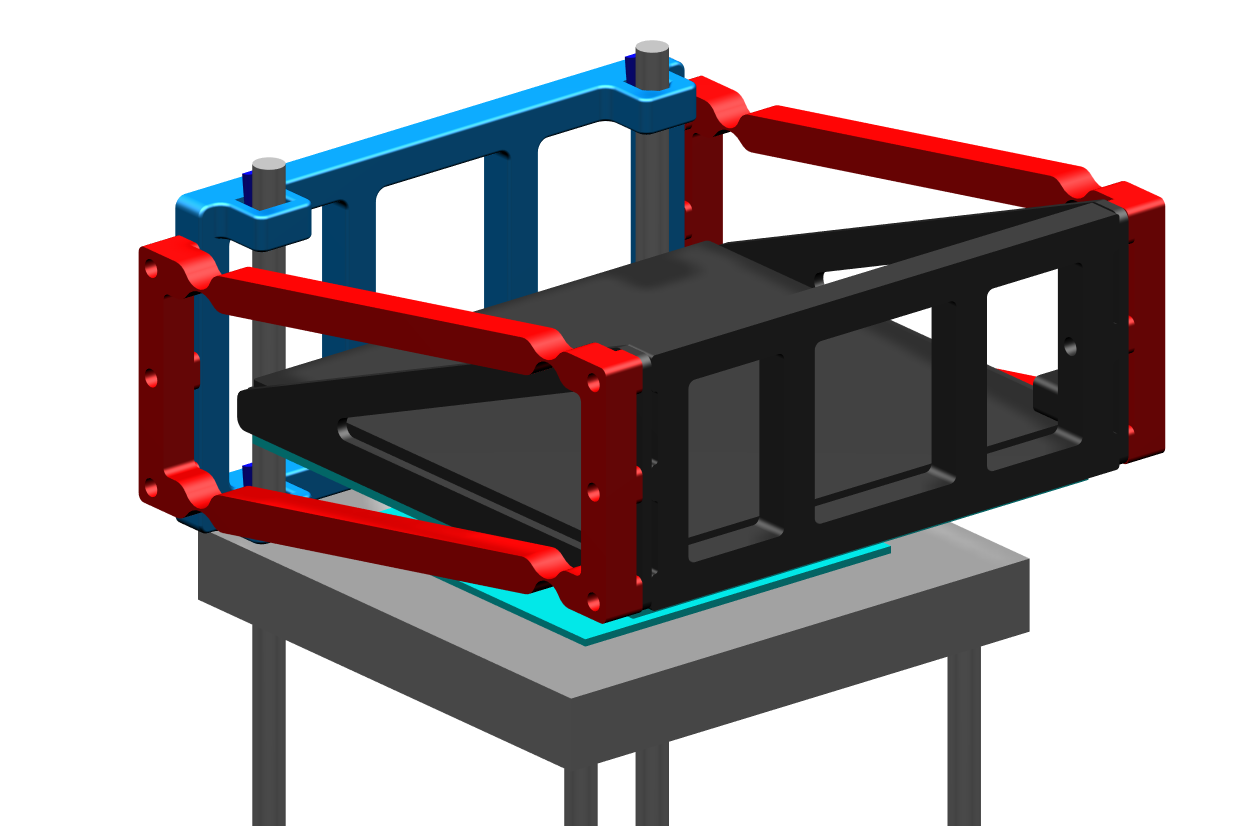}}
 \hfill
\subfloat{\includegraphics[width=0.25\textwidth]{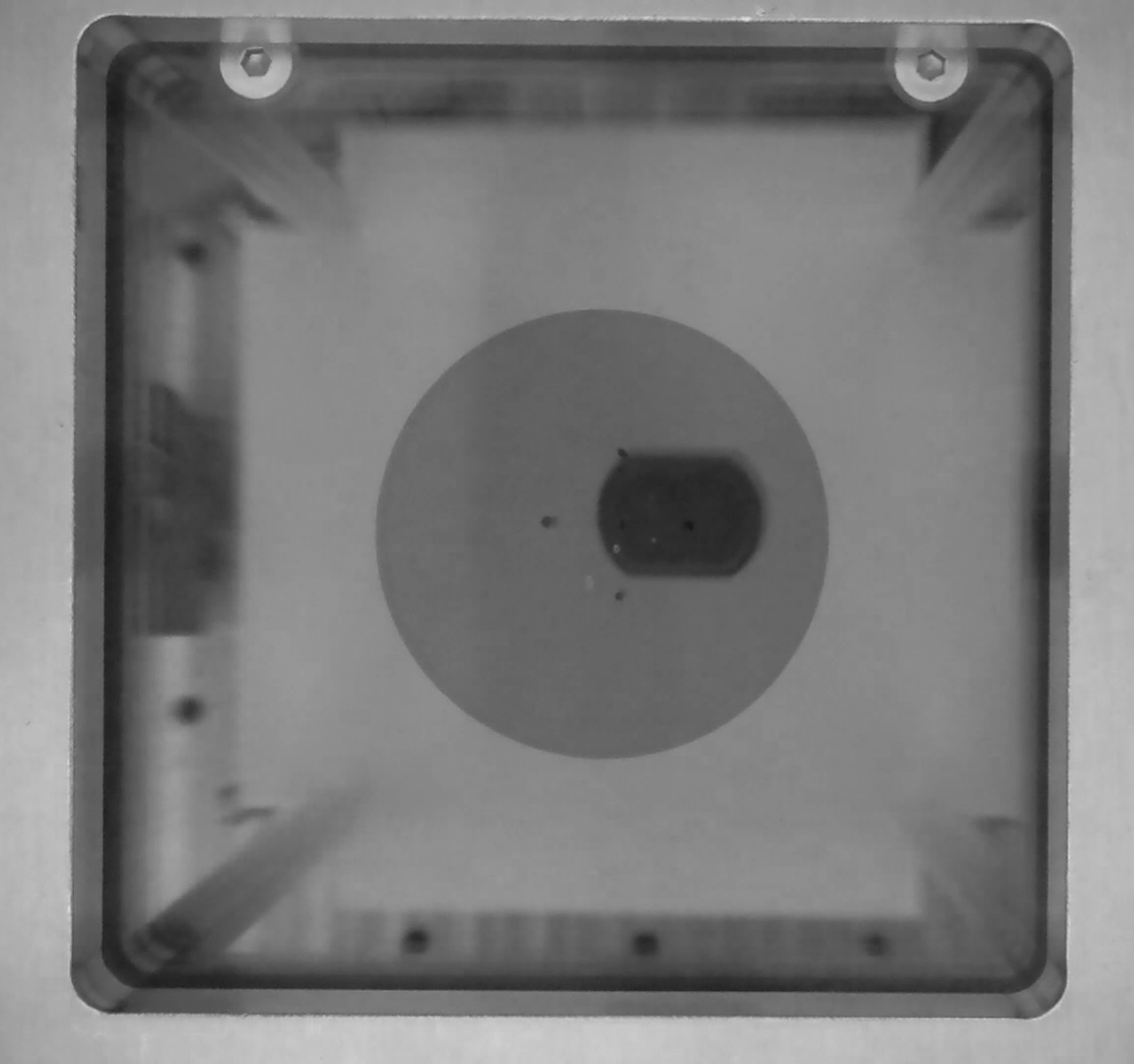}}
\hfill
\subfloat{\includegraphics[width=0.35\textwidth]{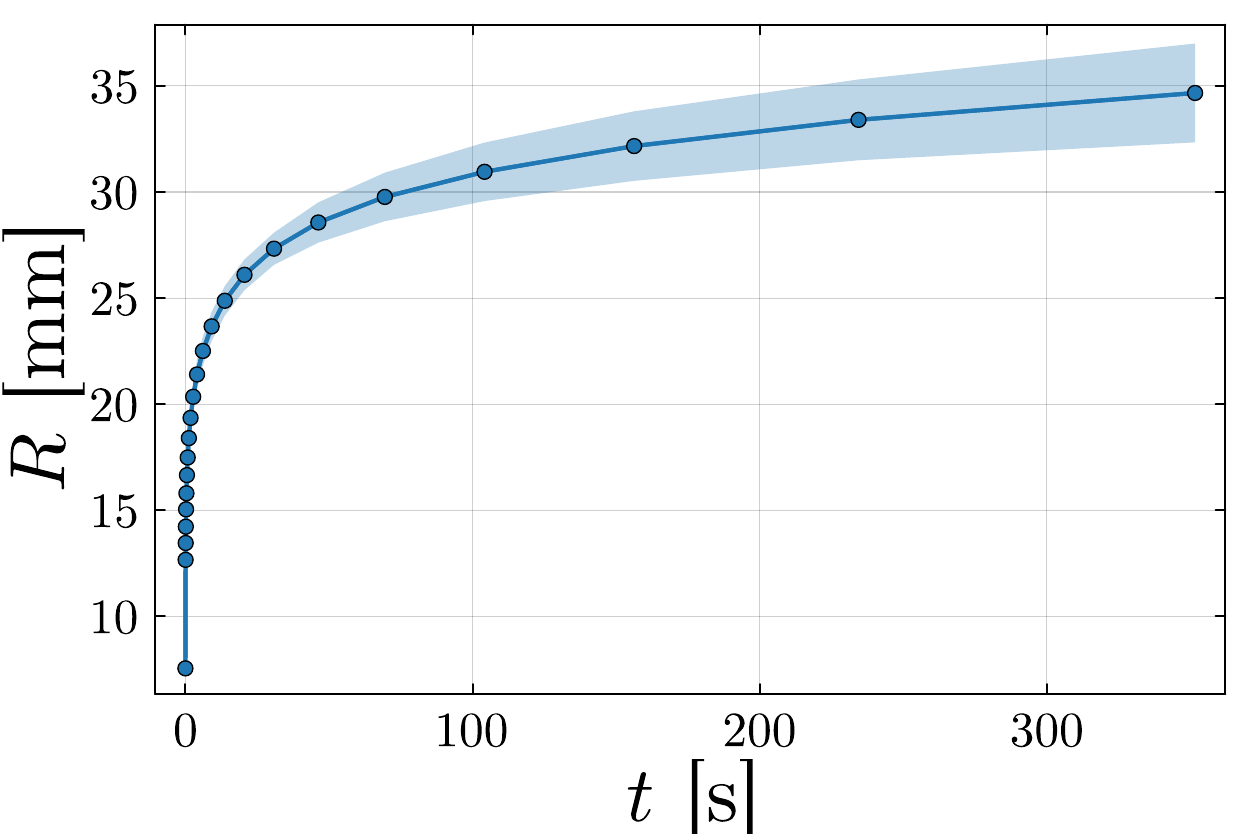}}
\caption{(\textbf{left}) Tailored experimental squeeze flow setup consisting of two transparent plates (cyan) and parallel leaf springs (red). (\textbf{middle}) An image of the fluid front captured by a camera that is positioned beneath the bottom plate. (\textbf{right}) The aggregated data obtained from repeated experiments, showing the mean value and the 95\% confidence interval. \textcolor{black}{Reproduced from \citet{rinkens2025comparison}}.\label{SF_setup_exp}}
\end{figure}

In \autoref{SF_setup_exp} we show \textcolor{black}{the experimental setup of \citet{Rinkens2023}}. The plates between which the fluid is compressed are made of transparent PMMA, allowing for observation of the fluid front \textcolor{black}{using a camera positioned underneath the bottom plate}. The parallel leaf spring structure in which the top plate is embedded is made of PLA. This structure has been designed such that the stiffness of the leaf springs is negligible compared to the other forces acting on the system. The fluid radius is extracted from these images using an image processing algorithm. Details regarding the setup and the image processing are given by \citet{Rinkens2023}.

\textcolor{black}{In this work, we use the glycerol experimental data previously presented in \citet{Rinkens2023} as case I. In this dataset, the evolution of the fluid front was recorded over a five‑minute interval. The raw video data were subsequently processed by extracting frames at exponentially increasing time instances, defined by $t_{i+1}=t_i + \Delta t_i$ with $\Delta t_i=1.5^i\Delta t_0$. The radii corresponding to these measurement points are collected in the observation array $\boldsymbol{y}$. The resulting measurements are presented in \autoref{SF_setup_exp}, where the mean value and associated confidence interval of 95\% were obtained by repeating the experiment ten times.}


\subsubsection{Model}
To model the viscous flow \cite{bird2002} between the parallel plates we consider the mass and momentum balance in absence of inertia, written as, respectively,
\begin{subequations}\label{eq:fluid:balancelaws}
    \begin{align}
        \nabla \cdot \boldsymbol{v} & = 0, \label{ch3_mass_bal_1}\\
        \nabla p - \nabla \cdot \boldsymbol{\tau} & = \mathbf{0}, \label{ch3_mom_bal_1}
    \end{align}
\end{subequations}
where $\boldsymbol{v}$ is the velocity field, $p$ the pressure field and $\boldsymbol{\tau}$ the extra stress tensor, which are all evolving in time. To relate the fluid's response to its deformation, we assume a Newtonian constitutive relation as, $\boldsymbol{\tau} = 2 \eta \boldsymbol{D}$, where $\eta$ is the viscosity and $\boldsymbol{D}=\frac{1}{2}(\nabla \boldsymbol{v} + (\nabla \boldsymbol{v})^T)$ is the rate-of-deformation tensor, i.e., the symmetric gradient of the velocity field. No-slip conditions are assumed on the plate boundaries, whereas the jump in pressure on account of capillary forces at the moving fluid front is set to
\begin{equation}
    \Delta p = 2 \gamma\kappa,
\end{equation}
where $\gamma$ is the surface tension of the fluid and $\kappa$ the curvature of the fluid front. The latter can also be defined as $\kappa=\frac{2\alpha}{H}$, with $\alpha \in [0,1]$, where the boundaries represent a straight interface ($\alpha=0$) and a maximum curvature interface ($\alpha=1$).

Assuming the solution to be axisymmetric around the vertical axis (gray line in \autoref{squeezeflow_schem}) and lubrication flow conditions to apply on account of the height of the fluid layer being much smaller than the radius of the fluid layer ($H\ll R$) \cite{szeri2010fluid}, the balance laws \eqref{eq:fluid:balancelaws} reduce to
\begin{subequations}
\begin{align}
    \frac{1}{r}\frac{\partial}{\partial r}\left(rv_r\right) - \frac{\partial v_z}{\partial z} & = 0, \\
    \frac{\partial}{\partial z}\tau_{rz} & = \frac{\partial p}{\partial r},
\end{align}
\end{subequations}
with $r$ and $z$ referring to the radial and vertical coordinate, respectively. For a Newtonian fluid the velocity of the upper plate $\dot{H}$ can then be expressed as \cite{Rinkens2023} 
\begin{equation}\label{eq:hdotN}
   \dot{H} = \frac{8}{3} \left( -\frac{FH^3}{4\pi \eta R^4} + \frac{\kappa \gamma H^3}{2\eta R^2} \right).
\end{equation}
Since the radius is directly related to the height of the fluid layer by $R(t) =\sqrt{V / (\pi H(t))}$, \autoref{eq:hdotN} is a nonlinear first-order differential equation. Using the initial condition $H(t=0)=H_0$, the forward Euler method \cite{Biswas2013AReview} is used to integrate this equation in time.
\section{Bayesian uncertainty quantification}\label{sec:3}
The idea of Bayesian uncertainty quantification is to calibrate the parameters of a model to a data set by determining their conditional probability distribution. To formalize this concept, we define a $p$-dimensional parameter domain as $\boldsymbol{\Theta}=\Theta_1 \times \Theta_2 \times  ... \times \Theta_p$ and $\boldsymbol{\theta} \in \boldsymbol{\Theta}$ as a specific parameter setting. Given \emph{prior} information on the parameters in the form of a probability distribution, $\textcolor{black}{\pi_{\rm prior}}(\boldsymbol{\theta})$, and a \emph{likelihood} function, $L(\boldsymbol{\theta}|\boldsymbol{y})$, that expresses the probability of observing the measured data, $\boldsymbol{y}$, given a certain parameter setting, the \emph{posterior} probability, i.e., the parameter probability given the data, follows from Bayes' rule as
\begin{equation}\label{eq:bayesrule}
        \pi_{\rm post}(\boldsymbol{\theta}|\boldsymbol{y}) = \frac{L(\boldsymbol{\theta}|\boldsymbol{y}) \textcolor{black}{\pi_{\rm prior}}(\boldsymbol{\theta})}{\pi_\text{evid}(\boldsymbol{y})},
\end{equation}
where $\pi_\text{evid}(\boldsymbol{y}) = \int_{\boldsymbol{\varTheta}}L(\boldsymbol{\theta}|\boldsymbol{y}) \textcolor{black}{\pi_\text{prior}}(\boldsymbol{\theta}) \,{\rm d}\boldsymbol{\theta}$ is a normalization constant referred to as the model \emph{evidence}.

The definition of the prior distribution and the likelihood function is an integral part of Bayesian uncertainty quantification. The prior distribution is typically selected based on expert opinion or available information regarding the parameters. The likelihood function is typically associated with a 
probabilistic model for the \emph{error} function
\begin{equation}
    \boldsymbol{\nu}(\bm{\theta},\boldsymbol{y})  = \boldsymbol{y} - \boldsymbol{d}(\boldsymbol{\theta}),
\end{equation}
which represents a combination of observation and model errors. That is, by assigning a probability distribution to $\boldsymbol{\nu}$, the probability of observing the data, $\boldsymbol{y}$, given a model result $\boldsymbol{d}(\boldsymbol{\theta})$ is modeled by
\begin{equation}\label{eq:likelihood_general}
    L(\boldsymbol{\theta}|\boldsymbol{y}) = \pi_{\rm error}( \boldsymbol{y} - \boldsymbol{d}(\boldsymbol{\theta}) ).
\end{equation}
To avoid arithmetic underflow  problems, in our implementation we consider the log-likelihood function
\begin{equation}\label{eq:log-likelihood_general}
    l(\boldsymbol{\theta}|\boldsymbol{y}) = \ln\left(L(\boldsymbol{\theta}|\boldsymbol{y})\right),
\end{equation}
instead of the likelihood function itself. In the remainder of this section, the specific choices made for the prior information and (log-)likelihood function are discussed for both physical systems.

\subsection{Thermal system}
\begin{figure}
    \centering
    \includegraphics[width=0.4\linewidth]{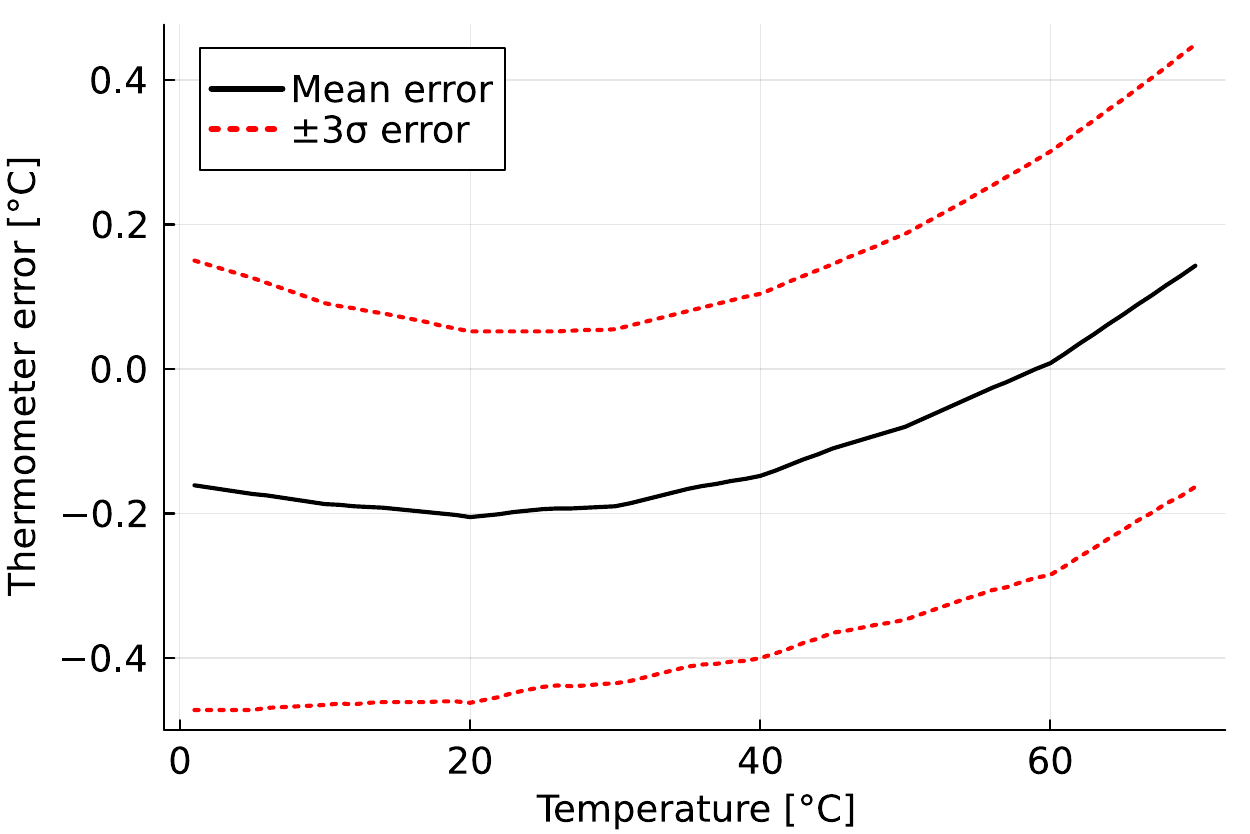}
    \caption{Temperature sensor error curve \cite{errorcurve2019}.}
    \label{fig:error-curve}
\end{figure}

For the thermal system we consider the calibration of the seven model parameters $\boldsymbol{\theta}=[k, \rho, c_p, \allowbreak h_{\rm source}, h_{\rm side}, h_{\infty}, T_{\rm source}]$ based on the experimental data $\boldsymbol{y}$. For each temperature measurement point we assume a Gaussian likelihood and set the variance based on the sensor error specification (\autoref{fig:error-curve}) to obtain the covariance matrix $\boldsymbol{\Sigma}_{\boldsymbol{y}}$, which is \textcolor{black}{assumed} diagonal on account of the temporal and spatial separation of the measurement points. The log-likelihood then follows as
\begin{equation}\label{eq:log-likelihood-thermal}
l\left(\boldsymbol{\theta}|{\boldsymbol{y}},\boldsymbol{\Sigma}_{\boldsymbol{y}} \right) \propto -\frac{1}{2} \left( \boldsymbol{d}(\boldsymbol{\theta}) - {\boldsymbol{y}} \right)^\mathrm{T} \left[ \boldsymbol{\Sigma}_{\boldsymbol{y}} \right]^{-1} \left( \boldsymbol{d}(\boldsymbol{\theta}) - {\boldsymbol{y}} \right).
\end{equation}

The prior distribution of all parameters except the source temperature are described using truncated normal distributions to ensure physically realistic values. The lower and upper truncation bounds are defined as 0.1 and 10 times the mean value, respectively.
The mean values and standard deviations for the density, specific heat and thermal conductivity of the paraffin wax are based on values reported in the literature \cite{kraiem2023thermophysical, sari2007thermal}. The standard deviations for these parameters are based on a 10\% coefficient of variation (the ratio of the standard deviation and the mean).
To obtain the mean values of the priors for the heat transfer coefficients, estimates are made based on a steady-state experimental result.
To acknowledge the significant uncertainties pertaining to this specification method, a coefficient of variation of 50\% is assigned to these priors.
The source temperature is described by a normal distribution, with the mean set to the temperature specified on the control unit.
The standard deviation is based on the tolerance specified for the controller (\SI{0.5}{\degreeCelsius}), which we equate to the 99\% confidence interval. The specific statistics of the prior distributions are listed in \autoref{tab:priorparamthermal}.
Note that the indicated statistical moments pertain to the distributions before truncation.

\begin{table} 
\centering
\caption{Statistics of the prior distributions for all model parameters of the thermal system.\label{tab:priorparamthermal}}
\begin{tabular}{rllccc}
\toprule
\multicolumn{2}{c}{Parameter} & Distribution & Mean & Std.~dev. & Coef.~of~var.~[\%] \\
\midrule
$k$ & [\si{W/(m.\degreeCelsius)}]		         & Truncated normal & 0.300	& 0.030  & 10.00 \\
$\rho$ & [\si{kg/m^3}]		                     & Truncated normal &  900.0	& 90.00  & 10.00 \\
$c_p$ & [\si{J/(kg. \degreeCelsius)}]		     & Truncated normal & 2500	& 250.0  & 10.00 \\
$h_{\rm source}$ & [\si{W/(m^2.\degreeCelsius)}] & Truncated normal & 100.0	& 50.00  & 50.00 \\
$h_{\rm side}$ & [\si{W/(m^2.\degreeCelsius)}]	 & Truncated normal & 1.000	& 0.500  & 50.00 \\
$h_{\infty}$ & [\si{W/(m^2.\degreeCelsius)}]	 & Truncated normal & 10.00  & 5.000  & 50.00 \\
$T_{\rm source}$ & [\si{\degreeCelsius}]		 & Normal                      & 40.00	& 0.190  & 0.475 \\
\bottomrule
\end{tabular}
\end{table}

\subsection{Viscous flow system}
For the considered Newtonian flow case we calibrate the six model parameters $\boldsymbol{\theta}=[F,V,R_0,\eta, \allowbreak \gamma,\alpha]$ based on the experimental data, $\boldsymbol{y}$. We assume a Gaussian likelihood and define the log-likelihood \eqref{eq:log-likelihood_general} based on ten squeeze flow measurements as
\begin{equation}\label{eq:log-likelihood_time}
l\left(\boldsymbol{\theta}|\boldsymbol{\mu}_{\boldsymbol{y}},\boldsymbol{\Sigma}_{\boldsymbol{y}} \right) \propto -\frac{1}{2} \left( \boldsymbol{d}(\boldsymbol{\theta}) - \boldsymbol{\mu}_{\boldsymbol{y}} \right)^\mathrm{T} \left[ \boldsymbol{\Sigma}_{\boldsymbol{y}} \right]^{-1} \left( \boldsymbol{d}(\boldsymbol{\theta}) - \boldsymbol{\mu}_{\boldsymbol{y}} \right),
\end{equation}
where $\boldsymbol{\mu}_{\boldsymbol{y}}$ and $\boldsymbol{\Sigma}_{\boldsymbol{y}}$ are the mean and covariance over the experiments.

The prior information for the parameters of interest is obtained through additional experiments:
\begin{itemize}
    \item The viscosity, $\eta$, is determined by measuring the simple shear response using steady rate sweep tests on a TA Instruments ARES rotational rheometer with a \SI{25}{mm} diameter cone-plate geometry. Using the mean and standard deviation of ten experiments, a log-normal prior distribution (Remark~\ref{rem:logtransform}) -- which acknowledges the non-negativity of the viscosity -- is constructed.
    \item The applied force, $F$, is measured using a spring suspension placed on the top part of the setup. We do not position additional weights on the setup, meaning that the applied force pertains to the own weight of the top part of the setup. We assign a normal prior distribution to the applied force and truncate it from below at zero to ensure physically realistic values. The mean and standard deviation of this distribution are based on ten measurements.
    \item The fluid volume, $V$, is measured by weighing a sample and multiplying it with the known density. A log-normal distribution based on ten measurements is used as a prior.
    \item The initial radius, $R_0$, is obtained through the squeeze flow experiment by extracting the initial frame and determining its radius. A log-normal distribution based on ten measurements is used as prior information.
    \item The surface tension, $\gamma$, is determined using a pendant drop experiment. Three measurements are used to construct a log-normal distribution.
\end{itemize}
Furthermore, based on the uncertainty propagation findings by \citet{Rinkens2023}, a beta distribution, $B(a,b)$, with shape parameters $a=3$ and $b=8$, is used for the curvature parameter $\alpha \in [0,1]$. The prior distributions are summarized in \autoref{tab:priorparam}.

\begin{remark}[Parameter log-transform]\label{rem:logtransform}
To ensure non-negativity of a parameter with a log-normal prior, $\theta$, in our implementation we represent it by its log-transform, $\hat{\theta}=\ln{(\theta)}$. This log-transformed parameter is then normally distributed, with a mean and standard deviation that can be computed from those of the original parameter \cite{Rinkens2023}.
\end{remark}


\begin{table} 
\centering
\caption{Statistics of the prior distributions for all model parameters of the viscous flow system.\label{tab:priorparam}}
\begin{tabular}{rllccc}
\toprule
\multicolumn{2}{c}{} & Distribution & Mean & Std.~dev. & Coef.~of~var.~[\%] \\
\midrule
$F$ & [\si{N}]                  & Truncated normal & 2.8      & 0.10     & 3.6  \\
$V$ & [\si{mL}]		            & Log-normal       & 0.20     & 0.022    & 11   \\
$R_0$ & [\si{mm}]	            & Log-normal       & 7.6      & 0.15     & 2.0  \\
$\eta$ & [\si{\pascal\second}]  & Log-normal       & 0.87     & 0.012    & 1.4  \\
$\gamma$ & [\si{\newton\meter}]	& Log-normal       & $4.54 \times 10^{-2}$  & $2.53 \times 10^{-3}$  & 5.57 \\
$\alpha$ &  [-]                 & Beta             & $2.73 \times 10^{-1}$  & $1.29 \times 10^{-1}$  & 47.1 \\
\bottomrule
\end{tabular}
\end{table}
\section{Markov Chain Monte Carlo sampling}\label{sec:4} 
The posterior distribution \eqref{eq:bayesrule} can be constructed using Markov Chain Monte Carlo (MCMC) sampling. For problems with more than a few parameters to be inferred, MCMC sampling is computationally the most practical inference method, in part because it does not require the normalization term (i.e., the evidence) to be calculated explicitly \cite{Kaipio2006}.

In Bayesian inference, MCMC sampling aims to explore the posterior distribution through a number of steps, resulting in a chain $\boldsymbol{\mathcal{C}}=\{ \boldsymbol{\theta}_n\}_{n=1}^{N}$ of length $N$, with $\boldsymbol{\theta}_n$ a vector containing the $p$ parameter values at step $n$. The next step in the chain, $\boldsymbol{\theta}_{n+1}$, is sampled from a transition distribution conditional to the current step, i.e., $\pi_{\rm trans}(\boldsymbol{\theta}_{n+1} | \boldsymbol{\theta}_{n})$, making the future steps in the chain independent of its history. If the transition distribution is properly defined and the chain is ergodic, the Markov chain converges to a limit distribution corresponding to the posterior. To design MCMC algorithms with a proper transition distribution, generally use is made of the global or detailed balance \cite{Kaipio2006,Gamerman2006}. 

In the remainder of this section, we introduce the metrics by which we compare sampler convergence and performance (\autoref{sec:mcmc:convergence}), as well as the MCMC samplers included in our comparison (\autoref{sec:mcmc:samplers}). 

\subsection{Convergence and performance indicators}\label{sec:mcmc:convergence}
We compare the convergence behavior of the considered sampling algorithms through heuristic criteria such as the Gelman-Rubin diagnostic, but also more rigorously through the Kullback-Leibler (KL) divergence. The latter is a statistical distance between the exact (reference) solution and the sampled solution. To compare the convergence criteria objectively, in \autoref{sec:geweke} we discuss the Geweke diagnostic \cite{Geweke1991} as a method to systematically exclude sampler start-up effects. The KL-divergence and heuristic criteria are subsequently discussed in \autoref{sec:kl} and \autoref{sec:heuristics}, respectively. The computational effort of the samplers is quantified by means of the number of model evaluations. This performance indicator choice is motivated and detailed in \autoref{sec:performance}.

\subsubsection{Burn-in removal using the Geweke diagnostic}\label{sec:geweke}
Markov Chain Monte Carlo sampling starts from an initial position (typically sampled from the prior distribution) and gradually moves toward the limit distribution, in this case the posterior. The path covered between the initial position and the sampling of the posterior is referred to as the burn-in period. This burn-in period, which depends on the initial position, is removed from the Markov chain before it is used for further statistical analysis. In the remainder of this section, we denote the Markov chain with $B$ steps removed by $\boldsymbol{\mathcal{C}}(B)=\{ \boldsymbol{\theta}_n\}_{n=B+1}^{B+N}$, so that the final chain is of size $N$.

To systematically discard the burn-in period for the samplers that do not automatically remove it, we use the Geweke diagnostic \cite{Geweke1991}. This method compares the first 10\% to the last 50\% of the chain for each parameter, $\mathcal{C}_i(B)$, using
\begin{align}\label{eq:zpscore}
    N^{\text{eff}}_{i,q} = \frac{N_{i,q}}{1+2 \sum_{t=1}^T \hat{\rho}_{i,t}},
    &&
    z_i(B) = \frac{\mu_{i,10} - \mu_{i,50}}{\sqrt{\frac{\sigma_{i,10}^2}{N^{\text{eff}}_{i,10}}+\frac{\sigma_{i,50}^2}{N^{\rm eff}_{i,50}}}},
    &&
    p_i(B) = 1 - \frac{2}{\sqrt{\pi}} \int_0^{|z_i|/\sqrt{2}} e^{-t^2}\,\mathrm{d}t.
\end{align}
\textcolor{black}{Here, $N^{\text{eff}}_{i,q}$ denotes the effective sample size computed from a fraction $q$ of the chain.
For $q=10$, this corresponds to the first 10\% of the samples, whereas for $q=50$, it corresponds to the last 50\% of the samples.
The effective sample size makes use of the autocorrelation time $\hat{\rho}_{i,t}$, defined as the correlation coefficient between two sampler steps with a lag $t$ \cite{sokal1997monte}.}
The maximum lag $T$ is determined by the criterion $\hat{\rho}_{i,T+1} + \hat{\rho}_{i,T+2} < 0$.
\textcolor{black}{Furthermore, $\mu_{i,q}$ and $\sigma^2_{i,q}$ are the mean and variance computed over the corresponding portion of the chain (first $q\%$ for $q=10$, last $q\%$ for $q=50$)}.
Finally, $p_i(B)$ is the probability of a standard Gaussian being larger than $z_i(B)$ in absolute sense.

To determine the burn-in size, $B$ is gradually increased until the Geweke diagnostic indicates convergence, meaning that $\min_i p_i(B) \geq \epsilon_p$. In our simulations we set the tolerance to $\epsilon_p=5\%$, which corresponds to a 95\% confidence interval, and we increase $B$ in steps of 10. The chain with the burn-in removed is then used for further statistical analysis. In the remainder, the argument of $\boldsymbol{\mathcal{C}}(B)$ is dropped for notational convenience. That is, unless indicated otherwise, $\boldsymbol{\mathcal{C}}$ refers to a chain of length $N$ from which the burn-in has been removed.

\subsubsection{Kullback--Leibler divergence}\label{sec:kl}
\begin{figure}
\centering
\subfloat{\includegraphics[width=0.4\textwidth]{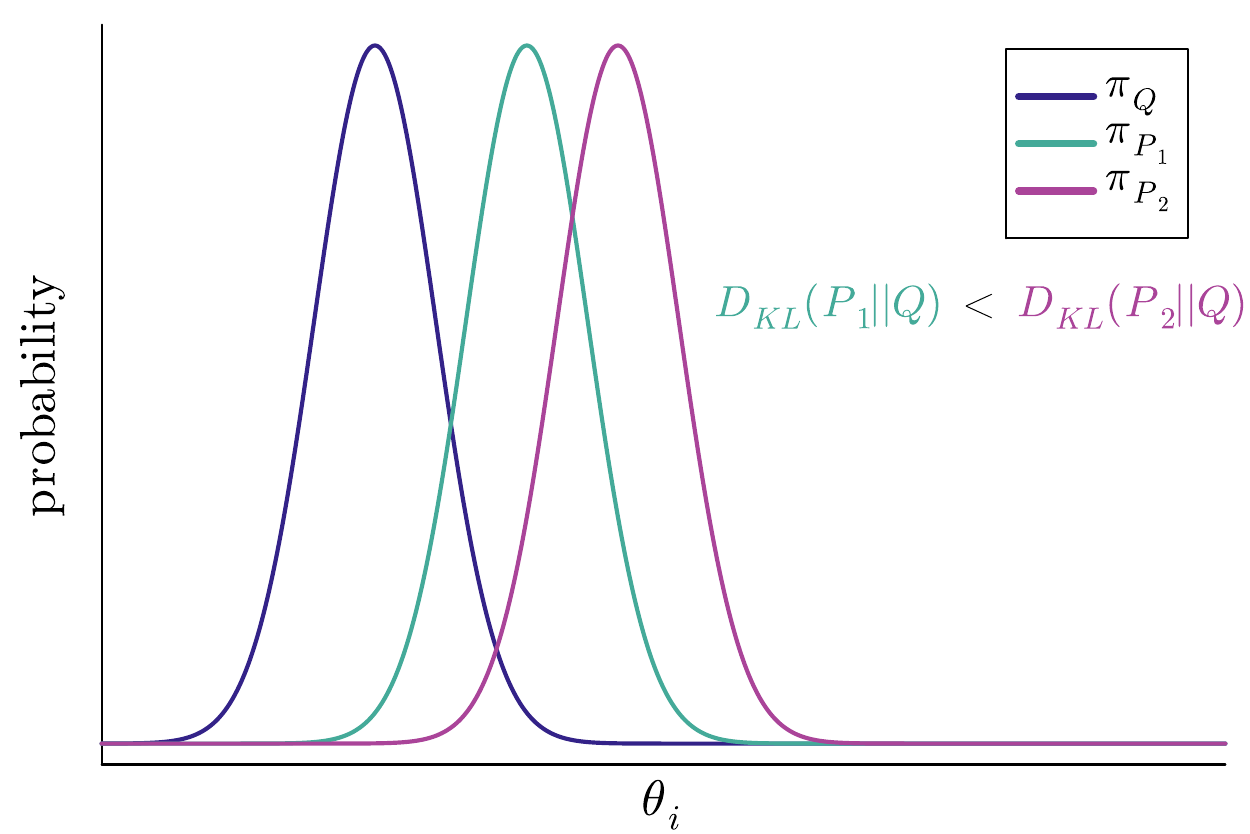}}
\hspace{5mm}
\subfloat{\includegraphics[width=0.4\textwidth]{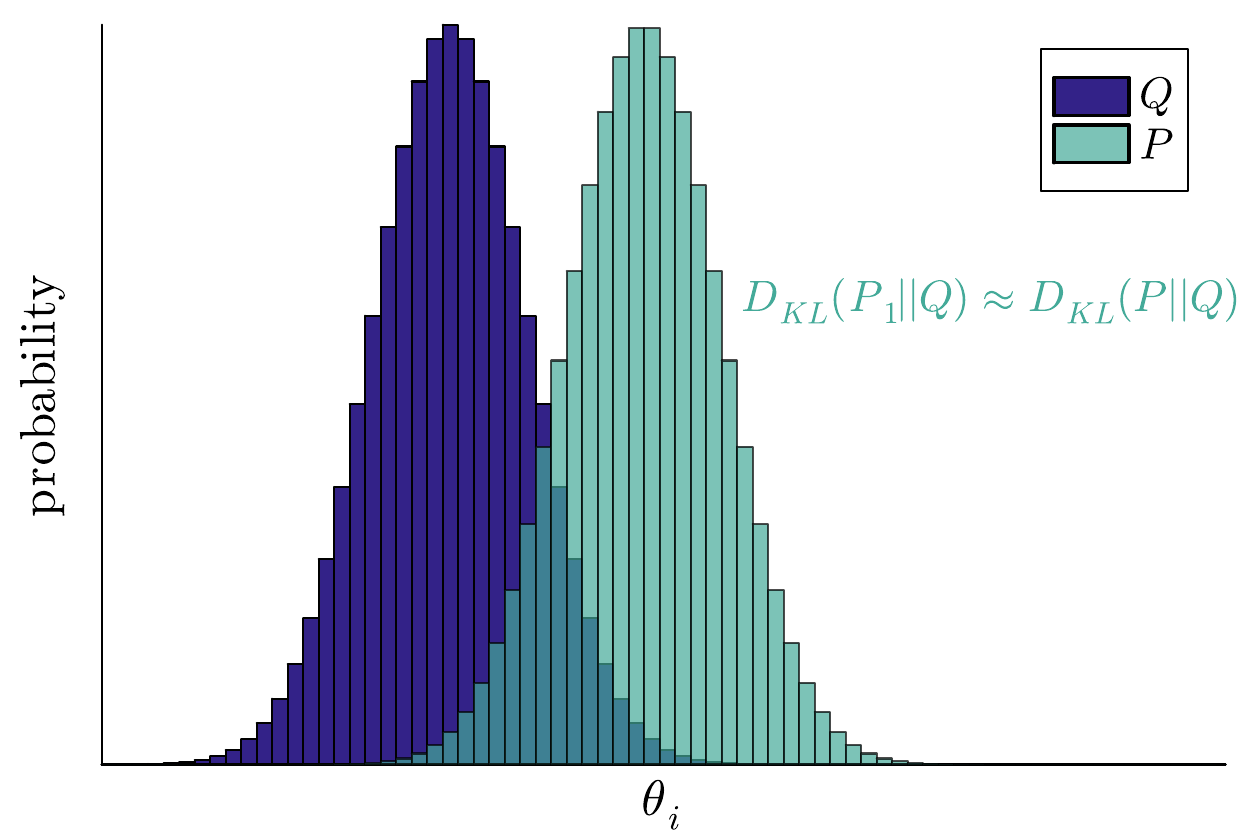}}
\caption{Illustration of the KL-divergence: (\textbf{left}) The  probability density function $\pi_{Q}$ is compared to two probability density functions $\pi_{P_1}$ and $\pi_{P_2}$. (\textbf{right}) The KL-divergence can also be computed between a distribution, $Q$, evaluated on a mesh, and a corresponding binned sample, $P$. \label{KLschem}}
\end{figure}

The KL-divergence quantifies the statistical distance between two continuous probability distributions $P$ and $Q$ as
\begin{equation}
    D_\text{KL} (P||Q) = \int \pi_P(\boldsymbol{\theta}) \log \left( \frac{\pi_P(\boldsymbol{\theta})}{\pi_Q(\boldsymbol{\theta})} \right)\,\mathrm{d}\boldsymbol{\theta},
    \label{eq:kl}
\end{equation}
where $\boldsymbol{\theta}$ is a vector-valued continuous random variable \cite{Kullback1951}. The KL-divergence cannot be negative, is equal to zero when the compared distributions are identical,  \textcolor{black}{and is not symmetric, i.e., $D_\text{KL} (P||Q) \neq D_\text{KL} (Q||P)$}. \autoref{KLschem} shows some probability density functions and histograms for a scalar-valued continuous variable that help illustrate the use of the KL-divergence. The probability density functions $\pi_{P_1}$ and $\pi_{P_2}$ are compared to the density $\pi_{Q}$. For these, the KL-divergence $D_\text{KL} (P_1||Q)$ is smaller than $D_\text{KL} (P_2||Q)$, because $P_1$ is closer to $Q$ than $P_2$. 

To use the KL-divergence as a convergence criterion for MCMC sampling, we let $P$ correspond to the sampled distribution, and $Q$ to a reference distribution computed on a uniform rectilinear mesh, $\mathcal{M}$. The KL-divergence is then approximated using the following discretization:
\begin{equation}\label{eq:KLdisc}
    D_\text{KL} (P||Q) \approx \sum_{E \in \mathcal{M}} \mathbb{P}_P(E) \log \left( \frac{\mathbb{P}_P(E)}{\mathbb{P}_Q(E)} \right).
\end{equation}
Here $\mathbb{P}_P(E)$ and $\mathbb{P}_Q(E)$ are the probabilities assigned to mesh bin $E$ for the distributions $P$ and $Q$, respectively. The probability of the sampled distribution is defined as the relative number of samples in a bin, i.e.,
\begin{equation}\label{eq:PE}
    \mathbb{P}_P(E) = \frac{ \# \{ \boldsymbol{\theta}_n \in E | n=1,\ldots N \} }{ \# \{ \boldsymbol{\theta}_n \in \cup \mathcal{M} | n=1,\ldots N \} }.
\end{equation}
To ensure that $P$ is a probability distribution, the normalization does not consider the samples that fall outside any bin. The probability of the reference distribution is computed as
\begin{equation}\label{eq:QE}
    \mathbb{P}_Q(E) = \frac{{\rm vol}(E) \, \pi_Q ( {\rm centroid}(E) )}{\sum_{F\in \mathcal{M}} {\rm vol}(F) \pi_Q ( \,{\rm centroid}(F) )},
\end{equation}
where the probability density function $\pi_Q$ is the posterior density \eqref{eq:bayesrule}. The normalization is required to ensure that $Q$ is a probability distribution.

{\color{black}
This definition of the KL-divergence should be interpreted as the sample-based average of the probability density mismatch between the sampled distribution and the mesh-based reference. An important limitation of this definition is that regions in the parameter domain that are not explored by the sampler do not contribute to the KL-divergence. We found the alternative definition, in which $P$ corresponds to the mesh and $Q$ to the sample, less suitable in the context of this work. For this choice, bins without samples lead to zero divisions in the KL-divergence definition. Alternative probability estimates are possible, but these lead to estimation errors in regions that are sparsely sampled, specifically the distribution tails. Because of this, in the context of this work, the alternative KL-divergence definition predominantly assesses how well the sampler covers the parameter domain, and less how well the probability density is approximated. In the considered KL-divergence definition, this coverage aspect is incorporated through the distribution of the samples over the mesh bins. Specifically, we define the coverage as the relative number of bins in which at least one sample resides.
}

To use the KL-divergence \eqref{eq:KLdisc} as a sampler convergence metric, it is essential that the rectilinear reference mesh $\mathcal{M}$ is defined appropriately. This means that the bounds of the mesh in each of the parameter directions are chosen such that the part of the posterior that falls outside of the mesh leads to errors that are insignificant in comparison to the sampler errors to be evaluated. In addition, the bin sizes should minimize the approximation error of the numerical integration of the reference solution (\autoref{eq:QE}) and the approximation error of the sample binning (\autoref{eq:PE}). Since the computational effort to evaluate equation \eqref{eq:KLdisc} scales with the number of bins, the bounds and bin sizes need to balance accuracy with computational effort. For problems with a significant number of parameters, finding a good balance is challenging. For this work we use the chains from the three sampling methods excluding the burn-in to determine the mean $\mu_i$ and standard deviation $\sigma_i$ per parameter direction. We then define the rectilinear reference mesh by the tensor product $\mathcal{M}= \mathcal{M}_1 \times \ldots \times \mathcal{M}_p$, with
\begin{equation}\label{eq:refgrid}
    \mathcal{M}_i = \bigcup \limits_{e=0}^{N^{{\rm bin}}_{i}-1} \left[ \mu_i - z \sigma_i + e h_i , \mu_i - z \sigma_i +  (e+1) h_i \right) 
\end{equation}
and $h_i = 2 z \sigma_i / N^{{\rm bin}}_{i}$. Here $N^{{\rm bin}}_{i}$ is the number of bins in the direction $i$, such that the total number of bins is equal to $N^{{\rm bin}} = \prod_{i=1}^p N^{{\rm bin}}_{i}$. In the remainder of the work we set $z$ to 4, which corresponds to a confidence interval of approximately 99.99\%. \textcolor{black}{Evidently, the size of the reference mesh and the number of bins influence the sample coverage.}

\subsubsection{Heuristics}\label{sec:heuristics}
Although the KL-divergence provides a \textcolor{black}{well-defined} error metric, its evaluation requires the computation of a computationally expensive reference solution. This makes it useful as a method to study convergence, as considered in this work, but limits its practical applicability. In practice, use is made of heuristic convergence criteria which can be evaluated without the need for a reference solution.

We herein consider $M$ chains with an average length of $N=\frac{1}{M}\sum_{m=1}^M N^m$, which are started from different initial positions and with the burn-in removed as discussed in \autoref{sec:geweke}. To assess the mixing of chains, i.e., they all sample from the posterior distribution, we consider the Gelman-Rubin diagnostic \cite{Gelman2014} for each parameter $\theta_i$ as
\begin{equation}\label{eq:rhat}
    \hat{R}_i = \sqrt{\frac{\widehat{\text{var}}^{+}_i}{W_i}},
\end{equation}
where
\begin{align}
    \widehat{\text{var}}^{+}_i &= \frac{N-1}{N}W_i + \frac{1}{N}B_i,
    \label{eq:varhat}
\end{align}    
with the within and in-between variances of the chains defined as
\begin{subequations}
\label{eq:WandB}
\begin{align}
     W_i & = \frac{1}{M}\sum_{m=1}^M (\sigma_i^m)^2, \label{eq:W} \\
     \frac{B_i}{N} & = \frac{1}{M-1} \sum_{m=1}^M (\mu_i^m-\mu_i)^2. \label{eq:B}
\end{align}    
\end{subequations}
In these expressions, $\mu_i^m$ and $\sigma_i^m$ are the mean and standard deviation of each chain, and $\mu_i = \frac{1}{M}\sum_{m=1}^M \mu_i^m$ is the mean of all chains\footnote{In previous sections, where only single chains are considered, the superscript $m$ has been omitted and hence $\mu_i$ in that case (also) indicates the mean of a single chain.}. We note that the estimator \eqref{eq:varhat} is unbiased under stationarity, but otherwise overestimates the posterior variance \cite{Gelman2014}. Since $\hat{R}_i$ converges to 1 upon increasing the sample size (Remark~\ref{rem:gelmanrubin}), the criterion $\hat{R}_i < 1 + \epsilon_{\hat{R}}$ is commonly used to assess chain convergence, e.g., with $\epsilon_{\hat{R}} = 10\%$ \cite{Gelman2014}.

\begin{remark}[Gelman-Rubin diagnostic]
\label{rem:gelmanrubin}
    Upon substitution of \autoref{eq:WandB} in \autoref{eq:varhat}, and subsequent substitution in \autoref{eq:rhat}, the Gelman-Rubin diagnostic can be expressed as:
    \begin{equation}
    \hat{R}_i = \sqrt{\frac{N-1}{N} + \frac{M}{M-1} \frac{ \sum_{m=1}^M (\mu_i^m-\mu_i)^2}{\sum_{m=1}^M (\sigma_i^m)^2}}.
    \end{equation}
    As, upon increasing the sample size, the means of the different chains converge relative to their variances, this diagnostic converges to 1.
\end{remark}

\begin{remark}[Multi-chain effective sample size]
    In the case of multiple chains, the effective sample size can still be computed as in \autoref{eq:zpscore}, but with the sample size set to $M \times N$. The multi-chain autocorrelation time in \autoref{eq:zpscore} is defined as
    \begin{equation}\label{eq:rhothat}
        \hat{\rho}_{i,t} = 1 - \frac{V_{i,t}}{2\widehat{\text{var}}^{+}_i},
    \end{equation}
    with $\widehat{\text{var}}^{+}_i$ as in \autoref{eq:varhat} and with
    \begin{equation}\label{eq:Vt}
        V_{i,t} = \frac{1}{M(N-t)} \sum_{m=1}^M \sum_{n=t+1}^{N^m} (\theta_{i,n}^m-\theta_{i,n-t}^m)^2.
    \end{equation}
\end{remark}

\subsubsection{Computational effort}\label{sec:performance}
In our application domain, the model evaluations comprise the overwhelming majority of the MCMC simulation time. That is, the computational overhead of the samplers is negligible. This makes the number of model evaluations an ideal indicator to compare the various samplers in terms of computational effort, as it is independent of the code implementation and computer architecture. Model evaluations pertaining to sampler configuration/optimization steps are excluded from the computational effort, as the plethora of options available for such purposes would further complicate the comparison. The number of model evaluations is also agnostic of parallel computing aspects. While this is favorable from the vantage point of comparison, the effectivity of parallelization depends on the sampler. With our indicator choice, this aspect is excluded from the comparison, which we reflect upon in our discussion.

An important aspect related to the model evaluations is whether gradients of the solution with respect to the model parameters are required by the sampler. When these gradients can be evaluated (semi-)analytically through automatic differentiation \cite{Baydin2018}, the computational effort associated with the gradient calculation is generally included in the model evaluation. In this case, the computational load associated with the gradient calculations is not reflected by the number of model evaluations. When multiple model evaluations are used to approximate the gradients, which \textcolor{black}{can be} necessary when models cannot be differentiated (semi-)analytically, the computational effort involved in computing the gradients is reflected by the number of model evaluations. This difference complicates the comparison of samplers that require gradient calculations with samplers that do not, in the sense that it becomes dependent on the automatic differentiation implementation. In our results and discussions we highlight the nuances of this aspect of the comparison.

\subsection{Samplers}\label{sec:mcmc:samplers}
\begin{figure}
\centering
\subfloat{\includegraphics[width=0.3\textwidth]{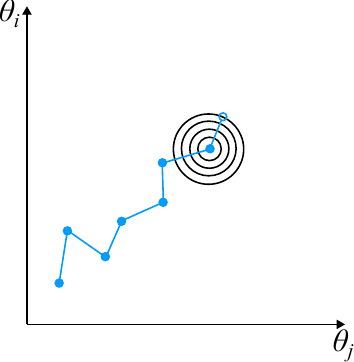}}
 \hfill
\subfloat{\includegraphics[width=0.3\textwidth]{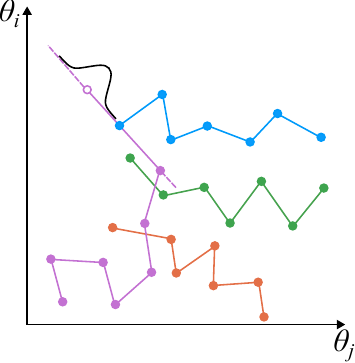}}
 \hfill
\subfloat{\includegraphics[width=0.3\textwidth]{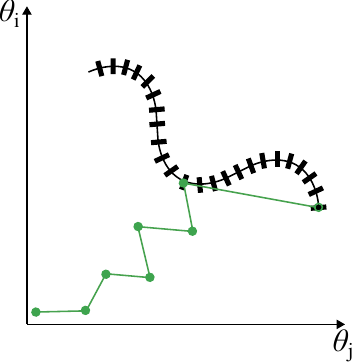}}
\caption{Illustration of the compared samplers: (\textbf{left}) Metropolis-Hastings (MH): The next step in the chain is sampled from a proposal distribution. (\textbf{middle}) Affine Invariant Stretch Move (AISM): For each walker, the next step is proposed using the latest step of a randomly selected different walker. (\textbf{right}) No-U-Turn Sampler (NUTS): The next step is determined using a leapfrog integration scheme that makes use of the gradient of the posterior. The leapfrog steps are visualized by the black rectangles. \textcolor{black}{The left and middle panels are reproduced from \citet{rinkens2025comparison}. The right panel is based on \citet{rinkens2025comparison}.}  \label{schemsamplers}}
\end{figure}

For our comparison, we consider three distinct commonly used samplers: the Metropolis-Hastings (MH) sampler, the Affine Invariant Stretch Move (AISM) sampler, and the No-U-Turn Sampler (NUTS).
\autoref{schemsamplers} illustrates the conceptual differences between these samplers.
MH is considered as the baseline sampler, AISM as a multi-walker sampler, and NUTS as a Hamiltonian Monte Carlo (HMC) sampler that makes use of gradient information.
\textcolor{black}{The MH and NUTS samplers are implemented using the \texttt{Turing.jl} package \cite{ge2018turing}, whereas the AISM sampler is implemented using the \texttt{AffineInvariantMCMC.jl} package \cite{AffineInvariantMCMCjl}, both in Julia.}

In the remainder of this section we discuss these samplers based on pseudo-code algorithms \textcolor{black}{(included at the bottom of this paper)}, aiming to clarify their distinctive algorithmic features. In particular, we highlight where in the algorithms the likelihood \eqref{eq:likelihood_general} is evaluated using the function $\texttt{EVAL\_LIKELIHOOD}$, as this function involves the evaluation of the model and therefore plays an important role in our performance assessment. In the case that also gradients of the likelihood are required, we highlight this using the function $\texttt{EVAL\_LIKELIHOOD\_GRADIENT}$. The presented pseudo-code algorithms are intentionally simplified to focus on the aspects that affect their computational effort. For more detailed algorithm expositions and implementation details we refer to the references included in the remainder of this section.

In this section, we omit the burn-in removal (\autoref{sec:geweke}) from the discussion, as this is an independent aspect. To keep notation compact, we here do not make a distinction between the chains generated by the samplers and those used for the further statistical analysis, i.e., those with the burn-in period removed.

\subsubsection{Metropolis--Hastings (MH) sampler}
The Metropolis-Hastings (MH) sampler \cite{Metropolis1953, Hastings1970} is generally considered as the baseline algorithm to sample from a target distribution \cite{Kaipio2006, Ozisik2018}. The pivotal idea of the MH algorithm is to propose a new step and then to accept or reject this step using a criterion that leads to the target distribution.

The MH algorithm, outlined in Algorithm~\ref{alg:mh}, starts with the definition of an initial guess for the vector of parameters, which we sample from the prior (L\ref{line-mh-sample-initial-guess}). The parameter space is then explored by performing a random walk, in which the new candidate for the vector of parameters is sampled from a proposal distribution (L\ref{line-mh-sample-proposal}). This proposal is typically chosen to be a multivariate normal distribution, with its mean centered at the current state (Remark~\ref{rem:metropolis}). The new candidate is then either accepted or rejected according to a probabilistic criterion (L\ref{line-mh-accept}--L\ref{line-mh-criterionendif}), where the acceptance probability (L\ref{line-mh-acceptance-prob}) depends on the likelihood (L\ref{line-mh-likelihood}). The calculation of this likelihood \eqref{eq:likelihood_general} involves the evaluation of the model, but not its gradients. As a result, in the MH algorithm every sampling step requires a single model evaluation.

The performance of the MH sampler is strongly influenced by the choice of the proposal covariance matrix (used inside the $\texttt{SAMPLE\_PROPOSAL}$ and $\texttt{EVAL\_PROPOSAL}$ functions on L\ref{line-mh-sample-proposal} and L\ref{line-mh-proposal}), making careful selection a necessity. To objectively and optimally determine the proposal covariance, we employ the adaptive MCMC algorithm \cite{Haario2001}. The idea behind this adaptive algorithm -- which is run as a precursor of Algorithm~\ref{alg:mh} -- is to calibrate the proposal covariance based on the acceptance ratio. \textcolor{black}{This proposal calibration step can involve a number of model evaluations similar to the sampling itself, and hence can have a significant impact on the computational effort of the MH-based analysis. However, since the computational effort relating to the calibration step depends on the employed tuning procedure and on the considered problem, we de not incorporate it in our comparison.} \textcolor{black}{We manually implemented this algorithm to obtain the Bayesian inference results for the two corresponding physical systems.}
    
\begin{remark}[Metropolis algorithm]
\label{rem:metropolis}
 In the specific case that a symmetric proposal distribution is considered, i.e., ${\pi_\text{prop}^{\bm{\theta}_{n-1} | \tilde{\bm{\theta}}}} =  {\pi_\text{prop}^{\tilde{\bm{\theta}} | \bm{\theta}_{n-1}}}$, the proposal fraction in the acceptance probability drops out. This specific case of the MH algorithm is referred to as the Metropolis algorithm \cite{Metropolis1953}.  
\end{remark}

\subsubsection{Affine Invariant Stretch Move (AISM) sampler}
The Affine Invariant Stretch Move (AISM) sampling method is based on the algorithm developed by \citet{Goodman2010}. This sampler uses an ensemble of Markov chains, referred to as walkers, to explore the parameter space. By basing the proposal for the next step of each walker on different walkers, a solver that is robust with respect to parameter scalings and correlations is obtained. This makes it particularly attractive in the case of anisotropic and strongly correlated distributions.

The AISM algorithm is outlined in Algorithm~\ref{alg:aism}. To compute the Markov chain, the sample size, $N$, and the number of walkers, $K$, must be specified. The sample size must be divisible by the number of walkers. We sample the initial position for each walker from the prior distribution (L\ref{alg:aism_forwalk}--L\ref{alg:aism_endwalk}). The next step of walker $k$ (L\ref{alg:aism_propsample} and L\ref{alg:aism_propwalk}) is proposed using the current step of a randomly selected different walker $j$. The proposal step size parameter $z$ is sampled from the probability density
\begin{equation}\label{eq:AISMdist}
    g(z) \propto \begin{cases} \frac{1}{\sqrt{z}} & \mathrm{if} \, z \in \left[ \frac{1}{a}, a \right], \\
    0 & \text{otherwise},\end{cases}
\end{equation}
where the parameter $a>1$. As in the MH algorithm, the densities of the prior and likelihood are evaluated for the proposal (L\ref{alg:aism_propstart}--L\ref{alg:aism_propend}), after which the proposal is accepted or rejected using a probabilistic criterion (L\ref{alg:aism_acceptpropstart}--L\ref{alg:aism_acceptpropend}). As such, every step in the sampler requires one model evaluation for the calculation of the likelihood.

The AISM sampler can be optimized by adjusting the number of walkers and the proposal scale parameter. Based on a parameter study, we set $a=2$ -- which is the typical default value \cite{Goodman2010} --  which provides a near optimal acceptance rate. Increasing the number of walkers reduces the dependence between any two chains caused by the affine invariant proposal, which is beneficial from an effective sample size point of view. However, the burn-in size increases with the number of walkers, leading to performance degradation when too many walkers are used \cite{foreman2013emcee}. For our simulations we have optimized the number of walkers to balance the effective sample size and the burn-in.

\begin{remark}[Ensemble averaging]
Since the walkers in the AISM sampler are not fully independent, their samples cannot be treated as separate Markov chains. Therefore, to determine the heuristic indicators in \autoref{sec:heuristics}, the ensemble of the walkers is considered as one Markov chain \cite{Goodman2010}. That is, for each step in the Markov chain, the average of all walkers is considered, resulting in an ensemble chain of length $N/K$. To objectively compare against the other samplers, the effective sample size is then defined as the product of the number of walkers and the effective sample size of the ensemble average \cite{Goodman2010}. Similarly, as the variance of the ensemble is inversely proportional to the number of walkers, this within-chain variance is also scaled with the number of walkers.
\end{remark}

\subsubsection{No-U-Turn Sampler (NUTS)}
\textcolor{black}{The No-U-Turn Sampler (NUTS) \cite{Hoffman2014} is a member of the Hamiltonian Monte Carlo (HMC) family of algorithms \cite{neal2011mcmc}, which navigate the parameter space by exploiting gradient information of the posterior.} One of the distinctive features of the NUTS is that it automatically adapts its step size while exploring the parameter space \cite{Hoffman2014}, avoiding the need for the user to specify sampler parameters.

To clarify the impact of model and gradient evaluations on the computational effort of the NUTS algorithm, we instead consider the HMC pseudo-code in Algorithm~\ref{alg:hamiltonian}. Although, in contrast to this HMC algorithm, NUTS automatically determines the number and size of the leapfrog steps, the HMC algorithm is representative of NUTS in terms of model and gradient evaluations. After defining a starting point (L\ref{line-hmc-sample-initial-guess}), the next step in the random walk is proposed by solving the Hamiltonian dynamics based on the likelihood function. To this end, in the current step, a starting momentum vector is sampled from a univariate standard normal distribution (L\ref{line-hmc-startingmomentum}). The proposal is then determined using $L$ leapfrog integration steps (L\ref{line-hmc-startleapfrog-loop}--L\ref{line-hmc-endleapfrog-loop}), after which the proposal is accepted using a probabilistic criterion similar to that used in the MH and AISM samplers (L\ref{line-hmc-startacceptancecriterion}--L\ref{line-hmc-endacceptancecriterion}). In addition to the model evaluation required to compute the acceptance probability in each sampling step (L\ref{line-hmc-alpha}), each leapfrog step also requires the evaluation of the gradient of the likelihood function (L\ref{line-hmc-gradienteval1} and L\ref{line-hmc-gradienteval2}). As discussed in \autoref{sec:performance}, this gradient evaluation is computationally demanding on account of the need to evaluate model gradients.
    
\textcolor{black}{ The NUTS is the only algorithm among the three aforementioned that requires gradient information of the model.} In this work, we obtain the gradient with the forward mode automatic differentiation method \cite{Baydin2018}. This method allows for automatic differentiation of the considered models. Although this increases the computational effort involved in the evaluation of the model, in terms of implementation no additional work is required.

\begin{remark}[NUTS burn-in]
    In the implementation of the NUTS sampler we use \cite{ge2018t}, the burn-in size is automatically set to $B=\min{(N/2,1000)}$.
\end{remark}
\section{Results}\label{sec:5}
In this section, we present the inference results for the thermal conduction system (\autoref{sec:results:thermal}) and for the viscous flow system (\autoref{sec:results:viscous}) obtained using the samplers discussed above. For each of the systems, the samplers are compared in terms of accuracy (KL-divergence and heuristics) and in terms of computational effort. In \autoref{sec:discussion} we discuss the similarities and differences between the two systems.

A key attribute of our solver comparison pertains to studying the influence of the sample size, $N$, on the performance metrics. To efficiently generate chains with different sample sizes, for each of the samplers we generate chains of length $N^{\rm max}$, from which the burn-in is removed (\autoref{sec:3}). Smaller chains are then obtained by considering the first $N$ steps of these chains.

\subsection{Thermal conduction system}\label{sec:results:thermal}
We infer the model parameters $\boldsymbol{\theta} = [k, \rho, c_p, h_\text{source}, h_\text{side}, h_\infty, T_\text{source} ]$ using $M=3$ chains for each sampling method with $N^{\rm max}=\num{179200}$. The calibration of the proposal covariance matrix for the MH sampler also uses \num{179200} steps. \textcolor{black}{These calibration samples, as well as the computationally effort involved in attaining them, are not included in our analysis.} For the AISM sampler we use 28 walkers, i.e., four walkers per parameter.

\begin{figure}
    \centering
    \includegraphics[width=\linewidth]{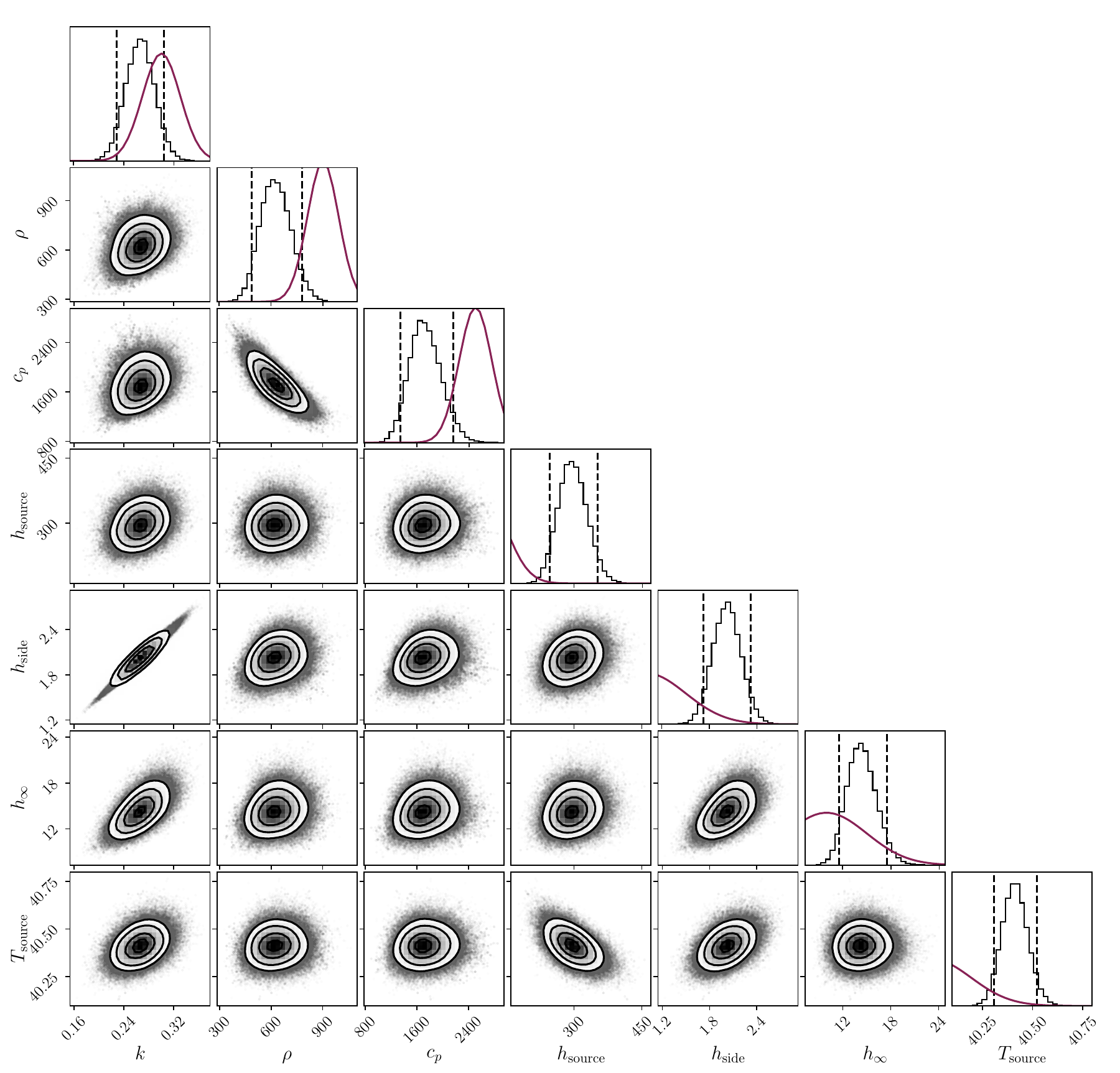}
    \caption{Inferred model parameters (see \autoref{tab:priorparamthermal} for units) and correlations for the thermal conduction system. The top part of each column compares the marginal prior (dark-red line) and posterior (histogram). \textcolor{black}{The vertical dashed lines in the marginal distributions represent the 95\% credible intervals.}}
    \label{fig:post_hc}
\end{figure}

In \autoref{fig:post_hc} we show the posterior distribution marginals and parameter correlations. From the posterior distributions, we observe a significant shift between the prior (red lines) and the posterior (histograms), indicating that the parameters are sensitive to the experimental data. A strong correlation between the conductivity, $k$, and the heat transfer coefficient, $h_{\text{side}}$, is observed. This is a consequence of the steady-state solution to \autoref{eq:thermal:balancelawoneD} -- which is representative for the considered data -- being dependent on the ratio between these parameters only. The heat transfer coefficient is similarly, although less pronounced, correlated to the conductivity via \autoref{eq:therma11DBC2}. For the boundary condition at the source \eqref{eq:therma11DBC1}, the heat transfer coefficient is so large that the paraffin obtains the source temperature at the interface, as a result of which the solution becomes insensitive to this heat transfer coefficient. Finally, since the dynamic term in \autoref{eq:thermal:balancelawoneD} scales with the product of the density and the conductivity, a correlation between these parameters is observed. 

\begin{figure}
    \centering
    \includegraphics[width=0.5\linewidth]{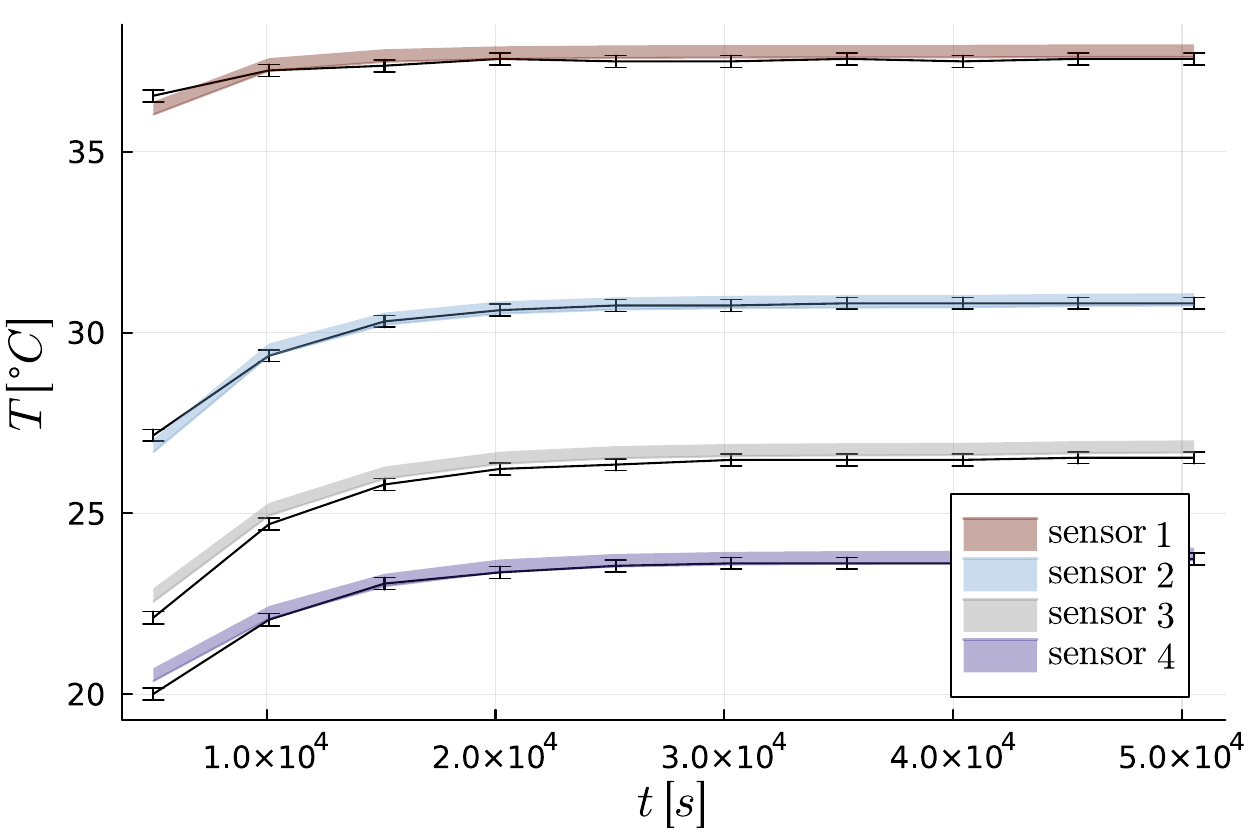}
    \caption{Comparison between the experimental data (error bars) and the posterior predictive distribution for the thermal conduction system per sensor.}
    \label{fig:pmodepred_hc}
\end{figure}

A comparison between the model predictions corresponding to the posterior distribution (colored 95\% credible intervals) and the experimental data (error bars corresponding to the 95\% confidence intervals) is shown in \autoref{fig:pmodepred_hc}. From the model predictions it is observed that the posterior leads to a prediction that matches the experimental data well, also in terms of the experimental uncertainty. The results in both figures are based on one of the AISM chains, but similar results are obtained using the other chains and sampling methods.

\begin{figure}
    \centering
    \subfloat[]{\includegraphics[width=0.48\textwidth]{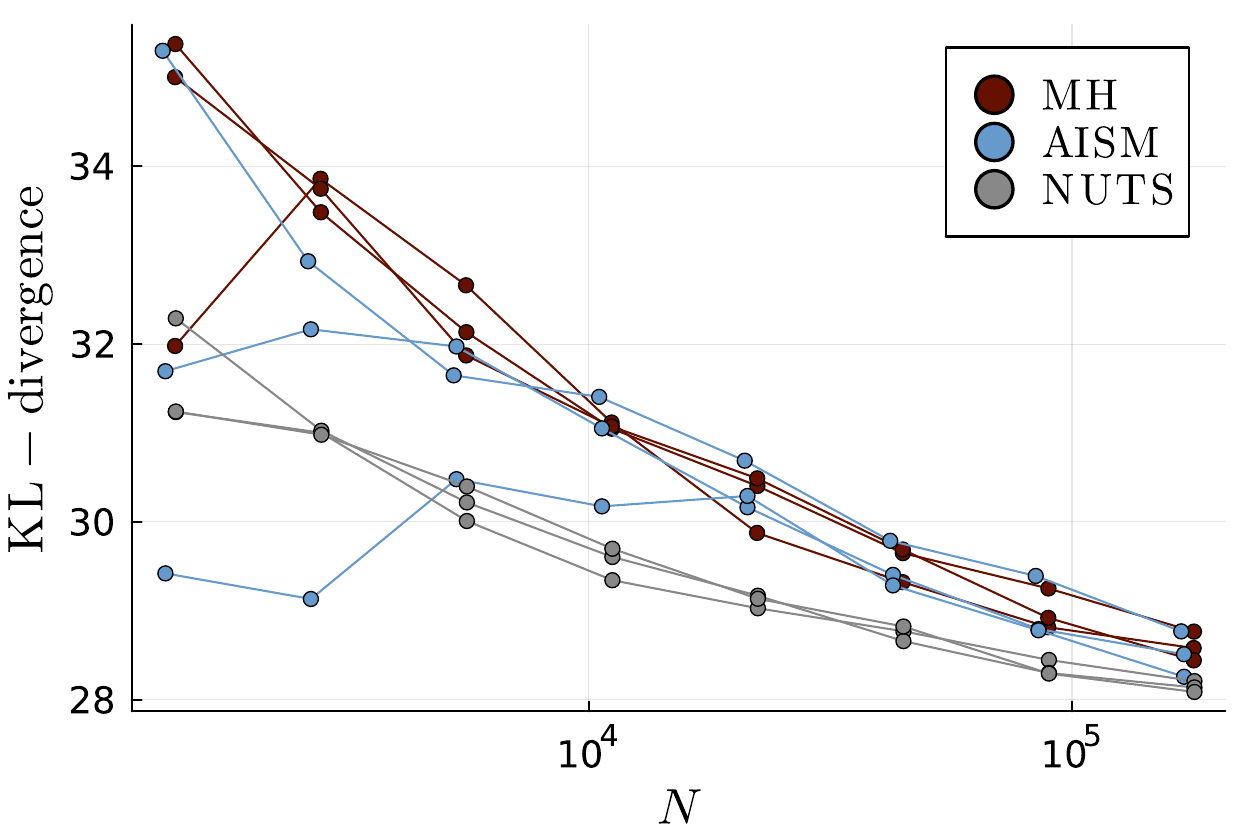}}
    \hfill
    \subfloat[]{\includegraphics[width=0.48\textwidth]{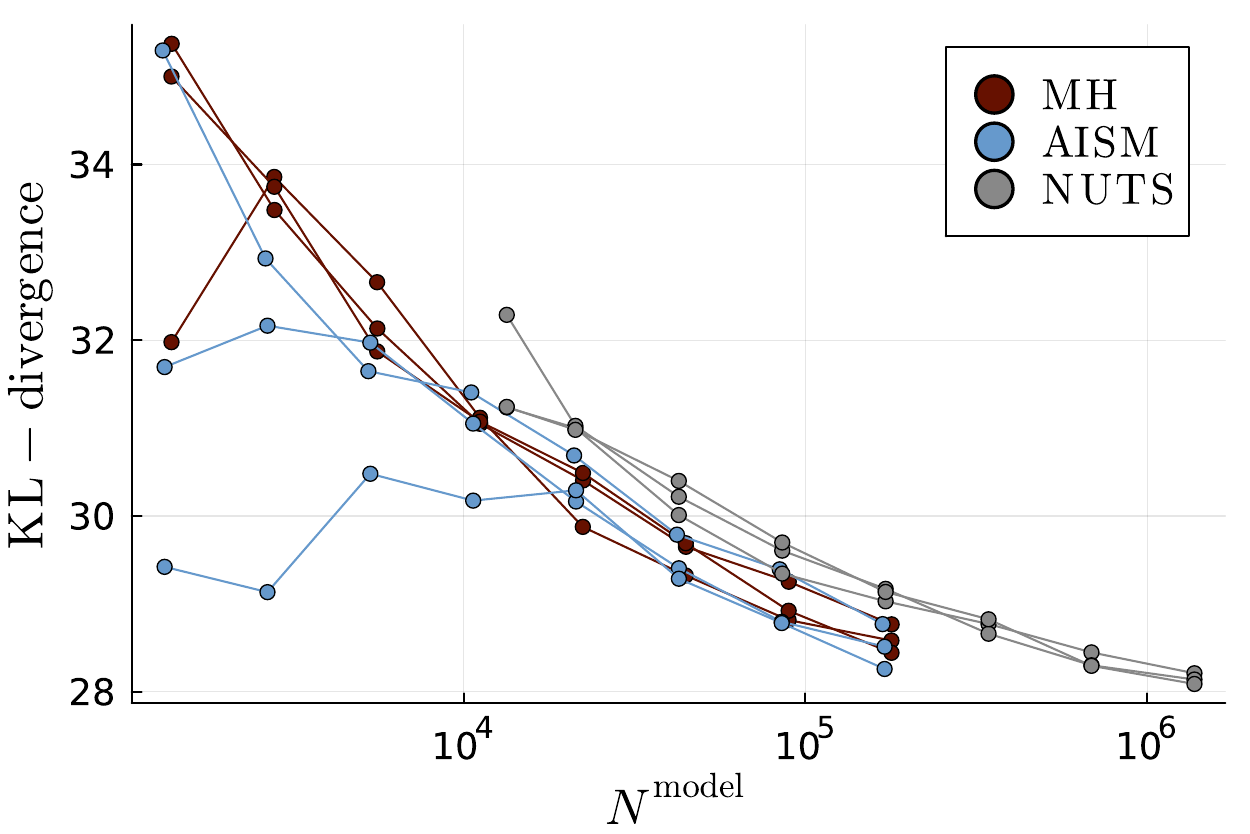}}
    \hfill
    \subfloat[]{\includegraphics[width=0.48\textwidth]{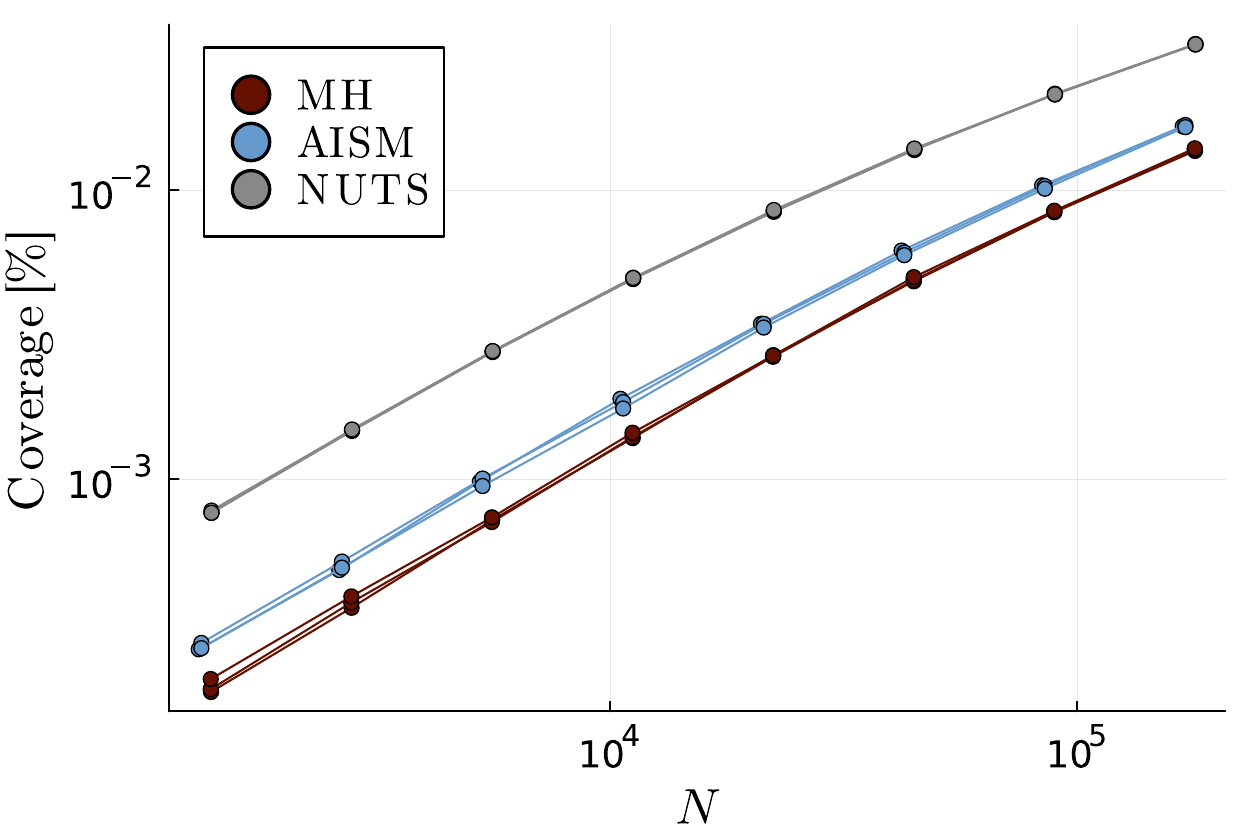}}
    \hfill
    \subfloat[]{\includegraphics[width=0.48\textwidth]{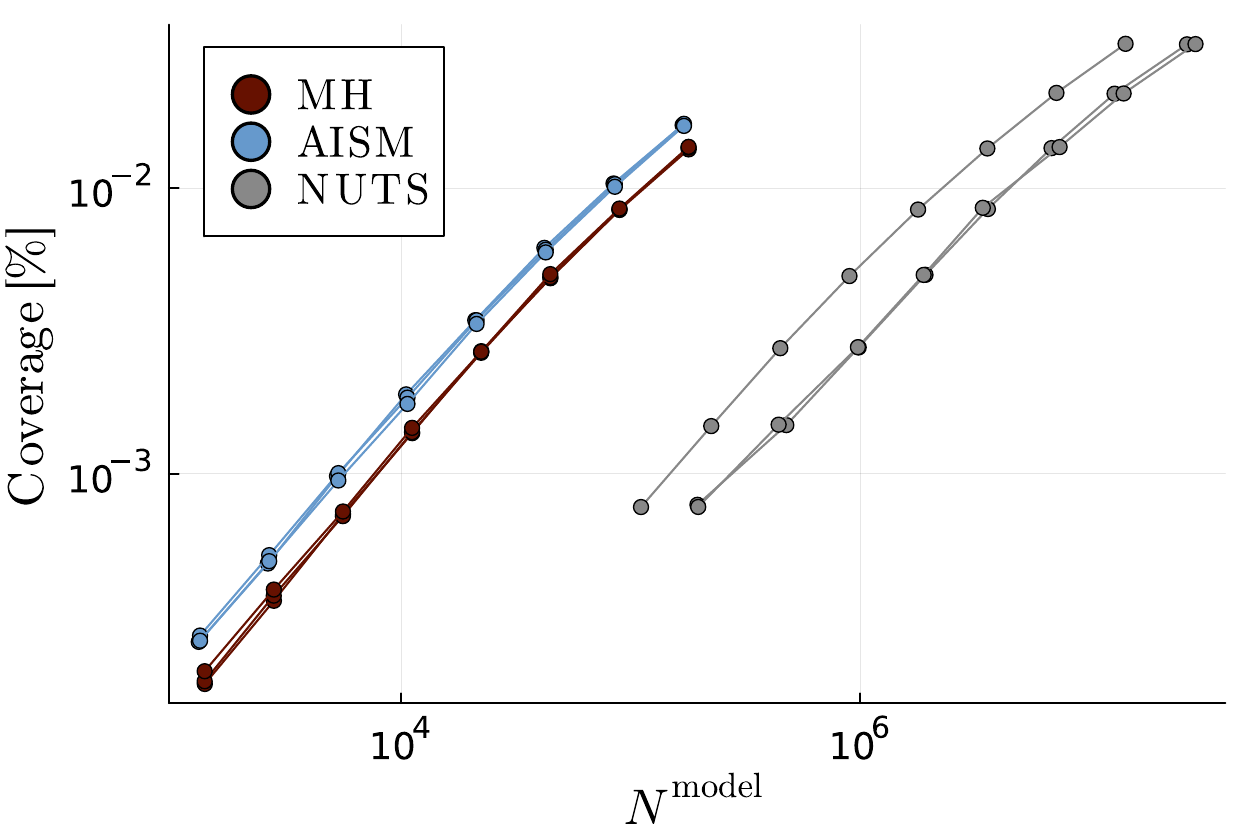}}
    \hfill
    \caption{{\color{black} {KL-divergence and coverage of the sampled posterior for the thermal conduction system.}}}
    \label{fig:KL_hc}
\end{figure}

\paragraph{KL-divergence}
To study the sampler accuracy, we evaluate the KL-divergence for each chain and each sampler by gradually increasing the sample size. To evaluate the KL-divergence, the reference solution is computed on a rectilinear mesh with 15 values per parameter ($N^{\mathrm{bin}}_i=15$, with $i=1,\ldots,7$). The results for the different chains and samplers are shown in \autoref{fig:KL_hc}. \textcolor{black}{Note that the sample size differs slightly between the chains because each chain has a distinct burn‑in period that is removed prior to the analysis.} {\color{black} The top row shows the KL-divergence and the bottom row the coverage (i.e., the percentage of mesh bins with at least one sample). The left column shows these quantities versus the sample size, whereas the right column plots them against the number of model evaluations.}

For all considered chains, the KL-divergence is almost two orders of magnitude lower than the KL-divergence of the prior with respect to the reference posterior distribution. This observation is in agreement with the significant shift between the prior and posterior observed in \autoref{fig:post_hc}. We observe that the KL-divergence for the largest sample size considered remains well above zero, which we attribute to the approximation of the reference solution (\autoref{eq:QE}). That is, the errors in the reference solution due to binning (\autoref{sec:kl}) are not yet negligible in comparison to the sampling errors. However, increasing the number of bins further is hindered by the involved computational effort.


Overall we observe the KL-divergence of the NUTS to be significantly lower than that of the MH and AISM samplers \textcolor{black}{when plotted versus the sample size}. No significant differences are observed between the MH and AISM samplers. At relatively small sample sizes we observe some outliers for the MH and AISM samplers, where in particular one of the AISM chains is observed to result in a lower KL-divergence than that obtained with the NUTS at a similar sample size. We attribute this observation to the variability in our KL-divergence estimator, which is corroborated by the observation that at relatively small sample sizes significant differences between the chains generated by the AISM sampler are observed. Upon increasing the sample size these differences diminish, resulting in consistently higher KL-divergences compared to the NUTS.

\textcolor{black}{When plotted versus the number of model evaluations, the NUTS results in higher KL-divergences compared to the other two samplers. This is a direct consequence of the NUTS requiring a significant number of model evaluations per Markov step. In terms of domain coverage, the NUTS outperforms the other two samplers on a per-step basis. However, on a per-model-evaluation basis the MH and AISM samplers outperform the NUTS.}


\begin{figure}
\centering
\subfloat[]{\includegraphics[width=0.48\textwidth]{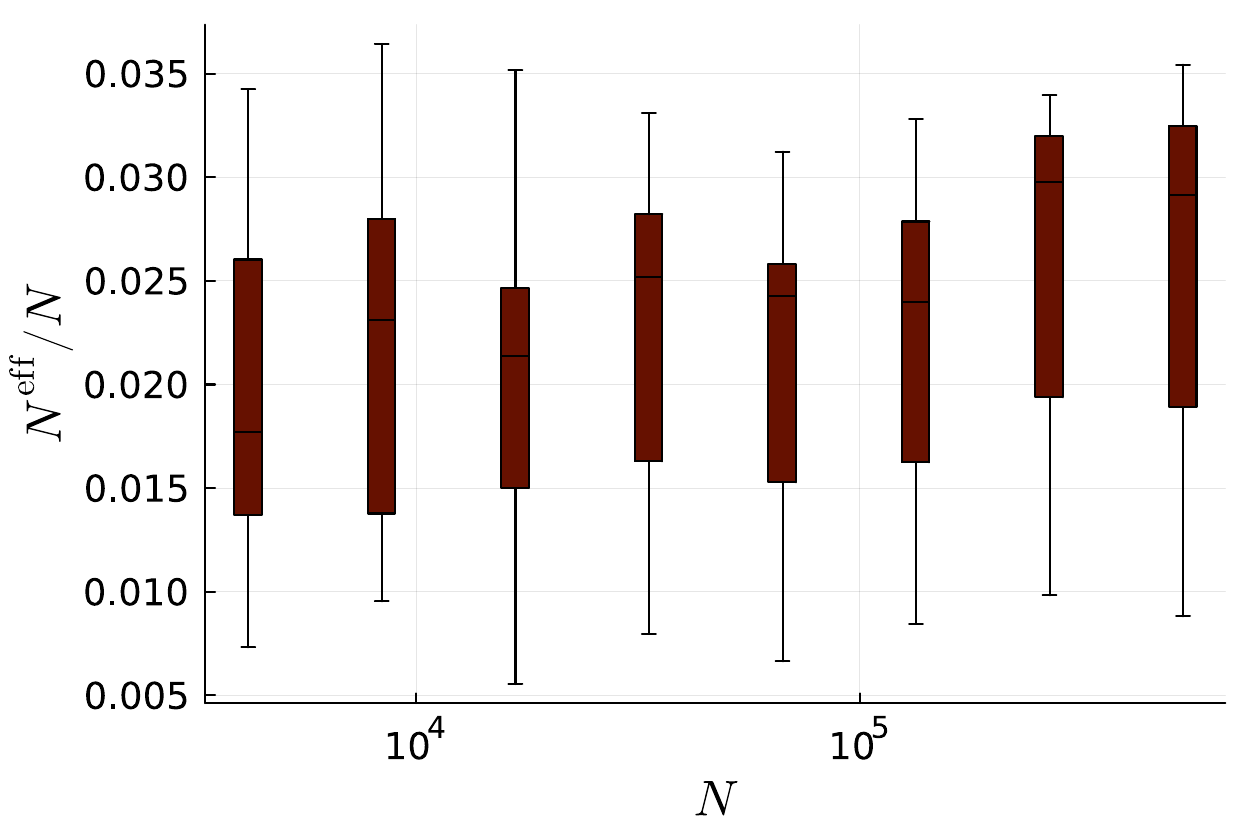}}
 \hfill
\subfloat[]{\includegraphics[width=0.48\textwidth]{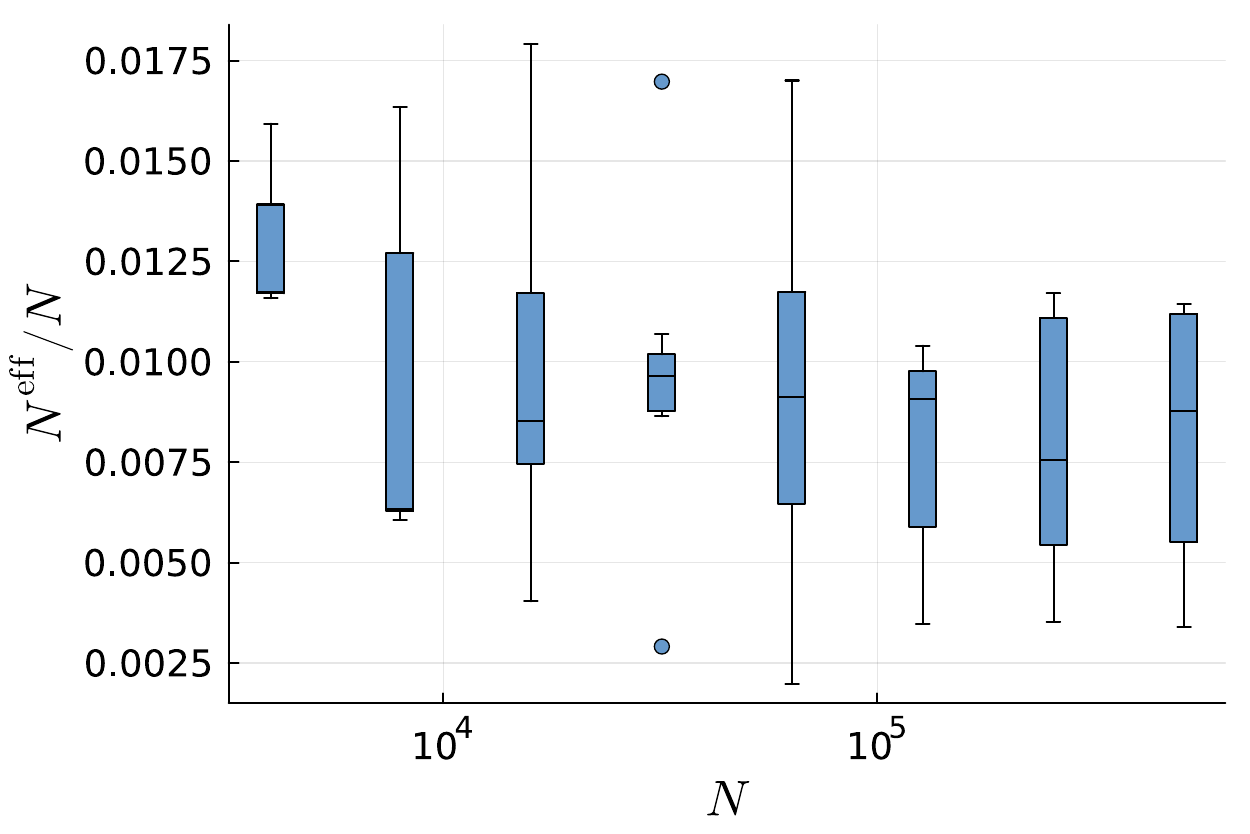}}
\hfill
\subfloat[]{\includegraphics[width=0.48\textwidth]{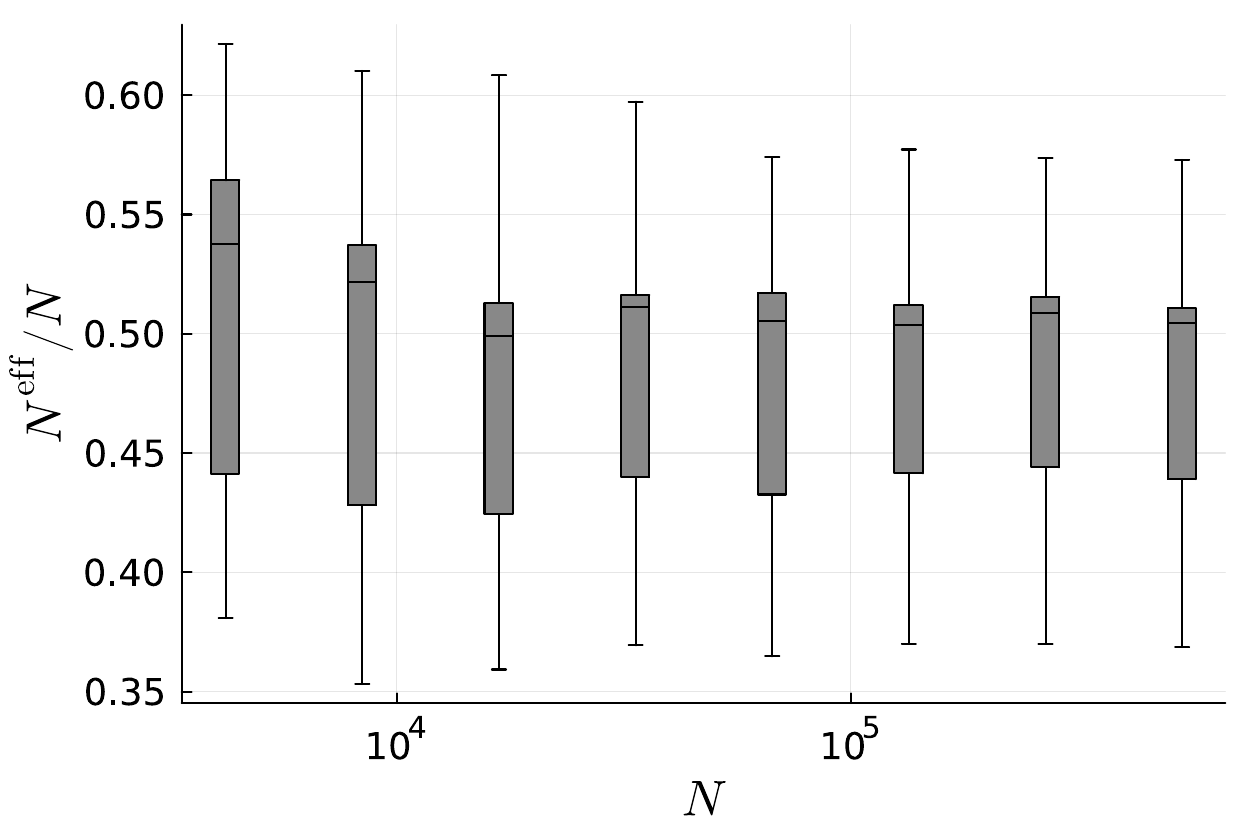}}
\hfill
\subfloat[]{\includegraphics[width=0.48\textwidth]{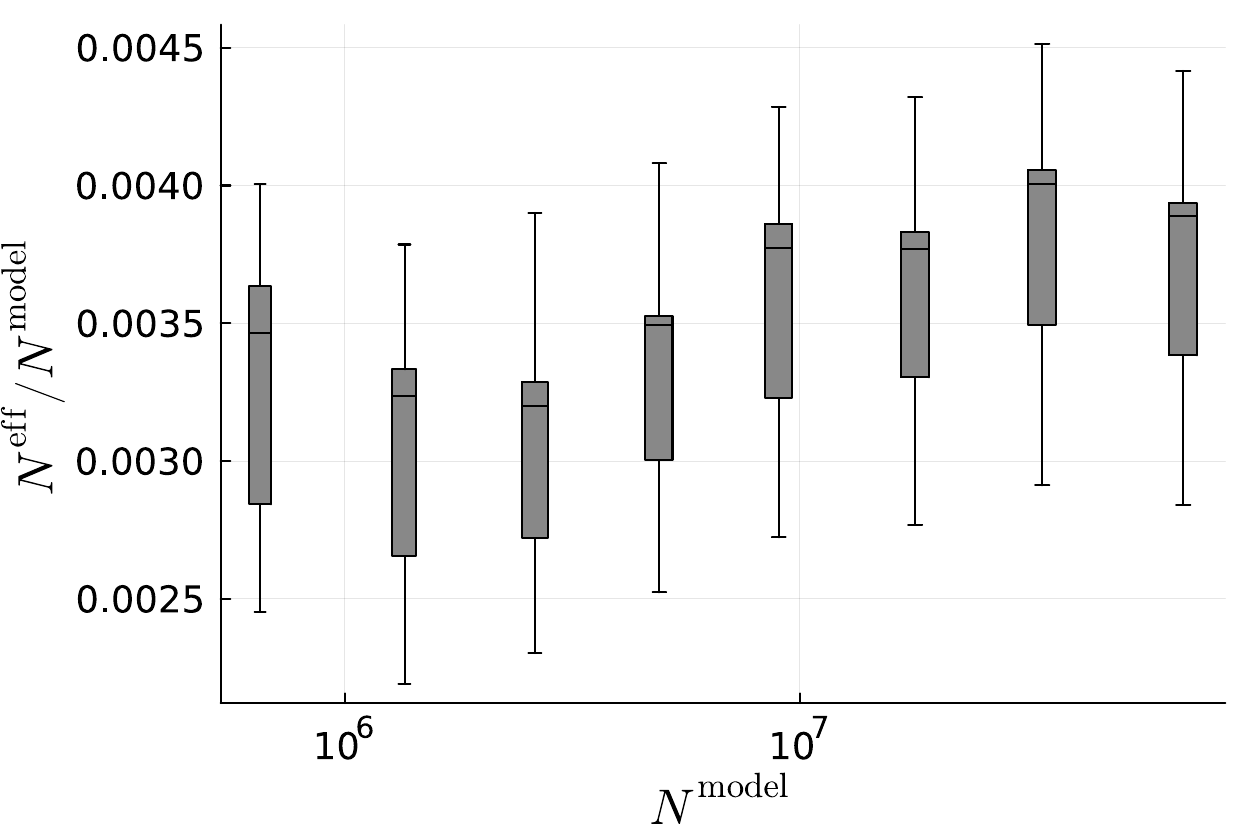}}
\caption{\color{black} {Normalized effective sample size for the thermal conduction system: (a) MH, (b) AISM, (c, d) NUTS. For the NUTS, normalization with respect to both the number of samples and the number of model evaluations is considered.}}
\label{fig:essN_hc}
\end{figure} 

\paragraph{Heuristics}
\autoref{fig:essN_hc} compares the normalized effective sample size for the samplers. \textcolor{black}{For the NUTS, which requires multiple model evaluations per sample, normalization with respect to both the number of samples and the number of model evaluations is shown.} The box plots visualize the distribution of the effective sample sizes for the different parameters and chains. Different vertical scales are used to clearly visualize the different samplers. For the NUTS, the majority of normalized effective sample sizes is observed to be in the range of 45\% to 55\% \textcolor{black}{of the sample size}, which is substantially higher than that observed for the MH sampler (1.5\%-3\%) and the AISM sampler (0.5\%-1.2\%). \textcolor{black}{When normalized with the number of model evaluations, the NUTS effective sample size is observed to be lower than the other two samplers (0.25\%-0.4\%).} This observation conveys that the NUTS more effectively explores the parameter space \textcolor{black}{when considering the Markov steps, but that this comes at the expense of additional model evaluations required to form the proposals}.

\begin{figure}
\centering
\subfloat[]{\includegraphics[width=0.48\textwidth]{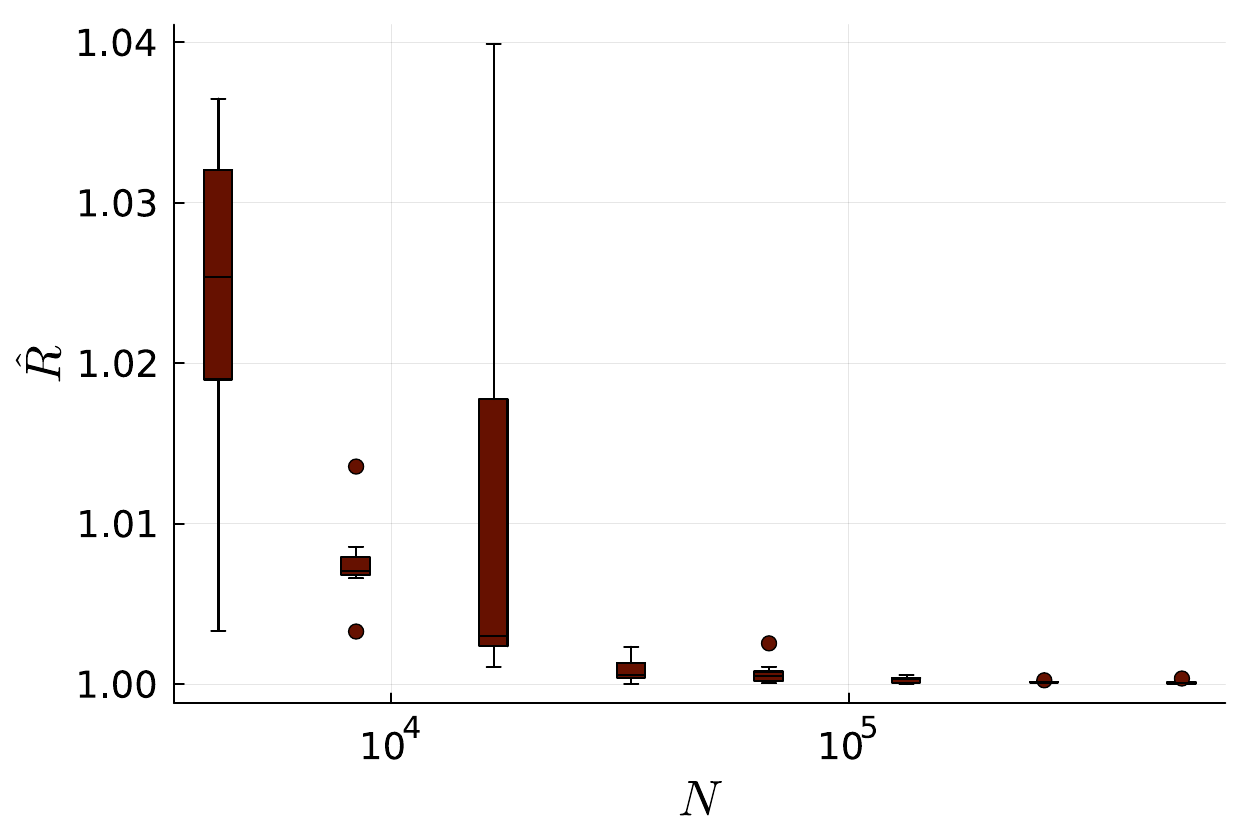}}
 \hfill
\subfloat[]{\includegraphics[width=0.48\textwidth]{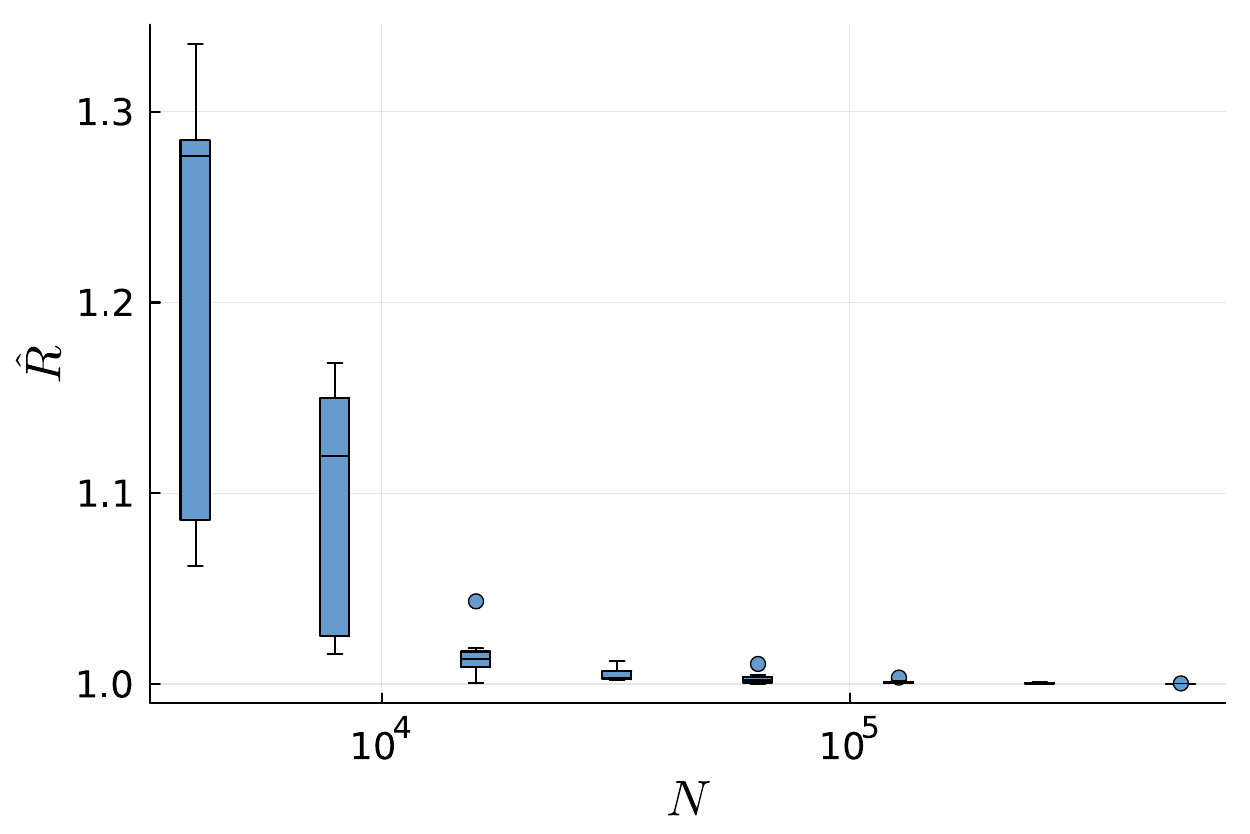}}
\hfill
\subfloat[]{\includegraphics[width=0.48\textwidth]{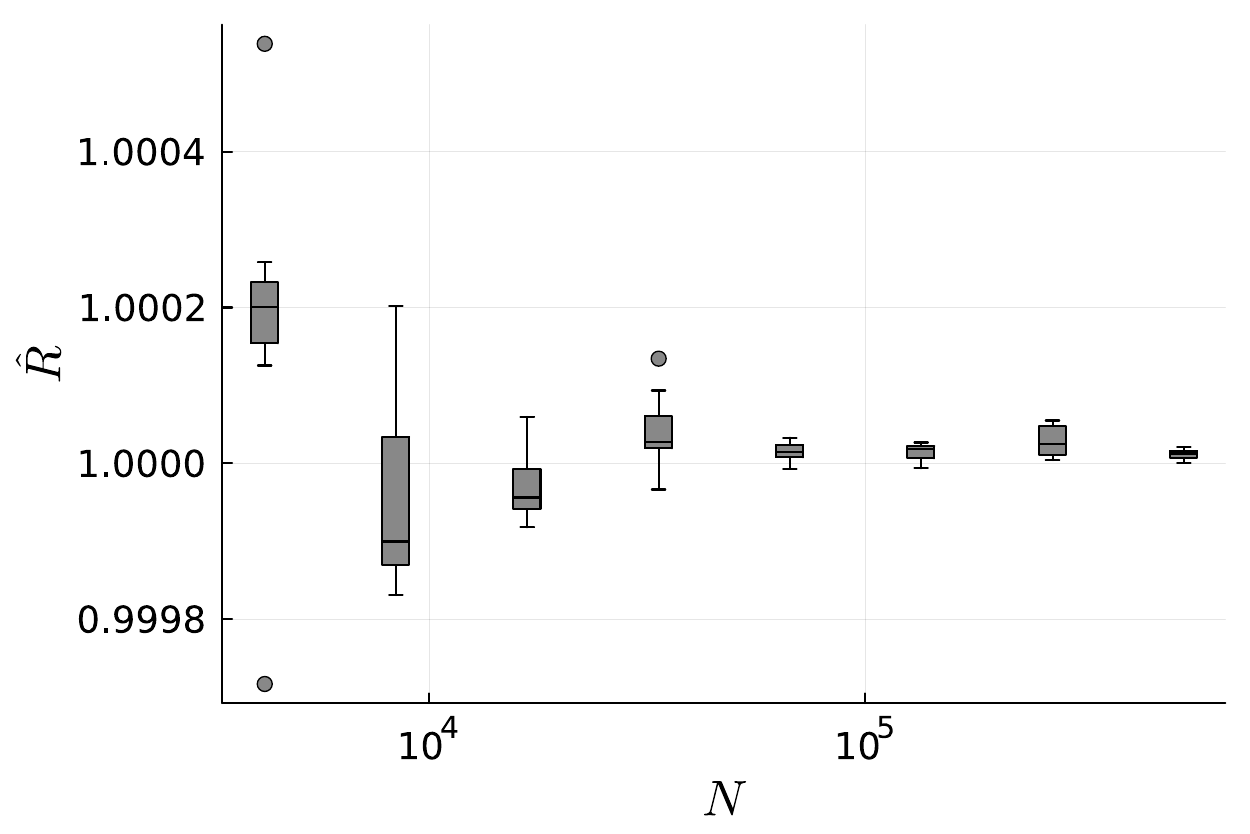}}
\hfill
\subfloat[]{\includegraphics[width=0.48\textwidth]{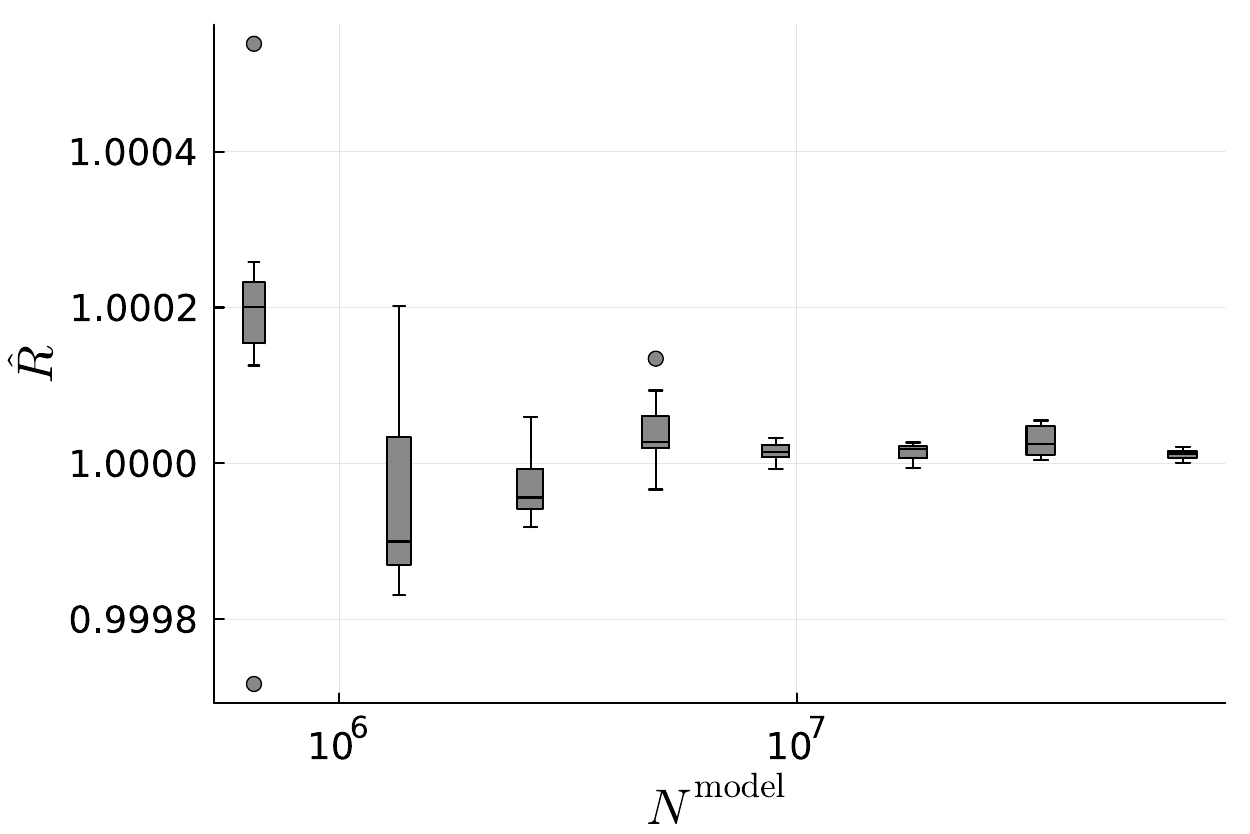}}
\caption{{\color{black} {Gelman-Rubin diagnostic for the thermal conduction system: (a) MH, (b) AISM, (c, d) NUTS.}}}
\label{fig:rhat_hc}
\end{figure}

\autoref{fig:rhat_hc} presents the Gelman-Rubin diagnostic for the considered samplers. The $\hat{R}$ is computed per chain and per parameter, which is represented by a box plot for different sample sizes. Different vertical axes are used to clearly visualize the different samplers. While for all samplers the $\hat{R}$ approaches one with a decreasing uncertainty upon increasing the sample size, a substantial difference in the deviations from unity is observed. While the MH and AISM samplers show similar $\hat{R}$ deviations (with MH being a factor 2-3 smaller for sufficiently large sample sizes), the NUTS shows a deviation from unity that is two orders of magnitude smaller than that for the other two samplers. \textcolor{black}{It is important to note, however, that a comparison between the samplers based on the number of samples does not account for the increased number of model evaluations required for the NUTS.} This observation is consistent with \textcolor{black}{these for the effective sample size} in \autoref{fig:essN_hc}, as the between-chain variance \eqref{eq:B} is inversely proportional to the effective sample size.


\begin{figure}
    \centering
    \includegraphics[width=0.5\linewidth]{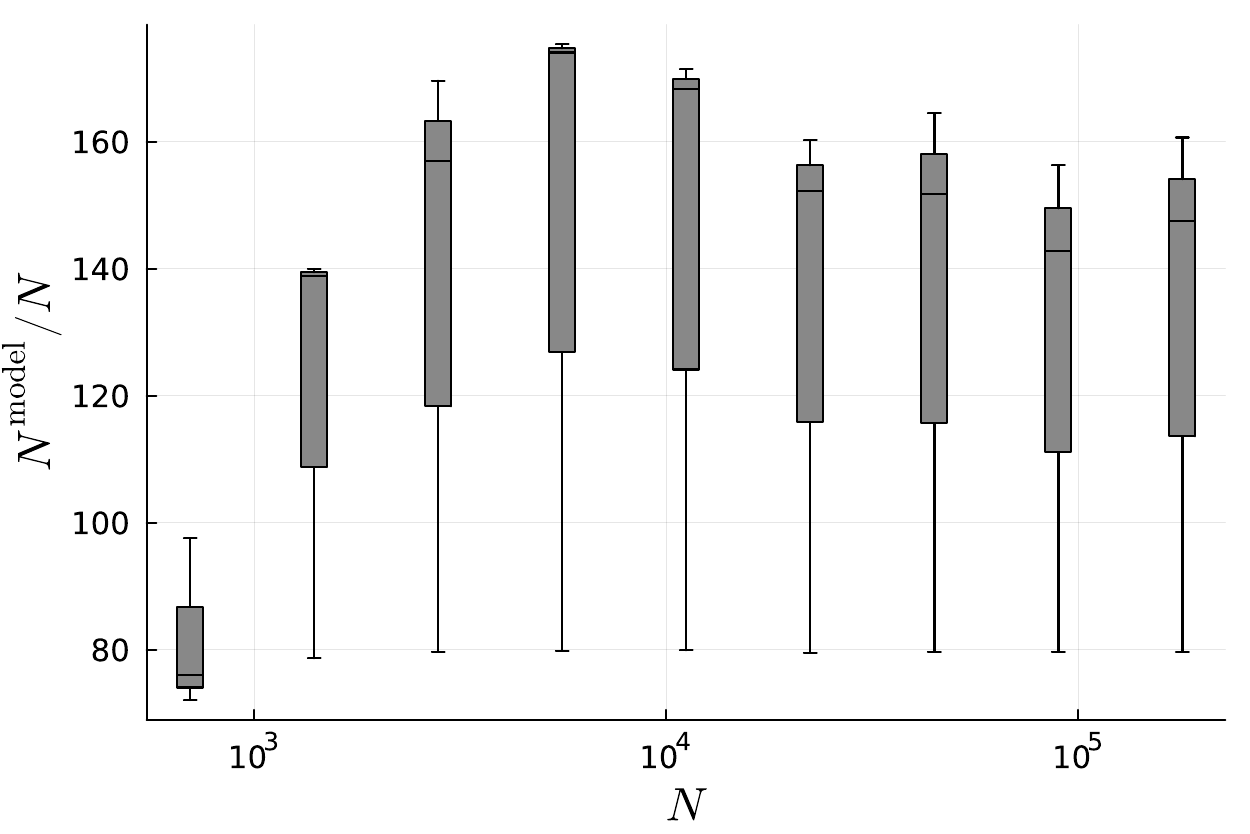}
    \caption{Number of NUTS model evaluations versus the sample size for the thermal conduction system, with the number of evaluations normalized by the sample size.}
    \label{fig:Nlike_hc}
\end{figure}

\begin{figure}
    \centering
    \includegraphics[width=0.5\linewidth]{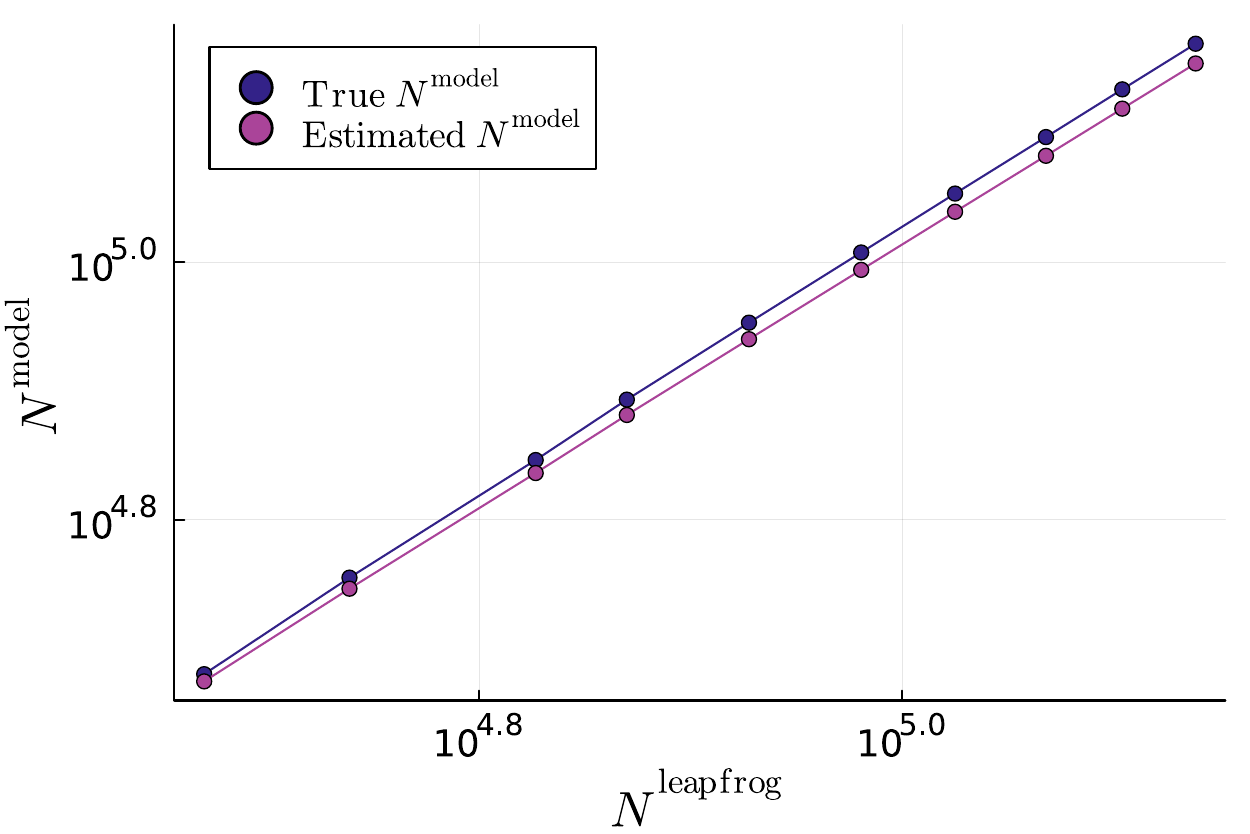}
    \caption{\textcolor{black}{Comparison between the estimated and measured number of model evaluations as a function of the number of leapfrog steps for the NUTS sampler applied to the heat conduction problem. Results are shown for different sample sizes, ranging from 500 to 5000 in increments of 500.}}
    \label{fig:leapfrog-vs-model-eval}
\end{figure}

\paragraph{Computational effort}
In \autoref{fig:Nlike_hc} we study the computational effort of the NUTS sampler based on the number of model evaluations as discussed in \autoref{sec:performance}.
The model evaluations, $N^{\text{model}}$, are normalized by the sample size, with the box plots representing the results obtained using three chains.
Since both the MH and AISM samplers require a single model evaluation per Markov chain step, their relative number of model evaluations is equal to one by definition, and hence these samplers are not plotted.
\textcolor{black}{In contrast, the NUTS requires significantly more model evaluations per step.
For a total number of samples $N$, the total number of model evaluations for the HMC algorithm (Algorithm~\ref{alg:hamiltonian}) can be estimated by $N^{\mathrm{model}} = N^{\mathrm{leapfrog}} + N$, where $N^{\mathrm{leapfrog}}$ denotes the total number of leapfrog steps.
This estimate relies on the assumption that gradients can be evaluated using a single model evaluation, which we achieve in our implementation using automatic differentiation based on dual numbers \cite{Baydin2018}.  The linear scaling between the number of leapfrog steps and the number of model evaluations is confirmed numerically in \autoref{fig:leapfrog-vs-model-eval}. The observed minor mismatch is attributed to additional model and gradient evaluations in the NUTS algorithm \cite{Hoffman2014} compared to the HMC algorithm.}

\textcolor{black}{The number of model evaluations per Markov chain step is therefore directly linked to the (adaptive) number of leapfrog steps, which varies across iterations in the NUTS algorithm.
In our experiments, this results in 80--180 model evaluations per Markov chain step.}
This increase in number of model evaluations per step implies that the computational effort of the NUTS per step is two orders of magnitude larger than for the other two solvers.
Of course, in terms of total computational effort this must be weighed against the substantial improvement in effective sample size as observed in \autoref{fig:essN_hc}, which is a factor 20 lower for the MH sampler and a factor 50 lower for the AISM sampler \textcolor{black}{when compared for a fixed sample size}. In terms of computational load per \textcolor{black}{effective sample}, the NUTS is in this case not beneficial.
 
\subsection{Viscous flow system}\label{sec:results:viscous}
We infer the six model parameters $\boldsymbol{\theta} = [F, V, R_0, \eta, \gamma, \alpha]$ using $M=3$ chains with $N^{\rm max}=\num{153600}$ samples. Prior to generating the MH sampler chains, we performed \num{153600} steps using the adaptive sampling algorithm to calibrate the proposal distribution. \textcolor{black}{These calibration samples, as well as the computational effort of the calibration procedure, are not taken into consideration in our analysis.} For the AISM sampling method we use 24 walkers, i.e., four walkers per parameter.

\begin{figure}
    \centering
   \includegraphics[width=\linewidth]{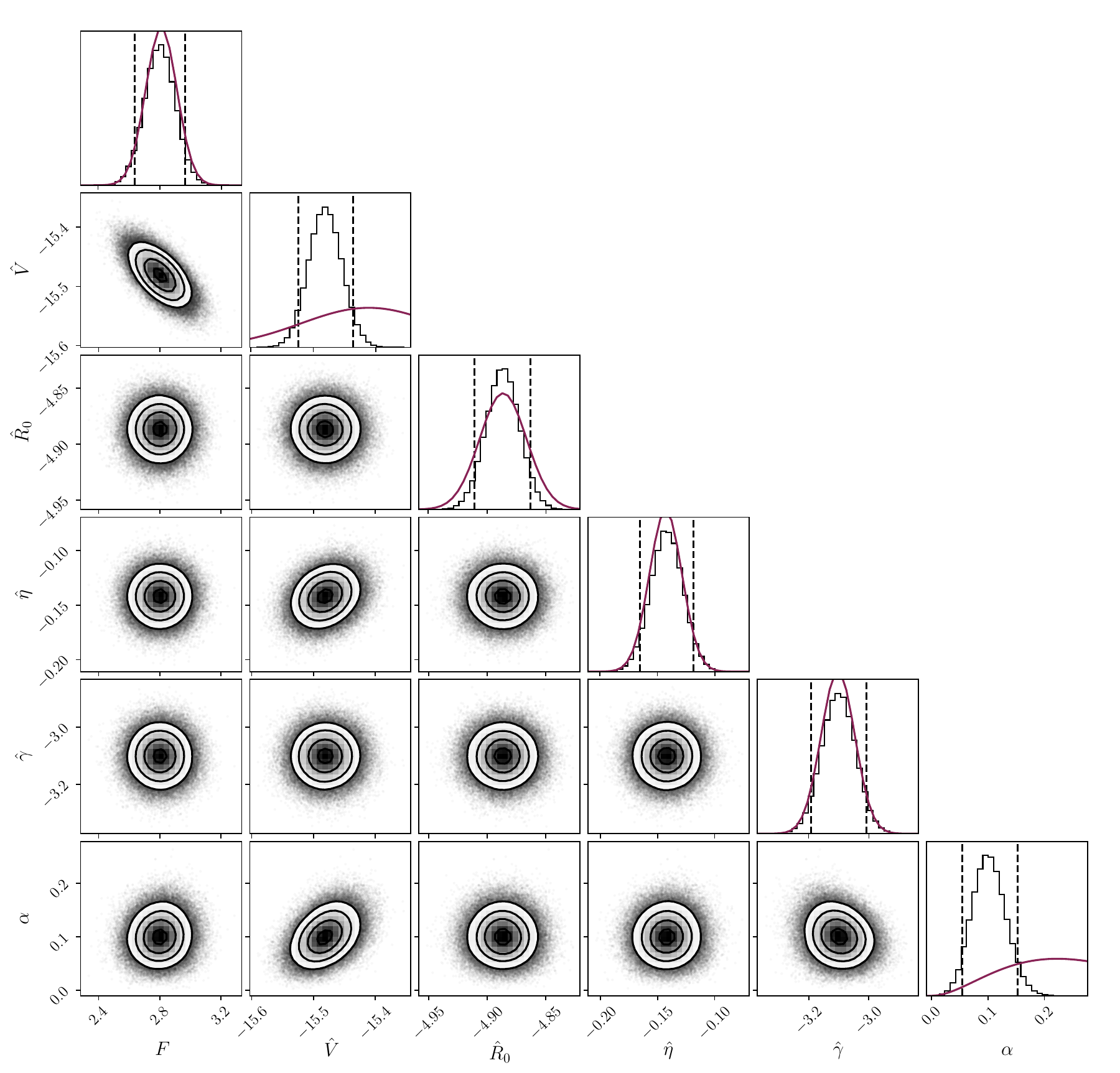}
    \caption{Inferred model parameters (see \autoref{tab:priorparam} for units) and correlations for the viscous flow system. Circumflexes are used to indicate log-transforms of the parameters. The top part of each column compares the marginal prior (dark-red line) and posterior (histogram). \textcolor{black}{The vertical dashed lines in the marginal distributions represent the 95\% credible intervals.}}
    \label{fig:post_sf}
\end{figure}

\autoref{fig:post_sf} shows the posterior results based on one of the chains of the AISM sampler. The posterior distribution for the volume, $V$, and interface curvature parameter, $\alpha$, are observed to shift substantially compared to the prior (red line in the marginal distributions). This is explained by the fact that it is difficult to obtain accurate prior information for the volume through weighing, and that virtually no prior information is available for the curvature parameter. For the other parameters, the priors are observed to align well with the posteriors, indicating limited sensitivity to the experimental data, in contrast to the thermal system. The observed correlation between the force, $F$, and the volume, $V$, is a consequence of both parameters influencing the radius evolution in a similar manner \cite{Rinkens2023}.

\begin{figure}
    \centering
   \includegraphics[width=0.5\linewidth]{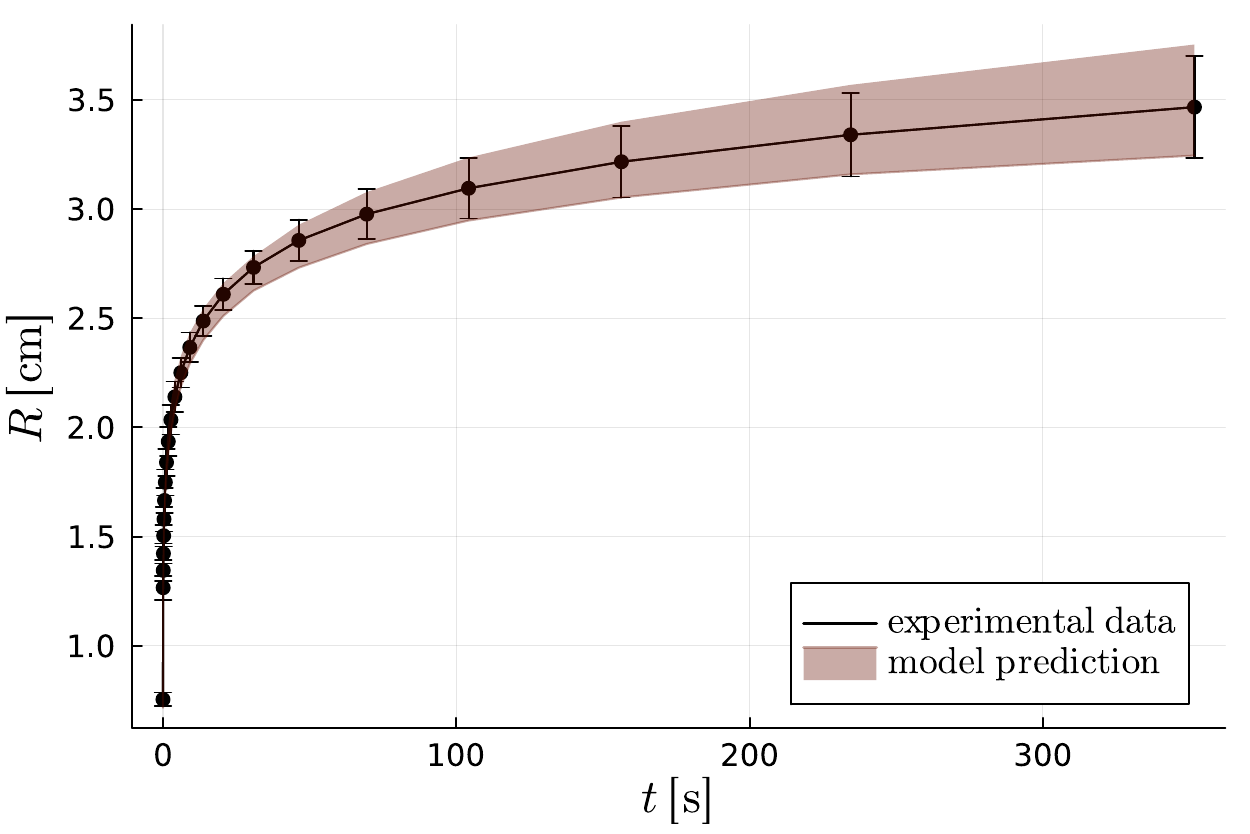}
    \caption{Comparison between the experimental data and the posterior predictive distribution for the viscous system.}
    \label{fig:modepred_sf}
\end{figure}

The predictive posterior is compared with the experimental data in \autoref{fig:modepred_sf} by examining the evolution of the radius of the fluid layer.   The 95\% credible interval of the predictive posterior is observed to closely resemble the 95\% confidence interval of the experimental data. Similar results for the posterior are obtained using the other chains and samplers.

\begin{figure}
    \centering
    \subfloat[]{\includegraphics[width=0.48\textwidth]{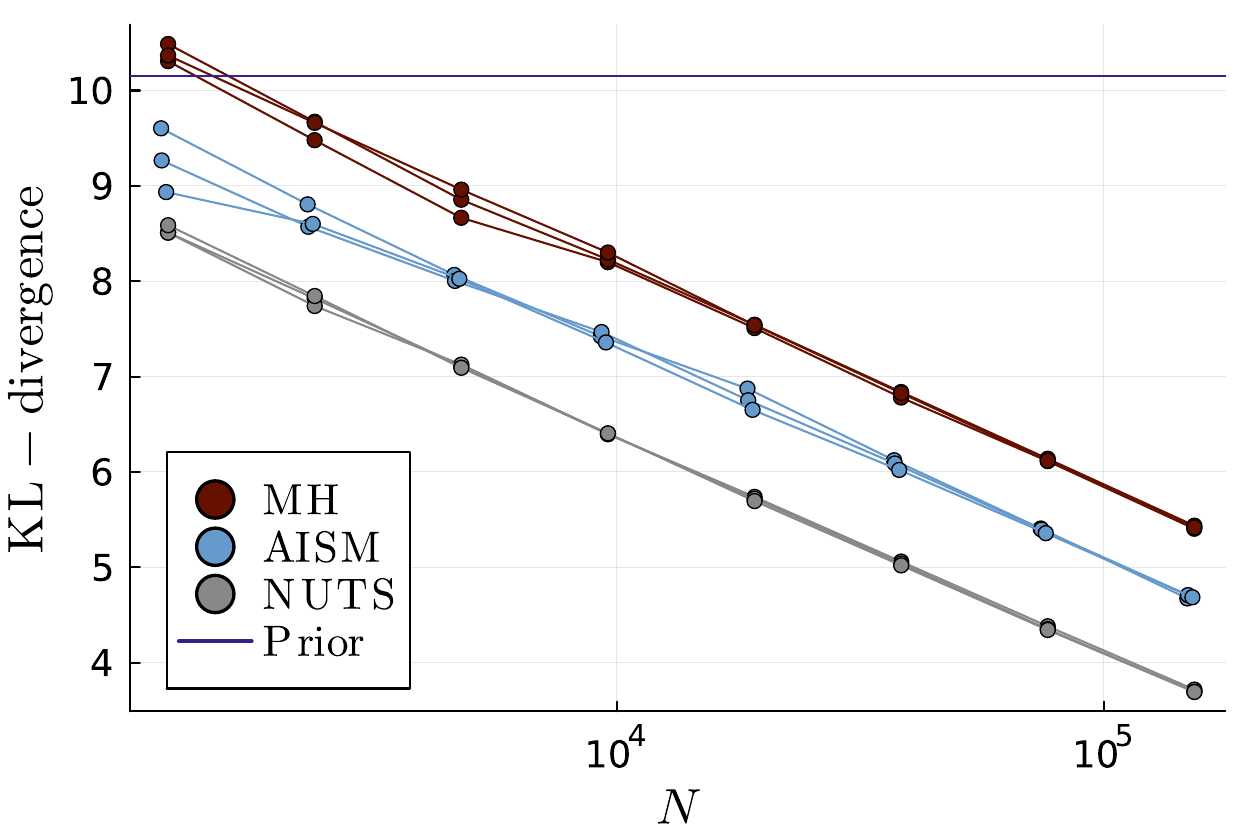}}
    \hfill
    \subfloat[]{\includegraphics[width=0.48\textwidth]{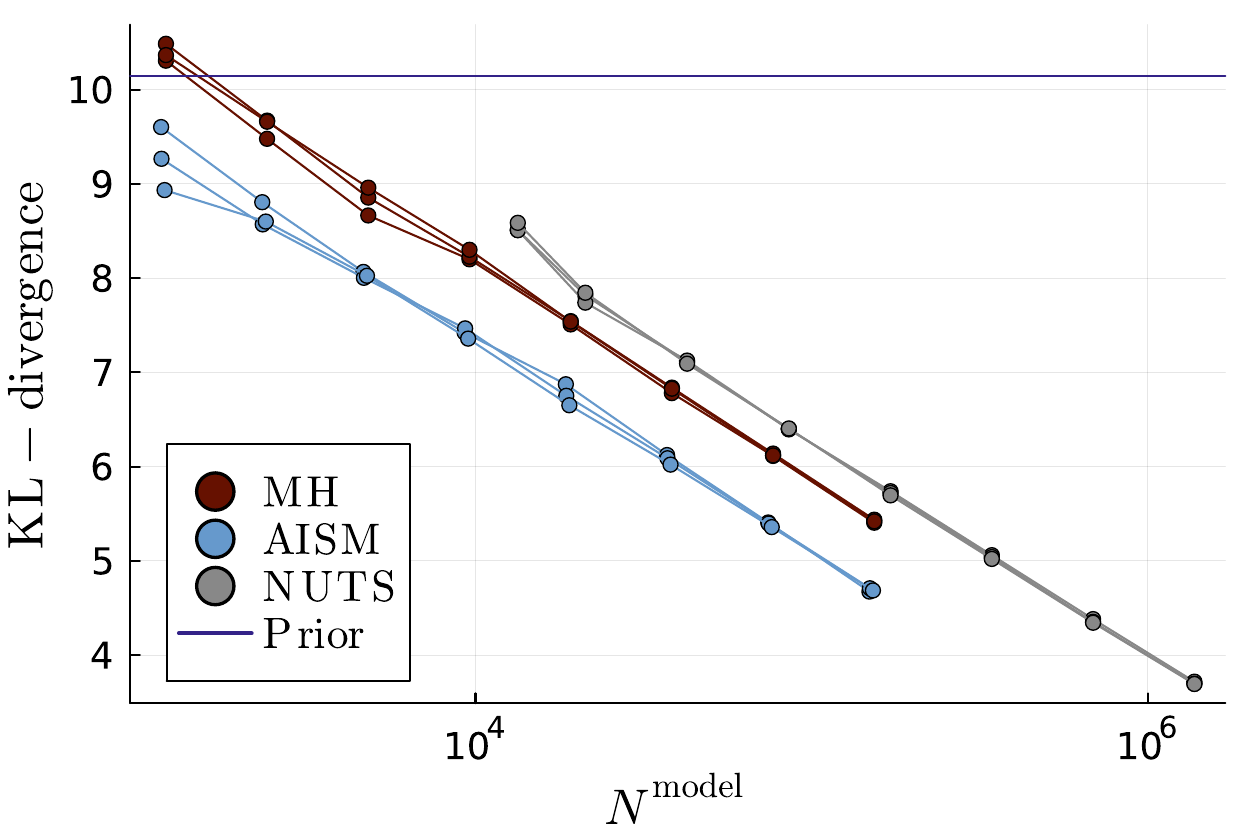}}
    \hfill
    \subfloat[]{\includegraphics[width=0.48\textwidth]{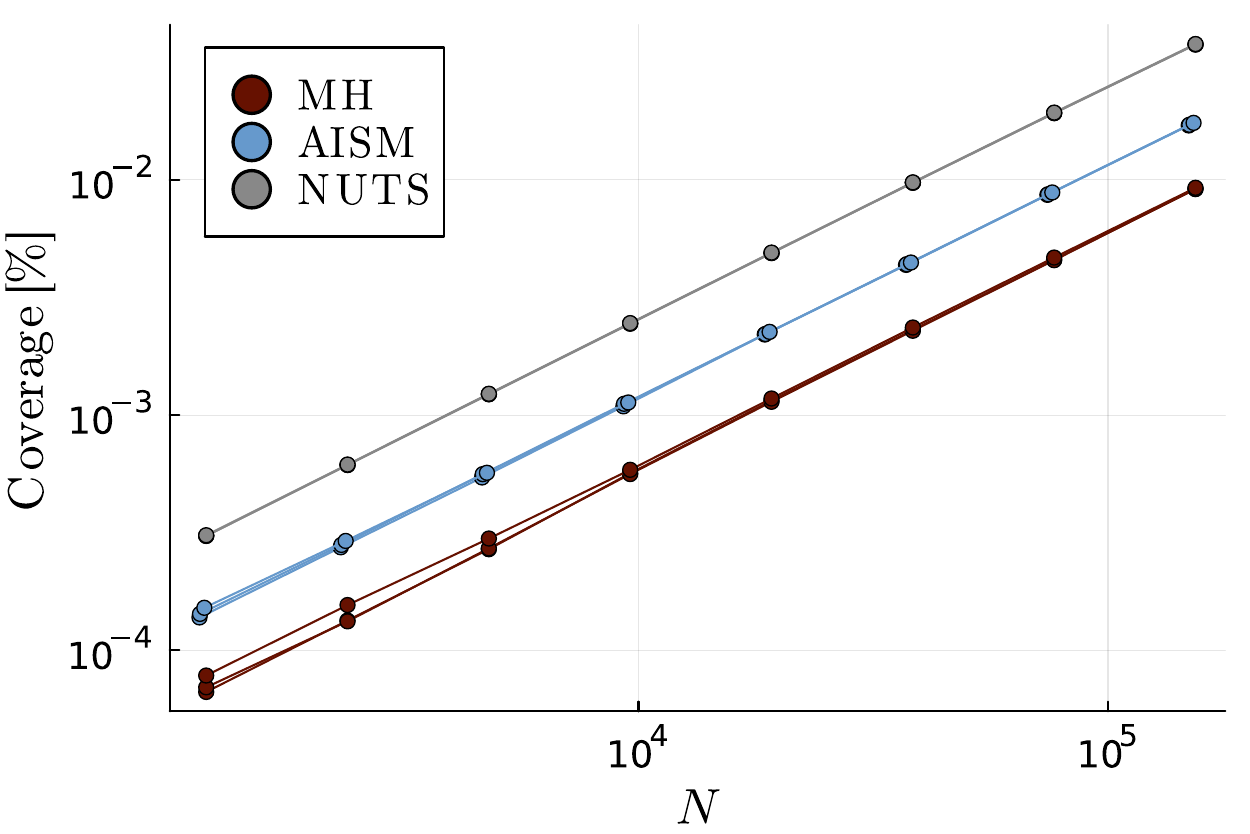}}
    \hfill
    \subfloat[]{\includegraphics[width=0.48\textwidth]{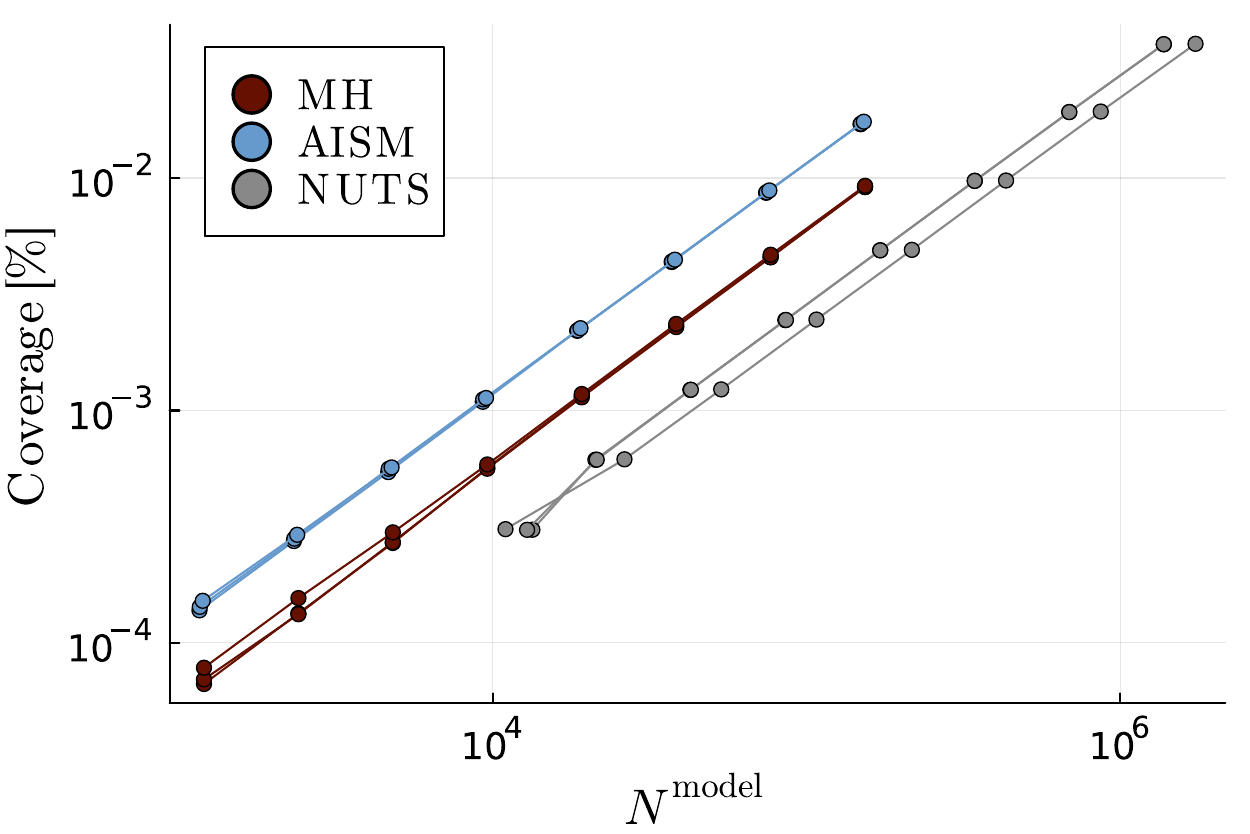}}
    \hfill
    \caption{{\color{black} {KL-divergence and coverage of the sampled posterior for the viscous flow system.}}}
    \label{fig:KL_sf}
\end{figure}

\paragraph{KL-divergence}
In \autoref{fig:KL_sf} we show the KL-divergence for the three samplers and the prior with respect to the reference solution evaluated on a rectilinear mesh (\autoref{eq:refgrid}) with $N^{\mathrm{bin}}_i=27$, with $i=1,\ldots,6$. The KL-divergence of the prior is observed to be approximately 10, which is comparable to the posterior obtained using the smallest sample sizes for the MH and AISM samplers. This is explained by the fact that the prior marginals for most of the parameters are well-aligned with the posteriors (\autoref{fig:post_sf}) and that hence a significant sample size is required to yield a further reduction in the KL-divergence. 

For all three samplers the KL-divergence is observed to converge at a fixed rate, but there is a notable difference in magnitude between the samplers. For a fixed KL-divergence, the AISM sampler and the NUTS require approximately 50\% and 80\% fewer samples, respectively, compared to the MH sampler. \textcolor{black}{However, when compared against the number of model evaluations, the MH sampler outperforms the the NUTS, which is a direct consequence of the NUTS requiring multiple model evaluations per Markov step. This similarly affects the coverage, which illustrates that on a per-step basis the NUTS outperforms the other two samplers, whereas this is the opposite on a per-model-evaluation basis.}


\begin{figure}
\centering
\subfloat{\includegraphics[width=0.48\textwidth]{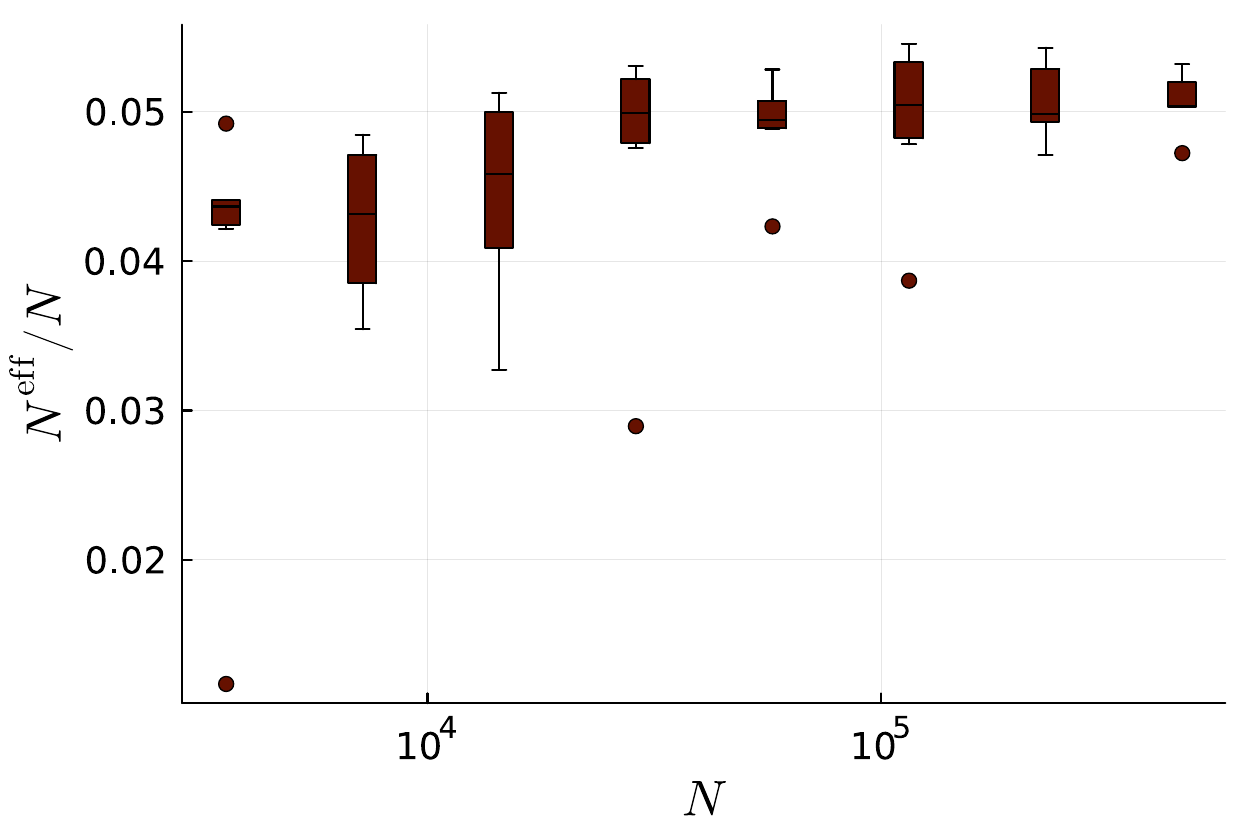}}
 \hfill
\subfloat{\includegraphics[width=0.48\textwidth]{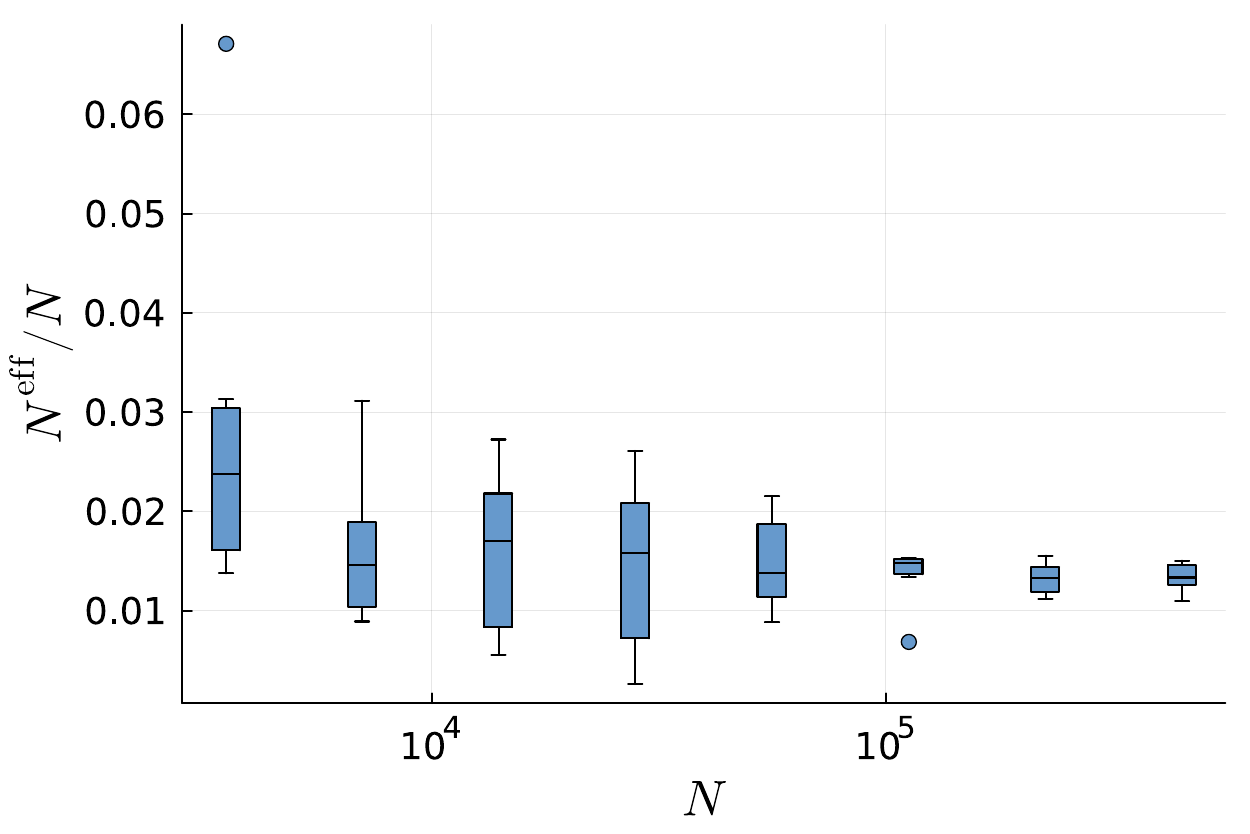}}
\hfill
\subfloat{\includegraphics[width=0.48\textwidth]{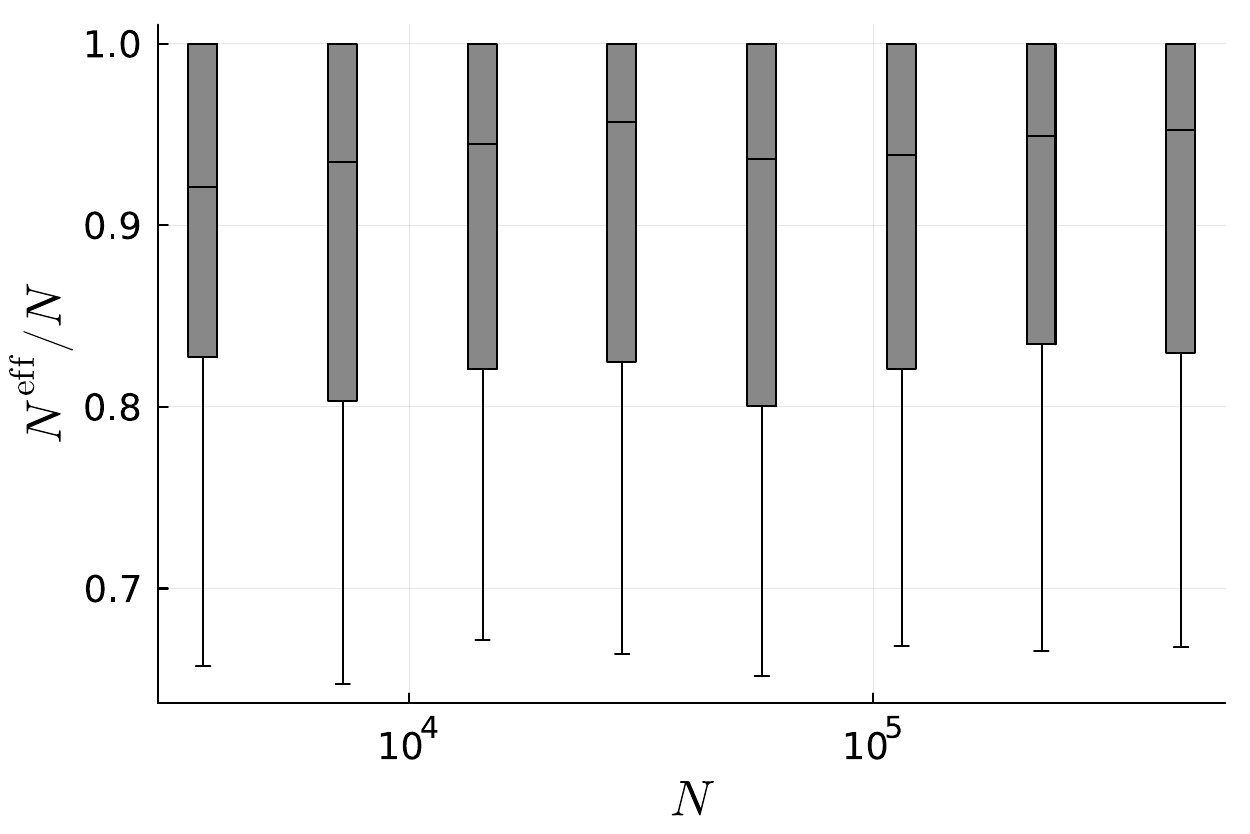}}
\hfill
\subfloat{\includegraphics[width=0.48\textwidth]{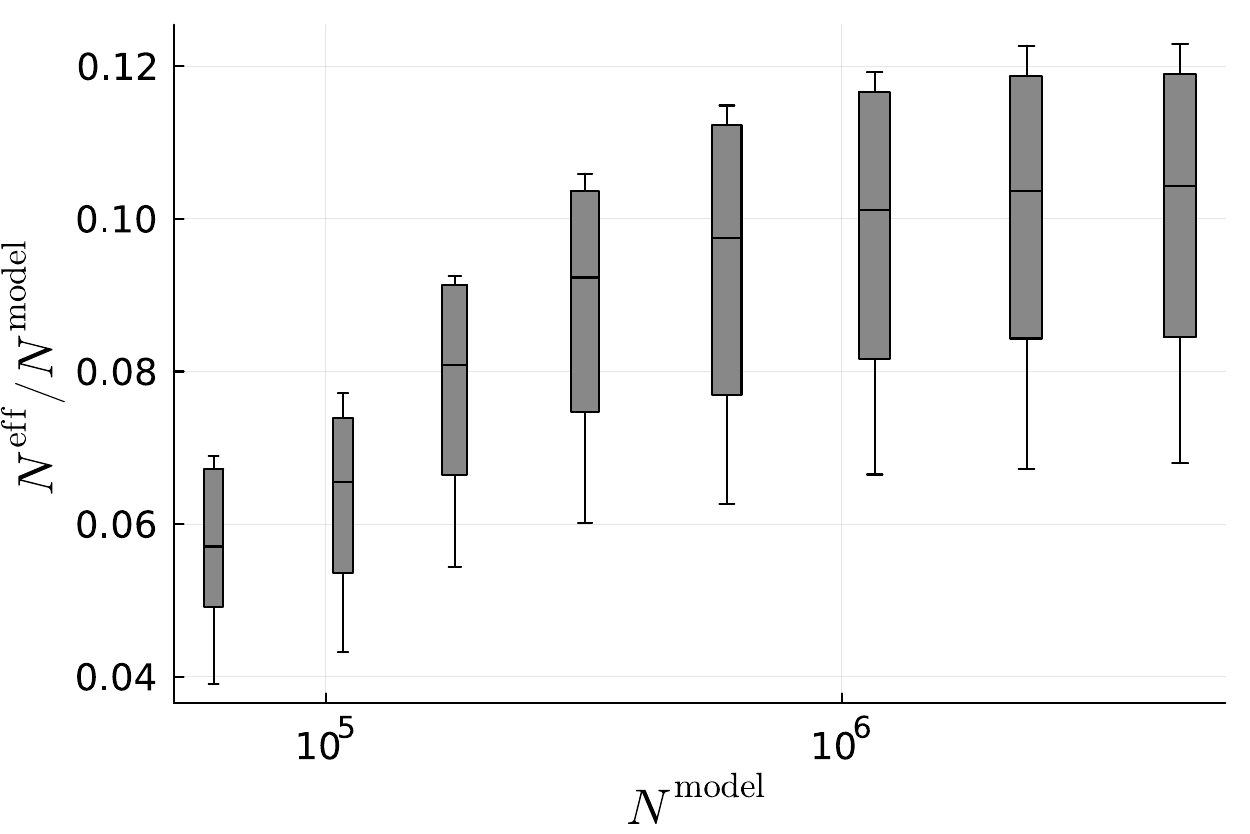}}
\caption{\color{black} {Normalized effective sample size for the viscous flow system: (a) MH, (b) AISM, (c, d) NUTS. For the NUTS, normalization with respect to both the number of samples and the number of model evaluations is considered.}}
\label{fig:essN_sf}
\end{figure} 

\paragraph{Heuristics}
The effective sample size for the different samplers is shown in \autoref{fig:essN_sf}. Different vertical scales are used to clearly visualize the different samplers. The effective sample size for the NUTS (85\%-100\%) is observed to be approximately 20 times larger than that for the MH sampler (4.5\%-5.5\%) and approximately 60 times larger than that of the AISM sampler (1\%-3\%) \textcolor{black}{when normalized by the sample size. When normalized by the number of model evaluations, the effective sample size of the NUTS (5\%-12\%) still outperforms the other two samplers, albeit marginally when compared to the MH sampler.}

\begin{figure}
\centering
\subfloat{\includegraphics[width=0.48\textwidth]{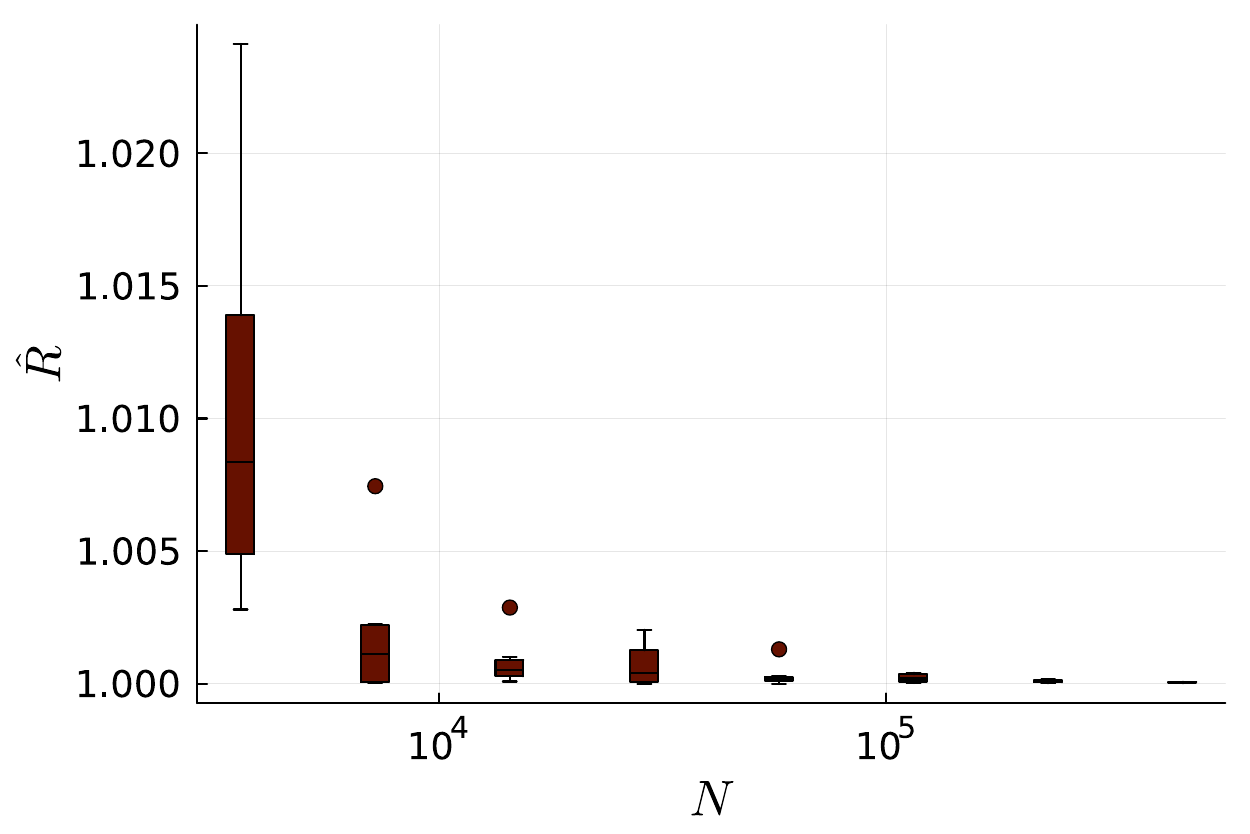}}
 \hfill
\subfloat{\includegraphics[width=0.48\textwidth]{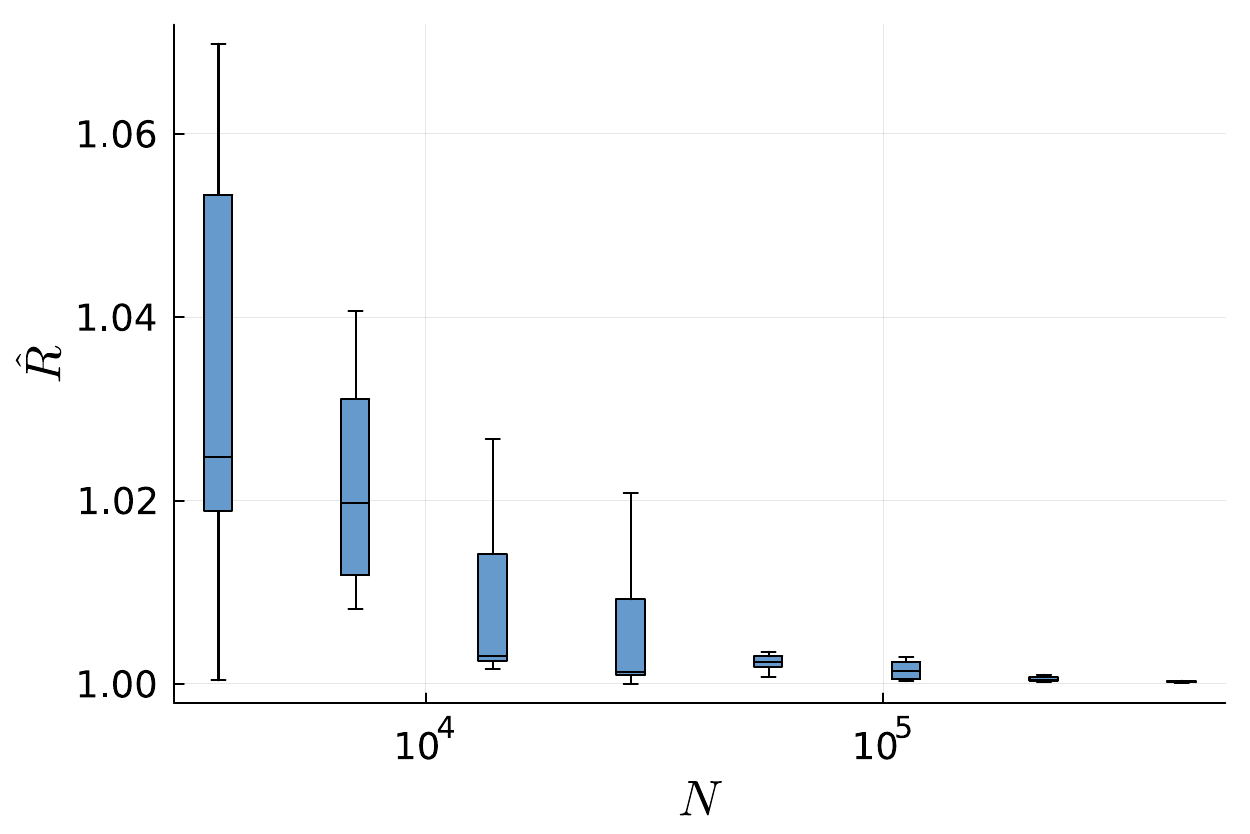}}
\hfill
\subfloat{\includegraphics[width=0.48\textwidth]{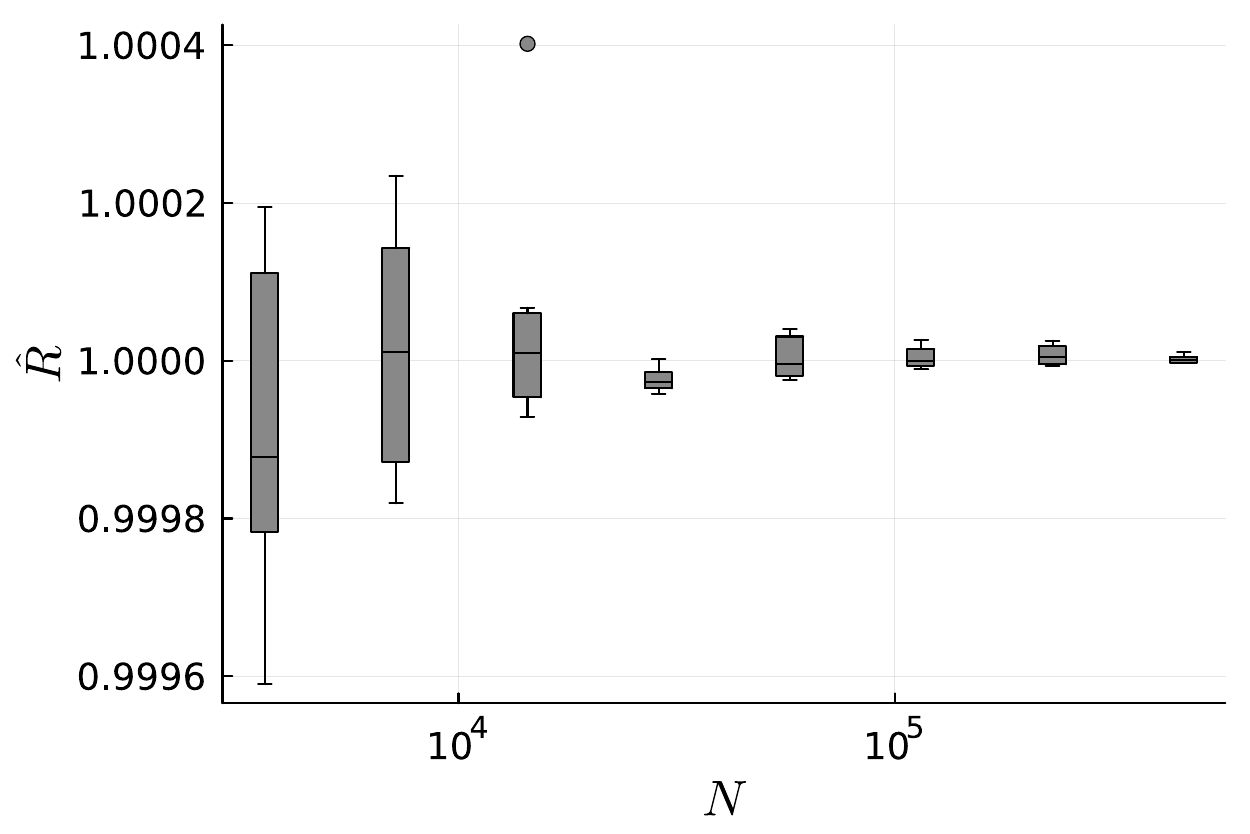}}
\hfill
\subfloat{\includegraphics[width=0.48\textwidth]{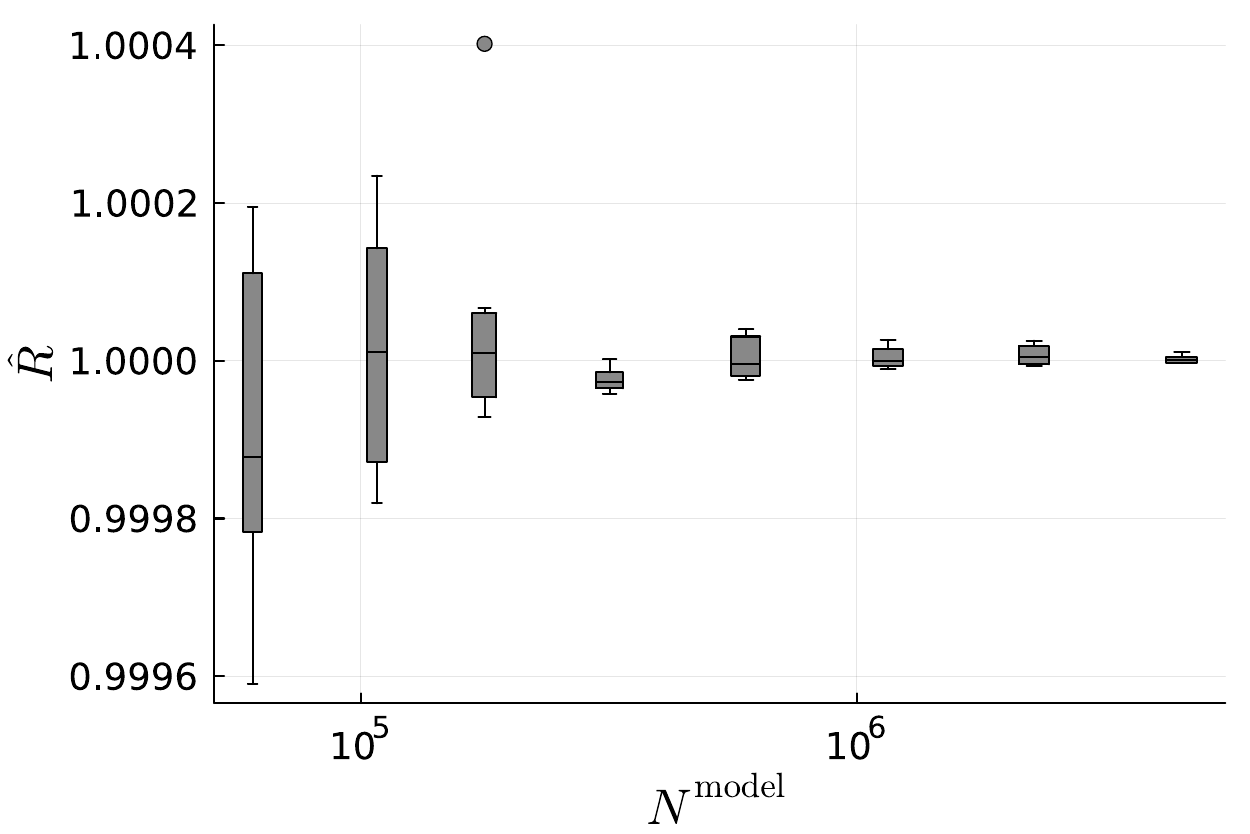}}
\caption{{\color{black} {Gelman-Rubin diagnostic for the viscous flow system: (a) MH, (b) AISM, (c, d) NUTS.}}}
\label{fig:rhat_sf}
\end{figure}

The Gelman-Rubin diagnostic for the three samplers is presented in \autoref{fig:rhat_sf}, with the box plot representing the $\hat{R}$ values for the different parameters and chains. Different vertical scales are used to clearly visualize the different samplers. The observed $\hat{R}$ values are consistent with the effective sample sizes in \autoref{fig:essN_sf}, in the sense that the NUTS performs best, and that the MH sampler outperforms the AISM sampler. This suggests that, for the considered sampler optimization parameters, the MH mixes better than the AISM sampler. As no strong correlations are observed in \autoref{fig:post_sf}, the AISM sampler is not expected to yield a substantial improvement compared to MH.


\begin{figure}
    \centering
    \includegraphics[width=0.5\linewidth]{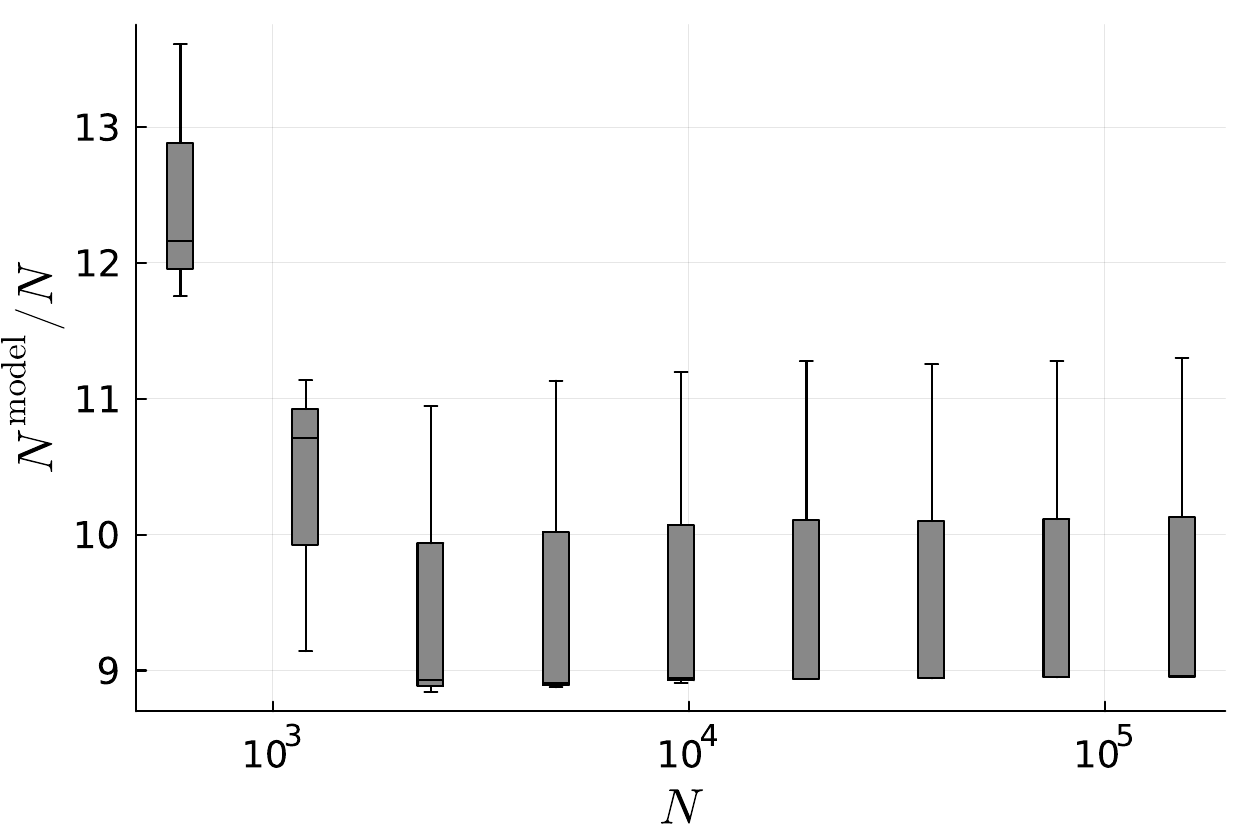}
    \caption{Number of NUTS model evaluations versus the sample size for the viscous flow system, with the number of evaluations normalized by the sample size.}
    \label{fig:Nlike_sf}
\end{figure}

\begin{figure}
    \centering
    \includegraphics[width=0.5\linewidth]{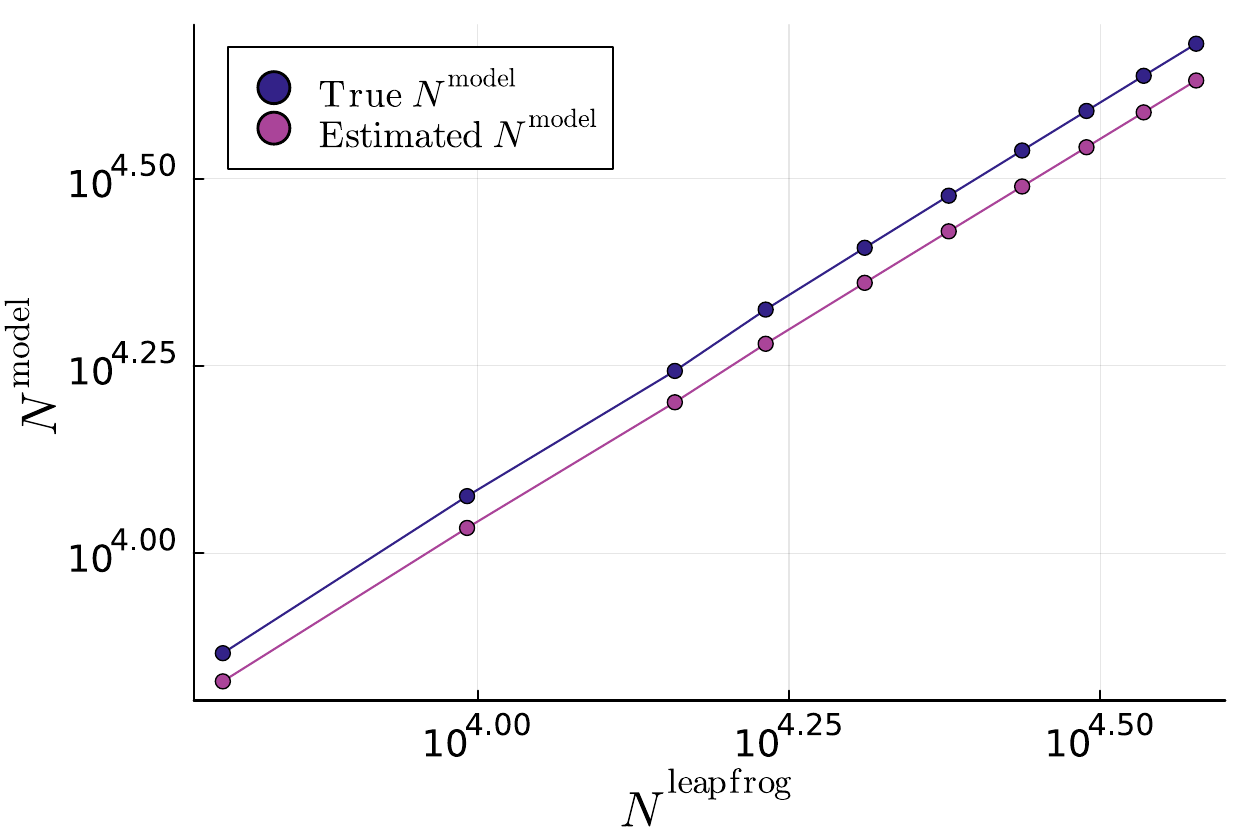}
    \caption{\textcolor{black}{Comparison between the estimated and measured number of model evaluations as a function of the number of leapfrog steps for the NUTS sampler applied to the squeeze flow problem. Results are shown for different sample sizes, ranging from 500 to 5000 in increments of 500.}}
    \label{fig:leapfrog-vs-model-eval_sf}
\end{figure}

\paragraph{Computational effort}
As discussed in \autoref{sec:performance}, the number of model evaluations serves as a metric to compare the computational effort between the samplers.
The number of model evaluations, normalized by the sample size, is presented in \autoref{fig:Nlike_sf} for the NUTS.
For the MH and AISM samplers, each step entails a single model evaluation, without the need to evaluate gradients.
\textcolor{black}{For the NUTS, the number of model evaluations again relates directly to the number of  leapfrog steps (\autoref{fig:leapfrog-vs-model-eval_sf}) when gradients can be evaluated in a single model evaluation using automatic differentiation.} For the viscous flow system we observe the number of model evaluations per step to decrease from approximately 12.5 to 10 when increasing the sample size from \num{1000} to above \num{10000}.
We attribute this decrease to the reduced influence of the starting position of the Markov chains for increasing sample sizes.
The increased number of model evaluations for the NUTS is evidently compensated by the increased effective sample size, when compared to the other two samplers.
With the effective sample sizes for the AISM and MH samplers being respectively a factor 60 and 20 lower than that for the NUTS \textcolor{black}{when compared for a fixed sample size}, for this problem, in terms of computational load per Markov chain step there is a net benefit for the NUTS.
This observation is explained by the fact that, for the semi-analytical model considered for this problem, the leapfrog scheme permits relatively large steps on account of the regularity/smoothness of the posterior distribution.

\subsection{Similarities and differences between the physical systems}\label{sec:discussion}
Based on the observed similarities and differences between the sampler comparisons for the two physical systems (\autoref{sec:results:thermal} and \autoref{sec:results:viscous}), in this section we discuss the overarching observations regarding the performance of the samplers.

\paragraph{KL-divergence}
The KL-divergence (\autoref{fig:KL_hc} and \autoref{fig:KL_sf}) for both systems \textcolor{black}{converges} at a fixed rate when the sample size becomes large enough. This constant rate of convergence is more pronounced for the viscous flow system. We attribute this difference to \textcolor{black}{three} factors. First, the thermal system overall results in a bigger posterior shift compared to the prior, which is used to sample starting points for the samplers. The apparent consequence of this is that larger sample sizes are required to reach the constant rate convergence regime in the KL-divergence. \textcolor{black}{The second contributing factor is that the posterior distribution following from the semi-analytical viscous flow model is expected to be more regular/smooth than that of the finite element based thermal model.} The \textcolor{black}{third} factor explaining the observed difference pertains to the quality of the reference solution, which uses smaller bins for the viscous flow problem due to the smaller number of parameters. This effect is also observed from the flattening of the KL-divergence curves for large sample sizes. \textcolor{black}{A more detailed investigation of how these contributing factors influence the KL-divergence is a topic of further study.}

In terms of the KL-divergence, the NUTS stands out as the \textcolor{black}{favorable} sampler for both problems \textcolor{black}{when compared against the sample size.} For the viscous flow problem the AISM sampler outperforms the MH sampler, whereas their performance is similar for the thermal conduction problem. \textcolor{black}{When plotted against the number of model evaluations, the NUTS is outperformed by the other two samplers for both cases. As reflected by the domain coverage, the NUTS steps explore the parameter domain effectively, but this comes at the expense of an increased number of model evaluations when compared against the other two samplers.}

\paragraph{Heuristics}
The Gelman-Rubin diagnostic (\autoref{fig:rhat_hc} and \autoref{fig:rhat_sf}) for both systems converges toward one for increasing sample sizes. For both cases a strong correlation with the effective sample size (\autoref{fig:essN_hc} and \autoref{fig:essN_sf}) is found on account of the inverse proportionality of the between-chain variance with the sample size. In terms of ranking, a clear performance gap between the NUTS and the other two samplers is observed \textcolor{black}{when normalized with the number of samples}. In contrast to the KL-divergence, the NUTS is followed by the MH sampler and finally the AISM sampler for both systems. \textcolor{black}{For both cases, the significant number of model evaluations per NUTS step affects this comparison when normalizing/comparing with the number of model evaluations. For the heat conduction problem, the effective sample size becomes smaller than that of the other two samplers, whereas NUTS marginally maintains its advantage for the viscous flow system.}

\paragraph{Computational effort}
The AISM and MH samplers compute one model evaluation per step by definition. In contrast, the NUTS requires a significant number of model evaluations per step on account of the leapfrog steps, including gradient evaluations (\autoref{fig:Nlike_hc} and \autoref{fig:Nlike_sf}). For the thermal problem, the number of model evaluations is observed to be an order of magnitude larger than for the viscous flow problem, which has been traced back to an increase in the number of leapfrog steps per Markov chain step. We attribute this difference to the complexity of the models considered, as a semi-analytical model is used for the viscous flow problem, whereas a time-dependent finite element problem is solved for the thermal case. This complexity indirectly affects the leapfrog step size through the regularity/smoothness of the posterior distribution. For the viscous flow problem, the increase in effective sample size for the NUTS compared to the other two samplers outweighs its increase in computational effort per Markov chain step. A more balanced comparison in terms of computational effort is observed for the thermal conduction problem. In this case, the MH sampler and AISM sampler perform similarly, but with the latter relying less on computationally demanding step size calibration.
\section{Conclusions and recommendations}\label{sec:6}
We investigated and compared the convergence behavior and performance of three prominent Markov Chain Monte Carlo sampling methods, viz., the Metropolis Hastings (MH) sampler, the Affine Invariant Stretch Move (AISM) sampler, and the No-U-Turn Sampler (NUTS). We did this in the context of two distinct Bayesian inference problems with tailor-made experimental setups, namely a thermal conduction problem modeled using finite elements and a viscous flow problem modeled using a semi-analytical expression. Our comparison is based on the Kullback-Leibler divergence, heuristic criteria in the form of the effective sample size and Gelman-Rubin diagnostic, and the number of model evaluations as a measure of the computational cost.

An innovative aspect of our comparison is the usage of the KL-divergence as a metric to study the convergence of the samplers. For both problems considered and for all three samplers, the KL-divergence is observed to converge as the sample size increases. The NUTS is observed to outperform the other two samplers when considering the KL-divergence at a fixed sample size, which is in agreement with the observed significantly higher effective sample size for this sampler. \textcolor{black}{For a fixed number of model evaluations, the NUTS is outperformed by the other two samplers on account of it requiring multiple model evaluations per Markov step.}

Since the evaluation of the KL-divergence is time consuming for high-dimensional problems, in practical situations convergence is instead assessed using heuristic criteria such as the Gelman-Rubin diagnostic. We have found the behavior of this diagnostic (and that of the effective sample size) indicative of the convergence of the KL-divergence. For each of the given samplers we observed that, upon increasing the sample size, the Gelman-Rubin statistic converges to one, the normalized effective sample size reaches a steady state, and the KL-divergence decreases. This qualitative correlation between the KL-divergence and the heuristics indicates \textcolor{black}{their} suitability to assess convergence for the considered cases. However, clear quantitative relations between the convergence criteria have not been observed. Most notably, although, \textcolor{black}{for a fixed sample size}, the NUTS also significantly outperforms the other two samplers in the heuristics, the performance ranking of the MH and AISM samplers was found to be different between the KL-divergence and the heuristics.

Although the NUTS outperforms the other two samplers for both problems and for all considered convergence criteria \textcolor{black}{when the sample size is fixed}, this performance comes at a significant computational cost. The NUTS -- as well as other Hamiltonian Monte Carlo methods -- performs leapfrog steps to form a proposal for the next Markov chain step, resulting in an increased computational effort per Markov Chain step. We have found the NUTS to be net beneficial in terms of computation time for the semi-analytical model, but not for the finite element model, which illustrates the problem- and implementation-dependent nature of the balance between effective sample size and computation time per step.

While often the NUTS (and other Hamiltonian Monte Carlo methods) may on paper be favorable, its practical utility can be limited by the necessity to evaluate gradients. In contrast, the MH and AISM samplers do not rely on gradient information, making them less computational demanding per Markov Chain step, and making them easier to implement. The AISM sampler, through the use of multiple walkers, can effectively sample complex posterior distributions, and reduces the chance that the Markov chain becomes trapped in local minima. In contrast to the MH sampler -- for which the calibration of the proposal covariance matrix \textcolor{black}{can require computational effort comparable to the sampling itself} -- the default configuration of the AISM sampler is suitable for a broad range of problems. In addition, although not considered in this work, the AISM sampler can be parallelized efficiently \cite{foreman2013emcee}, giving it an additional competitive edge compared to the other samplers.
\begin{algorithm}[H]
    \caption{Metropolis-Hastings (MH) algorithm}\label{alg:mh}
    \hspace*{\algorithmicindent} \textbf{Input:} data $\bm{y}$, sample size $N$
    \\
    \hspace*{\algorithmicindent} \textbf{Output:} Markov chain $\boldsymbol{\mathcal{C}}=\{ \boldsymbol{\theta}_n\}_{n=1}^N$
    \begin{algorithmic}[1]
        \State $\boldsymbol{\theta}_0 \gets \Call{sample\_prior}{ }$ \label{line-mh-sample-initial-guess}
        \Comment{Starting point}
        \State $\pi^{\bm{\theta}_0}_\text{prio} \gets \Call{eval\_prior}{\bm{\theta}_0}$
        \State $L^{\bm{\theta}_0 | \bm{y}} \gets \Call{eval\_likelihood}{\bm{\theta}_0, \bm{y}}$
        \For{$n=1,\ldots,N$}
            \State $\tilde{\boldsymbol{\theta}} \gets \Call{sample\_proposal}{\bm{\theta}_{n-1}}$ \label{line-mh-sample-proposal}
            \Comment{Proposal}
            \State $\pi^{\tilde{\bm{\theta}}}_\text{prio} \gets \Call{eval\_prior}{\tilde{\bm{\theta}}}$
            \State $L^{\tilde{\bm{\theta}} | \bm{y}} \gets \Call{eval\_likelihood}{\tilde{\bm{\theta}}, y}$\label{line-mh-likelihood}
            \State $\pi_\text{prop}^{\bm{\theta}_{n-1} | \tilde{\bm{\theta}}},~\pi_\text{prop}^{\tilde{\bm{\theta}} | \bm{\theta}_{n-1}} \gets \Call{eval\_proposal}{\bm{\theta}_{n-1}, \tilde{\bm{\theta}}}$ \label{line-mh-proposal}
            \State $\alpha \gets {\rm min} \left( 1,
            \dfrac{L^{\tilde{\bm{\theta}} | \bm{y}}}{L^{\bm{\theta}_{n-1} | \bm{y}}}
            \dfrac{\pi^{\tilde{\bm{\theta}}}_\text{prio}}
            {\pi^{\bm{\theta}_{n-1}}_\text{prio}}
            \dfrac{\pi_\text{prop}^{\bm{\theta}_{n-1} | \tilde{\bm{\theta}}}}
            {\pi_\text{prop}^{\tilde{\bm{\theta}} | \bm{\theta}_{n-1}}}
            \right)$ \label{line-mh-acceptance-prob}
            \Comment{Acceptance probability}
            \If{$\alpha \geq \Call{sample\_std\_uniform}{ }$} \label{line-mh-accept}
                \Comment{Accept the proposal with probability $\alpha$}
                \State $\bm{\theta}_{n} \gets \tilde{\bm{\theta}}$
            \Else
                \Comment{Reject the proposal with probability $1-\alpha$}
                \State $\boldsymbol{\theta}_{n} \gets \boldsymbol{\theta}_{n-1}$
            \EndIf \label{line-mh-criterionendif}
        \EndFor
    \end{algorithmic}
\end{algorithm}

\begin{algorithm}[H]
    \caption{Affine-Invariant Stretch Move (AISM) algorithm}\label{alg:aism}
    \hspace*{\algorithmicindent} \textbf{Input:} data $\bm{y}$, sample size $N$, number of walkers $K$
    \\
    \hspace*{\algorithmicindent} \textbf{Output:} Markov chain $\boldsymbol{\mathcal{C}}=\{ \{\boldsymbol{\theta}_n^k\}_{n=1}^{N/K} \}_{k=1}^K$
    \begin{algorithmic}[1]
        \For{$k=1,\ldots,K$} \label{alg:aism_forwalk}
            \State $\bm{\theta}_0^k \gets \Call{sample\_prior}{ }$
            \Comment{Starting points}
            \State $\pi_\text{prio}^{\bm{\theta}_0^k} \gets \Call{eval\_prior}{\bm{\theta}_0^k}$ 
            \State $L^{\bm{\theta}_0^k | \bm{y}} \gets \Call{eval\_likelihood}{\bm{\theta}_0^k, \bm{y}}$
        \EndFor \label{alg:aism_endwalk}
        \For{$n=1,\ldots,N/K$}
            \For{$k=1,\ldots,K$}
                \State $j, z \gets \Call{sample\_proposal\_parameters}{ }$ \label{alg:aism_propsample}
                \State $\tilde{\bm{\theta}}^k \gets z \bm{\theta}_{n-1}^k + (1-z) \bm{\theta}_{n-1}^j$
                \label{alg:aism_propwalk}
                \State $\pi_\text{prio}^{\tilde{\bm{\theta}}^k} \gets \Call{eval\_prior}{\tilde{\bm{\theta}}^k}$ \label{alg:aism_propstart}
                \State $L^{\tilde{\bm{\theta}}^k | \bm{y}} \gets \Call{eval\_likelihood}{\tilde{\bm{\theta}}^k, \bm{y}}$
                \State $\pi_\text{prop}^{\bm{\theta}^k_{n-1} | \tilde{\bm{\theta}}^k},~ \pi_\text{prop}^{\tilde{\bm{\theta}}^k | \bm{\theta}^k_{n-1}} \gets \Call{eval\_proposal}{\bm{\theta}^k_{n-1}, \tilde{\bm{\theta}}^k}$
                \State $\alpha \gets {\rm min} \left( 1,
                \dfrac{L^{\tilde{\bm{\theta}}^k | \bm{y}}}{L^{\bm{\theta}^k_{n-1} | \bm{y}}}
                \dfrac{\pi^{\tilde{\bm{\theta}}^k}_\text{prio}}
                {\pi^{\bm{\theta}^k_{n-1}}_\text{prio}}
                \dfrac{\pi_\text{prop}^{\bm{\theta}^k_{n-1} | \tilde{\bm{\theta}}^k}}
                {\pi_\text{prop}^{\tilde{\bm{\theta}^k} | \bm{\theta}^k_{n-1}}}
                \right)$ \label{alg:aism_propend}
                \Comment{Acceptance probability}
                \If{$\alpha \geq \Call{sample\_std\_uniform}{ }$} \label{alg:aism_acceptpropstart}
                    \Comment{Accept the proposal with probability $\alpha$}
                    \State $\bm{\theta}_{n}^k \gets \tilde{\bm{\theta}}^k$
                \Else
                    \Comment{Reject the proposal with probability $1-\alpha$}
                    \State $\bm{\theta}_{n}^k \gets \bm{\theta}_{n-1}^k$
                \EndIf \label{alg:aism_acceptpropend}
            \EndFor
        \EndFor
    \end{algorithmic}
\end{algorithm}

\begin{algorithm}[H]
    \caption{Hamiltonian Monte Carlo algorithm}
    \label{alg:hamiltonian}
    \hspace*{\algorithmicindent} \textbf{Input:} data $\bm{y}$, sample size $N$, step size $\bm{\epsilon}$, number of leapfrog steps $L$
    \\
    \hspace*{\algorithmicindent} \textbf{Output:} Markov chain $\boldsymbol{\mathcal{C}}=\{ \bm{\theta}_n\}_{n=1}^N$
    \begin{algorithmic}[1]
        \State $\bm{\theta}_0 \gets \Call{sample\_prior}{ }$ \label{line-hmc-sample-initial-guess}
        \Comment{Starting point}
        \State $\pi^{\bm{\theta}_0}_\text{prio} \gets \Call{eval\_prior}{\bm{\theta}_0}$
        \State $L^{\bm{\theta}_0 | \bm{y}} \gets \Call{eval\_likelihood}{\bm{\theta}_0, \bm{y}}$
        \For{$n = 1, ..., N$}
            \State $\bm{r}_0 \gets \Call{sample\_std\_mv\_normal}{ }$ \Comment{Starting momentum}\label{line-hmc-startingmomentum}
            \State $\tilde{\bm{\theta}} \gets \bm{\theta}_{n-1}$, $\tilde{\bm{r}} \gets \bm{r}_0$
            \State $\nabla_{\bm{\theta}} \log ( \pi_{\text{prio}}^{\tilde{\bm{\theta}}} ) \gets \Call{eval\_prior\_gradient}{\tilde{\bm{\theta}}, \text{ log = true}}$
                \State $\nabla_{\bm{\theta}} \log ( L^{\tilde{\bm{\theta}} | \bm{y}} ) \gets \Call{eval\_likelihood\_gradient}{\tilde{\bm{\theta}}, \bm{y}, \text{ log = true}}$ \label{line-hmc-gradienteval1}
            \For{$i = 1, ..., L$}
            \label{line-hmc-startleapfrog-loop}
            \Comment{Leapfrog integration}
                \State $\tilde{\bm{r}} \gets \tilde{\bm{r}} + \frac{1}{2} \bm{\epsilon} \odot \left[ \nabla_{\bm{\theta}} \log ( \pi_{\text{prio}}^{\tilde{\bm{\theta}}} ) + \nabla_{\bm{\theta}} \log (  L^{\tilde{\bm{\theta}} | \bm{y}} ) \right]$
                \State $\tilde{\bm{\theta}} \gets \tilde{\bm{\theta}} + \bm{\epsilon} \odot \tilde{\bm{r}}$
                \State $\nabla_{\bm{\theta}} \log ( \pi_{\text{prio}}^{\tilde{\bm{\theta}}} ) \gets \Call{eval\_prior\_gradient}{\tilde{\bm{\theta}}, \text{ log = true}}$
                \State $\nabla_{\bm{\theta}} \log ( L^{\tilde{\bm{\theta}} | \bm{y}} ) \gets \Call{eval\_likelihood\_gradient}{\tilde{\bm{\theta}}, \bm{y}, \text{ log = true}}$ \label{line-hmc-gradienteval2}
                \State $\tilde{\bm{r}} \gets \tilde{\bm{r}} + \frac{1}{2} \bm{\epsilon} \odot \left[ \nabla_{\bm{\theta}} \log ( \pi_{\text{prio}}^{\tilde{\bm{\theta}}} ) + \nabla_{\bm{\theta}} \log (  L^{\tilde{\bm{\theta}} | \bm{y}} ) \right]$
            \EndFor\label{line-hmc-endleapfrog-loop}
            \State $\pi^{\tilde{\bm{\theta}}}_\text{prio} \gets \Call{eval\_prior}{\tilde{\bm{\theta}}}$
            \State $L^{\tilde{\bm{\theta}} | \bm{y}} \gets \Call{eval\_likelihood}{\tilde{\bm{\theta}}, \bm{y}}$
            \State $\alpha = \min \left\{ 1, \dfrac{\exp \left[ \log ( \pi_{\text{prio}}^{\tilde{\bm{\theta}}} ) + \log ( L^{\tilde{\bm{\theta}} | \bm{y}} ) - \frac{1}{2} \tilde{\bm{r}} \cdot \tilde{\bm{r}} \right]}{\exp \left[ \log ( \pi^{\bm{\theta}_0}_\text{prio} ) + \log ( L^{\bm{\theta}_0 | \bm{y}} ) - \frac{1}2 \bm{r}_0 \cdot \bm{r}_0 \right]} \right\}$
            \label{line-hmc-alpha}
            \Comment{Acceptance probability}
            \If{$\alpha \geq$ \Call{sample\_std\_uniform}{ }}\label{line-hmc-startacceptancecriterion}
                \State $\bm{\theta}_n \gets \tilde{\bm{\theta}}$
                \Comment{Accept with probability $\alpha$}
            \Else
                \State $\bm{\theta}_n \gets \bm{\theta}_{n-1}$
                \Comment{Reject with probability $1 - \alpha$}
            \EndIf\label{line-hmc-endacceptancecriterion}
        \EndFor
    \end{algorithmic}
\end{algorithm}

\section*{Acknowledgements}
This research was partially conducted as part of the DAMOCLES project within the EMDAIR program of the Eindhoven Artificial Intelligence Systems Institute (EAISI).

This article is an expanded version of a paper presented at 
\textcolor{black}{the 43$^\text{rd}$ International Workshop on Bayesian Inference and Maximum Entropy Methods in Science and Engineering} on July 1, 2024 in Ghent, Belgium. \textcolor{black}{The conference paper is available at \href{https://doi.org/10.3390/psf2025012004}{https://doi.org/10.3390/psf2025012004}}.
\section*{Data availability statement}
The experimental data sets are publicly available \cite{rinkens_2025_15784954}.

\bibliographystyle{unsrtnat}
\bibliography{references}

\end{document}